\documentclass[11pt]{article}

\usepackage{natbib}

\usepackage{graphicx, amsmath, amsthm, amssymb,float,setspace,color}
\usepackage[margin=0.8in]{geometry}
\usepackage{cleveref}

\usepackage{bbm}
\usepackage{multirow}
\newtheorem{theorem}{Theorem}%

\newtheorem{lemma}{Lemma}
\newtheorem{proposition}{Proposition}

\newtheorem{assumption}{Assumption}[section]
\newtheorem{cor}{Corollary}
\theoremstyle{definition}
\newtheorem{Remark}{Remark}%
 
\newcommand{\Y}{\mathbb Y} 
\newcommand{\X}{\mathbb X} 
\newcommand{\Z}{\mathbb Z} 
\newcommand{\R}{\mathbb R} 
 
\newcommand{\bepsilon}{\boldsymbol {\epsilon}}
\newcommand{\mf}{\mathbf} 
\newcommand{\GG}{\mathcal G} 
\newcommand{\FF}{\mathcal F} 
\usepackage{autonum}
\makeatletter
\autonum@generatePatchedReferenceCSL{Cref}
\makeatother
\usepackage{etoolbox} %

\newcommand\hoaxref[1]{%
}
\makeatletter
\autonum@generatePatchedReference{hoaxref} %
\makeatother

\renewcommand{\baselinestretch}{\vv}

\usepackage{setspace}
\title{A Random Graph-based Autoregressive Model for Networked Time Series}
\author{Weichi Wu and Chenlei Leng
  \footnote{Wu is with TCSS, Tsinghua University.  Leng is with Department of Statistics, 
    University of Warwick (C.Leng@warwick.ac.uk). The authors thank Prof. Eric Auerbach for sharing the code in \cite{auerbach2022identification} %
    and Xierui Song and Dexun Shang for collecting the Sina Weibo data.  } 
}
\date{\today}
\begin{document}
\renewcommand{\baselinestretch}{1}
\maketitle

\def\baselinestretch{1}\selectfont
\begin{abstract}
Contemporary time series data often feature objects connected by a social network that naturally induces temporal dependence involving connected neighbours. The network vector autoregressive model is useful for describing the influence of linked neighbours,  while recent generalizations aim to separate influence and homophily. Existing approaches, however, require either correct specification of a time series model or accurate estimation of a network model or both, and rely exclusively on least-squares for parameter estimation. This paper proposes a new autoregressive model incorporating a flexible form for latent variables used to depict homophily. We develop a first-order differencing method for the estimation of influence requiring only the influence part of the model to be correctly specified. When the part including homophily is correctly specified admitting a semiparametric form, we leverage and generalize the recent notion of neighbour smoothing for parameter estimation, bypassing the need to specify the generative mechanism of the network. We develop new theory to show that all the estimated parameters are consistent and asymptotically normal.  The efficacy of our approach is confirmed via extensive simulations and an analysis of a social media dataset.
\end{abstract}
\noindent
{\bf Key Words:} {\it Homophily; Latent variables; Networks;  Semiparametrics; Social influence.}

\section{Introduction}
In many modern, contemporary time series data, entities are often tied in relationships best represented by a network. For example, in social media, we observe at different times the characteristics of users that are linked by leader-follower networks, with a concrete example provided in Section \ref{sec:data}. In neuroscience, responses to stimuli are recorded multiple times for different brain regions of interest that are connected by a functional connection network  \citep{bullmore2009complex,zalesky2010network}. Intuitively, the dependence information encoded by linking patterns in this type of data can be exploited for the modelling of quantities of interest. This paper is about a new model to achieve this.

To fix notation, consider a multivariate networked time series $Y_{i,t} \in \mathbb{R}$, the observed response of entity $i$ at time $t$. These entities are linked via a time-invariant network represented as an adjacency matrix $A=(A_{ij})$. As in a typical regression setup, $Y_{i,t}$ is likely influenced by the covariate at the same node denoted as $U_i \in \mathbb{R}^p$. In the social media data example in Section \ref{sec:data} that motivated this paper, $U_i$ denotes the gender and number of social labels of each account.

For time series, in addition to $U$, the evolving pattern of $Y_{i,t+1}$ may also depend on its predecessors $Y_{i,t-k}, k=0, 1, ...$, and neighbors.  Towards this,  \cite{zhu2017network} first proposed the network autoregression model (NAR) that admits the following form
\begin{equation}
 Y_{i, t+1} = \alpha Y_{i,t} +
	\beta\frac{\sum_{j}{\left( Y_{j,t}A_{ij}\right)}}{\sum_{j}{A_{ij}}} +
	 \gamma^\top U_i + \epsilon_{i,t+1},\label{eq:nar}
\end{equation}
with parameters $\alpha\in\mathbb{R}, \beta\in \mathbb{R}$ and $\gamma \in \mathbb{R}^p$, where $\epsilon_{i,t+1}$ is a white noise. An attractive feature of NAR is that in addition to the autoregressive process $\alpha Y_{i,t}+\gamma^\top U_i$, the \textit{influence} of those objects connected to node $i$ is explicitly accounted for in the term $\frac{\sum_{j}{\left( Y_{j,t}A_{ij}\right)}}{\sum_{j}{A_{ij}}}$ that assesses the effect of neighbours in a lag-one manner along the temporal dimension. Because of this, the parameter $\beta$ will be referred to as the {\it influence} parameter. Since $\alpha$ models the dependence of $Y_{i,t+1}$ on its immediate predecessor, following \cite{zhu2017network}, we will refer to it as the \textit{momentum} parameter. On the other hand, $\gamma$ is the usual covariate parameter for explicitly modelling the effect of the covariate at the node level.

Despite the explicit characterization of social influence in $\beta$, NAR overlooks \textit{homophily}, an important effect of networks. Homophilous effect arises when network ties are more likely to form between objects with similar nodal features, of which the response $Y_{i,t}$ itself is a part. In the context of our time series data with a network structure, that is, on one hand, $Y_{i,t}$ may depend on its linked neighbours such that the responses of linked nodes become similar; on the other hand,  this similarity in $Y_{i,t}$, which can be seen as a trait of node $i$, is the reason that nodes similar in $Y_{i,t}$ are connected.  In this sense, incorporating network and covariates effects alone as in \eqref{eq:nar} while leaving out the network forming mechanism may not fully explain the response. However, it is found that developing models to identify influence and homophily in general is not possible in a single snapshot of the network, unless all the nodal attributes relevant to tie-formation and characteristics of interest are observed \citep{shalizi2011homophily}. In networked time series, \cite{mcfowland2021estimating} found that these two effects can be separated with proper assumptions. The key idea 
is to dissect network forming process with the use of latent variables  to account for homophily. 

Formally, \cite{mcfowland2021estimating} augmented NAR as
\begin{equation}	Y_{i,t+1} = \alpha Y_{i,t} +
	\beta\frac{\sum_{j}{\left(Y_{j,t}A_{ij}\right)}}{\sum_{j}{A_{ij}}} +
	\gamma^\top U_{i} +\delta^\top C_{i} + \epsilon_{i,t+1},\label{eq:ms}
\end{equation}
where $C_i$ is the latent variable associated with node $i$. These nodal variables $C_i$ are constructed such that the propensity of $A_{ij}=1$, or nodes $i$ and $j$ making a connection, depends on $C_i$ and $C_j$. The idea underlying the augumented NAR model is that homophily of players in a network can be often captured by the proximity of their latent positions. By feeding these homophily-informative latent variables into NAR, we will be able to separate the effect of influence due to connected neighbours as a parameter $\beta$ and the effect of homophily as a parameter $\delta$, once homophily is accounted for by $C_i$'s. 

In \eqref{eq:nar} and \eqref{eq:ms}, the influence of the neighbours on any node is their average. While this is relatively easy to justify and intuitively appealing, the linear form $\gamma^\top U_{i} +\delta^\top C_{i}$ in  \eqref{eq:ms} imposed  may not fully capture the dependence of a characteristic on homophilous effects. Moreover, for many latent variable models used for modelling networks for example the graphon model, these latent variables are unidentifiable \citep{bickel2009nonparametric} or often of secondly interest. 
In view of these shortcomings, we propose the following new model
\begin{equation}	Y_{i,t+1} =  \alpha Y_{i,t} +
	\beta\frac{\sum_{j}{\left( Y_{j,t}A_{ij}\right)}}{\sum_{j}{A_{ij}}} +
	 \tilde{f}(C_i, U_i)+ \epsilon_{i,t+1}, \label{singlebeta}
\end{equation}  
with an unspecified function $\tilde{f}$ documenting the effects of observed covariates $U_i$ and latent traits $C_i$. To distinguish our model from NAR, we will refer to the model in \eqref{singlebeta} as the Random Graph-based Autoregressive Model (RGAM), due to the fact that our approach explicitly exploits the randomness of the underlying network for model formulation, parameter estimation and statistical inference.

\subsection{Literature review}
Networked time series data  are becoming increasingly available. Intuitively, connected individuals in these data will influence each other's characteristics. Such influence is often referred to as social influence, contagion, social interaction, interference, peer effects and spillover effects, to name just a few. To incorporate such influence in a time series model, \cite{zhu2017network} first proposed the network autoregression model (NAR) that relates momentum, influence and observed covariates using an autoregressive type formulation as in \eqref{eq:nar}. This line of research has attracted substantial attention in recent years.  
\cite{zhu2019network} extended NAR to study the conditional quantiles. \cite{chen2022community} enlarged the notion of influence by connection to influence based on communities of nodes. \cite{zhu2020grouped} developed a mixture NAR model. \cite{huang2020two} proposed a NAR model for bipartite networks. \cite{zhou2020network} studied a GARCH model utilizing the network structure for describing conditional variance, while \cite{li2023inference} considered time-varying NAR by allowing $\alpha$ and $\beta$ in \eqref{eq:nar} to change with time.  For regression analysis of single snapshots of network-linked data in various contexts, we refer readers to \cite{li2019prediction,su2019testing,le2022linear}. None of these models considered how network forming process affects a response of interest.

In addition to the potential dependence brought by connectivity in a network, the  presence of ties brings another major challenge to regression analysis with a network structure.  It is well known that social ties are more likely to form between objects with similar traits, a hallmark of many networks known as homophily  \citep{mcpherson2001birds}. To separate influence from homophilous effect in networked time series, \cite{mcfowland2021estimating} explicitly included the term $C_i$. See also \cite{peixoto2022disentangling,cristali2022using,leung2019distinguishing}. Inevitably, this approach requires  the modelling of network forming process, thus relying on the large literature on  statistical analysis of social networks  \citep{goldenberg2010survey,kolaczyk2014statistical,de2017econometrics,newman2018networks}. In particular, many network data are assumed to be randomly generated from probabilistic models that include the stochastic block models \citep{holland1983stochastic}, the latent space model \citep{hoff2002latent},  graphon models \citep{aldous1981representations, lovasz2012large}, and other general exchangeable models  \citep{orbanz2014bayesian,caron2017sparse}. 

An attractive feature of the time series models in \eqref{eq:nar} and \eqref{eq:ms} 
 lies in the fact that one can empirically evaluate the dependence of $Y_{i, t+1}$ on  $Y_{j, t}$ when $i$ are $j$ are connected.  However, NAR in \cite{zhu2017network} ignores homophily altogether, while the model in 
  \cite{mcfowland2021estimating} relies on a correctly specified network model. Particularly, the latter paper requires  plugging in an estimator of the latent variables in \eqref{eq:ms}, for which a generative model for the network is required.  Moreover,  both paper depend exclusively on least-squares for the estimation of the parameters that critically depends on correct specification of their models.
 
 We remark that the inclusion of $C_i$ in \eqref{eq:ms} and \eqref{singlebeta} is motivated by a fundamental result in graph theory. The network structure captured by the adjacency matrix corresponds to a graph denoted as $G_N=(V_N, E_N)$, where $V_N=\{1, ..., N\}$ is the node set and $E_N=\{(i,j): A_{ij}=1 \}$ is the set of vertices. For simplicity, we assume that $G_N$ is an undirected simple graph without self-loops; that is, $A_{ii}=0$ and $A_{ij}=A_{ji}$. 
 For the adjacency matrix  $A$, a discrete object, to have a limit in a continuous space independent of the network size, a suitable generative representation for $A$ is that it is exchangeable, in the sense that the distribution of edges in $A$ remain invariant to any permutation of the labelling of its indices. The Aldous-Hoover representation of exchangeable networks  \citep{lovasz2012large,aldous1981representations,hoover1979relations} fully characterizes this exchangeability as
\begin{equation}\label{graphon}
	A_{ij}=\mathbbm 1( \eta_{ij}\leq g(C_i,C_j) ),
\end{equation}
where $\mathbbm 1(\cdot)$ is the usual indicator function, $g(\cdot, \cdot)\in [0,1]$ is an unknown symmetric measurable function known as graphon, and $C_i, C_j$ and $\eta_{ij}$ are independent  random variables. We remark that graphons  as a unified framework for dealing with network data have gained  substantial interest in recent years \citep{bickel2009nonparametric,airoldi2013stochastic, gao2015rate, li2020network,gao2021minimax}.

 \subsection{Contributions}
 In view of the shortcomings on the existing approaches outlined above, the methodological contribution of this paper is the RGAM in \eqref{singlebeta} which is a flexible model featuring homophilous effect via latent variables. Compared to the simpler model \eqref{eq:ms}, our model  leaves the functional form $\tilde{f}$ of $U_i$ and $C_i$ totally unspecified.  
 
 For our new model, we develop new methods for the estimation of the parameters in RGAM. Specifically, noting that first-order differencing $Y_{i,t+1}-Y_{i,t}$ cancels out $\tilde{f}(C_i,U_i)$, we employ  instrumental variables for the estimation of the influence parameter $\beta$ and the momentum parameter $\alpha$. This estimation does not depend on the specific form of $\tilde{f}(C_i,U_i)$ and thus is more robust than the least-squares estimators in  	\cite{zhu2017network}  and  \cite{mcfowland2021estimating}. One consequence of our estimation strategy is that our method gives asymptotically normal estimators even when $N$ is fixed as long as $NT \rightarrow \infty$, while  \cite{zhu2017network} and \cite{mcfowland2021estimating} require $N\rightarrow \infty$. 
 When $\tilde{f}(C_i,U_i)$ admits the following semiparametric form
 \begin{equation}
  \tilde{f}(C_i, U_i) =f(C_i)+\gamma^\top U_i, \label{eq:plm}
 \end{equation}
 where $f$ is  some unknown function, we propose the estimation of the covariate parameter $\gamma$ and the homophily effect $f(C_i)$ as a single parameter based on the notion of neighbourhood smoothing \citep{zhang2017estimating},  particularly inspired by the recent result in \cite{auerbach2022identification}  that  connects a graphon model and a partial linear model. 
 We show that all our estimators are consistent and asymptotically normal. 
  
 In studying the properties of our estimators, we develop new theory based on new assumptions. One new set of assumptions on the network connect to its geometric properties, via the bottleneck ratio of the weighted adjacency matrix that measures the total number of length-two paths for example. These assumptions are somewhat more transparent than the ones in \cite{zhu2017network}. 
 
 To establish the properties of our estimators, we substantially extend and generalize existing technical tools. Specifically,  
  for estimating $\alpha$ and $\beta$,  the martingale difference structure utilized by \cite{zhu2017network} for their proofs is no longer applicable \citep{hall1980martingale} and tools developed in the field of dynamic panel data do not consider cross-sectional dependence in networks \citep[cf.]{anderson1981estimation,arellano1991some} {allowing both $N$ and $T$ to diverge}. On the other hand, for the estimation of $f(C_i)$ and $\gamma$, we find that the estimation errors for estimating $\alpha$ and $\beta $ propagate. Therefore the theory of $U$-statistics, for example developed in Lemma A.3 of \cite{ahn1993semiparametric} used in \cite{auerbach2021identification}, is invalid in our situation.  Even though there are a few works on residual-based $U$-statistics related to our estimators, for example in \cite{randles1984tests},  those methods are designed for $i.i.d.$ samples not suitable for our setup. To make progress,   we develop new $m$-dependent approximation to establish the central limit theory (CLT) based on $m$-dependent series \citep{janson2021central}. {Our results provide a useful device for the analysis for large dynamic panels with latent  and network structures, allowing growing $N$ and $T$}.

 In validating our estimation method, we conduct extensive simulation study and a data analysis to verify our methodology. We remark that neither \cite{mcfowland2021estimating}  nor \cite{auerbach2021identification} provided data examples to guide the use of their models.

\subsection{Notations and organization of the paper}
We collect the main notions used in the paper here. For a symmetric matrix $B\in \mathbb{R}^{d\times d}$, we arrange its eigenvalues in the decreasing order as $\lambda_1(B)\ge \lambda_2(B) \ge \cdots \ge \lambda_d(B)$. We sometimes denote the smallest eigenvalue as $\lambda_{min}$ for better clarity. We denote the $d\times d$ dimensional identity matrix as $I_d$. From the graphon representation of a network, we write $\mathcal G_N=\{\{C_i, U_i\}_{1\leq i\leq N}, \{\eta_{ij}\}_{1\leq i<j\leq N}\}\}.$ 
Denote $n_i=\sum_{j=1}^NA_{ij}$ as the degree of the $i$th node, $w_i=(A_{ij}/n_i, 1\leq j\leq N)^\top\in \mathbb R^{N}$ as the $i$th row of $A$ weighted by the degree of $i$, and $f_i=f(C_i)+\gamma^\top U_i$ if $\tilde{f}$ assumes a partial linear form as in \eqref{eq:plm} or with some abuse of notation $f_i= \tilde{f}(C_i, U_i)$ for the general case in \eqref{singlebeta}. Write  $\Y_t=(Y_{1,t}, ..., Y_{N,t})^\top$ as the response vector at time $t$,  $W_N=(w_1,...,w_N)^\top$ as the row-normalized adjacency matrix,  $\bepsilon_t=(\epsilon_{1,t},....,\epsilon_{N,t})^\top\in \mathbb R^N$ as the random error vector at time $t$,  and $\mf f=(f_1,....,f_N)^\top$. Using these notation, we can write \eqref{singlebeta} as
\begin{align}\label{Y_(t+1)}
	\Y_{t+1}=G_0+G_1\Y_{t}+ \bepsilon_{t+1},~~~~0\leq t\leq T-1,
\end{align}
where $G_0=\mf f$ and $G_1=\alpha I_N+\beta  W_N$.  Notice that the coefficients $G_0$, $G_1$ depend on $N$, but we omit this dependence in the notation below for brevity if no confusion is caused. In the sequel, we write $A$ sometimes as $A_N$ to indicate its  dependence on $N$. Denote $e_i$ as an $N\times 1$ vector with the $i${th} element $1$ and $0$ otherwise.  For random variable $x$, let $\|x\|_v=(E|x|^v)^{1/v}$ be the $\mathcal L^v$ $(v>0).$%
The notation $\|\cdot\|_2$ is also used for the $\mathcal L^2$ norm of functions in $L^2[0,1]$, i.e.,  if $f(\cdot)\in L^2[0,1]$ then $\|f\|_2=(\int_0^1 f^2(x)dx)^{1/2}$. For convenience, if $f(\cdot)$ is a random function such that $\int_0^1 f^2(x)dx<\infty$ almost surely, we still denote $\|f\|_2=(\int_0^1 f^2(x)dx)^{1/2}$ in which case 
$\|f\|_2$ is a random variable.
For any matrix $A$, let $|A|_F$ be its Frobenius norm.

The rest of the paper is organized as follows. We outline our estimation approach in Section \ref{sec:estimation} and discuss the assumptions in Section \ref{sec:assumption}. The properties of the estimators are presented in Section \ref{sec:asymptotic}. Simulation is provided in Section \ref{sec:simulation} and a data analysis is presented in Section \ref{sec:data}. Section \ref{sec:conclusion} outlines future work. All the proofs are relegated to the Supplementary Materials.

\section{Estimation Method} \label{sec:estimation}

We first state the stationarity of the network time series. Given the network  $A_N$, $\Y_t$ in \eqref{singlebeta} is strictly stationary as long as $|\alpha|+|\beta|<1$. To see this, notice that by \cite{banerjee2003hierarchical}, $\max_{N}|\lambda_{1}(W_N)|\leq 1$.  Therefore the condition $|\alpha|+|\beta|<1$ implies 
$\max_{N}|\lambda_1(G_1)| \leq |\alpha|+|\beta||\lambda_1(W)| < 1$. 
Following the proof of Theorem 1 of \cite{zhu2017network}, we have 
\begin{proposition}\label{prop1}
	Assume $|\alpha|+|\beta|<1$ and $E f_i^2<\infty$ for $1\leq i\leq N$. Then given $\mathcal G_N$ and if $( \bepsilon_i)_{i \in \mathbb Z}$  is independent of $\GG_N$, equation \eqref{singlebeta} has a unique strictly stationary solution that can be written as
	\begin{align}\label{stationary-solution}
		\Y_t=(I-G_1)^{-1}G_0+\sum_{j=0}^\infty G_1^j\bepsilon_{t-j}, ~~a.s..
	\end{align}
\end{proposition}
The following theoretical results build on the stationary solution in \eqref{stationary-solution}.  By \Cref{lemma4} in the Supplementary Materials, \Cref{prop2} and iteration argument, it can be shown actually that the form of $\Y_0$ is not important as long as the time series is long, if $\max_{1\leq i\leq N}E Y^2_{i,0}<\infty$. 

\subsection{Estimation of $\alpha$ and $\beta$}\label{sec:alpha and beta}
Denote the first-order differencing operator as $\Delta$ and write
\begin{align} \label{modelvector}
	\Delta Y_{i,t}=Y_{i,t}-Y_{i,t-1}, ~~\Delta \Y_t=(\Delta Y_{1,t},...,\Delta Y_{N,t})^\top,~~
	X_{i,t}=(\Delta Y_{i,t}, w_i^\top\Delta \Y_t)^\top.
\end{align}
By first-order differencing \eqref{singlebeta} we obtain
\begin{align}\label{deffierencing}
	\Delta \Y_{t+1}=\X_t\theta+\Delta \bepsilon_{t+1},~~~~ 1\leq t\leq T-1,
\end{align}
where  $\Delta \bepsilon_t=\bepsilon_t-\bepsilon_{t-1}$,  $\X_t=(X_{1,t},...,X_{N,t})^\top\in \R^{N\times 2}$, and $\theta=( \alpha, \beta)^\top$. 
Since differencing induces endogeneity due to the correlation between $\Delta\bepsilon_{t+1}$ and $\Y_t$, the usual least-squares estimator is inconsistent. An approach to overcome this is via the use of  instrumental variables.    Towards this, write the historical observations $Z_{i,t}=( Y_{i,t-1}, w_i^\top\Y_{t-1})^\top$ and $\Z_t=(Z_{1,t},...,Z_{N,t})^\top\in \R^{N\times 2}$, where $Z_{i,t}$ is  independent of $\Delta \bepsilon_{t+1}$. Using $Z_{i,t}$ as an instrument, we propose the following estimator of $\theta$ 
\begin{align}\label{eq9theta}
	\hat \theta:=(\hat \alpha, \hat \beta)=(\sum_{t=1}^{T-1}\Z_t^\top\X_t)^{-1}\sum_{t=1}^{T-1}\Z^\top_t\Delta \Y_{t+1}=\theta+(\sum_{t=1}^{T-1}\Z_t^\top\X_t)^{-1}\sum_{t=1}^{T-1}\Z^\top_t\Delta \bepsilon_{t+1}.
\end{align}

\subsection{Estimation of $f(C_i)$ and $\gamma$}\label{sec:C and gamma}
After $\alpha$ and $\beta$ are estimated, we profile them out of the model by defining
\begin{align}\label{tildee}
	\tilde e_{i,t}=Y_{i,t}-\hat \alpha Y_{i,t-1}-\hat \beta \frac{\sum_j(Y_{j,t-1}A_{ij})}{\sum_j A_{ij}}.
\end{align}
Write $\hat e_{i}=\sum_{t=1}^T \tilde e_{i,t}/T$ and likewise $e_{i}=\sum_{t=1}^T  e_{i,t}/T$, where $e_{i,t}$ is defined by replacing the estimators above by their true values. From our model \eqref{singlebeta} when $\tilde{f}(C_i,U_i)=f(C_i)+\gamma^\top U_i$,
 we have
\[  e_{i} = f(C_i)+\gamma^\top U_i +\sum_{t=1}^T \epsilon_{i,t}/T,
\] 
which is a partial linear model \citep{robinson1988root, hardle2000partially}. What is different to the usual partial linear model, however, is that $C_i$ is not observed. One way to proceed is to estimate $C_i$ first assuming a network model, as is done in \cite{mcfowland2021estimating}. However, this estimation approach ties  to the specification of a network model that can be difficult to justify. Instead, we assume that $C_i$'s are latent variables in a general graphon model that underpins many network processes \citep[cf.]{bickel2009nonparametric}, but our estimation procedure does not require their estimation as seen below.
 
 Immediately we note that by taking expectation we have 
 \[\mathbb{E} (e_i-e_j|C_i=C_j, U_i, U_j) = \gamma^\top (U_i-U_j), \]
 which motivates our estimation method. Since typically the number of $C_i$'s taking the same values does not grow, an idea is to aggregate those $C_i$ and $C_j$ close in a certain sense. For those $C_i$'s, with some assumption on $f$, roughly we have $f(C_i) \approx f(C_j)$ which enables one to eliminate $f(C_i)$ by taking pairwise differences as
\[ \hat e_{i} -\hat e_{j} \approx \gamma^\top(U_i-U_j),
\] as long as $\alpha$ and $\beta$ are consistently estimated. 
Thus, an estimator of $\gamma$ can be obtained by the least-squares estimator of $\hat e_{i} -\hat e_{j}$ on $U_i-U_j$ known as the pairwise difference estimator \citep{honore1997pairwise}. 
Formally, this least-squares estimation will be weighted by a suitable measure of closeness between $C_i$ and $C_j$. Towards this, define the following measure 
\begin{align}\label{estimate-deltaij}
	\hat \delta_{ij}=\left(\frac{1}{N}\sum_{t=1}^N\left(\frac{1}{N}\sum_{s=1}^NA_{ts}(A_{is}-A_{js})\right)^2\right)^{1/2},
\end{align} 
which assesses the closeness of the rows of $A$ in some sense. A more detailed discussion of its meaning is deferred until Section \ref{sec:regression identification}. 
Following \cite{auerbach2022identification} and \cite{honore1997pairwise},  we estimate $\gamma$ using kernel-weighted least-squares and $f(C_i)$ respectively as
\begin{align}\label{estimate-gamma}
	\hat \gamma=\left(\sum_{i=1}^{N-1}\sum_{j=i+1}^N(U_i-U_j) (U_i-U_j)^\top K(\frac{\hat \delta_{ij}^2}{h_N})\right)^{-1}
	\left(\sum_{i=1}^{N-1}\sum_{j=i+1}^N(U_i-U_j) (\hat e_i-\hat e_j) K(\frac{\hat \delta_{ij}^2}{h_N})\right),\\
	\label{estimate-f}
	\widehat{f(C_i)}=(\sum_{t=1}^NK(\frac{\hat \delta_{it}^2}{h_N}))^{-1} \left(\sum_{t=1}^N(\hat e_t-\hat \gamma^\top U_t)K(\frac{\hat \delta_{it}^2}{h_N})\right),\end{align}
where $K(\cdot)$ is a kernel function with bandwidth $h_N\rightarrow 0$.

\section{Theoretical Assumptions}\label{sec:assumption}

\subsection{Assumptions on the network}
For the estimation of $\alpha$ and $\beta$, 
we discuss assumptions on the random network $A$ or $W_N$ equivalently.
 For any given $N\geq 2$, $W_N$ is a transition matrix for simple random walk on a graph with adjacency matrix $A$ \citep[Section 1.4]{levin2017markov}, and its stationary distribution is 
\begin{align}\label{barpi}
	\bar \pi_N=(\pi_1,....,\pi_N), ~~ \pi_i=\frac{n_i}{2|E_N|},
\end{align}
where $|E_N|$ is the number of edges of the network since $\bar \pi_N W_N=\bar \pi_N$. As shown in Example 1.21 of \cite{levin2017markov}, $W_N$ is a reversible transition matrix. {In the following, we state the assumptions on network when its size diverges. The conditions when the size is fixed are simpler and are listed in \Cref{Theorem1}.}

\begin{assumption}[Connectivity]\label{Graph1}
	As $N$ diverges, with probability approaching 1, the  graph $G_N$ for the network  is connected, and is not a bipartite graph. 
\end{assumption}
Since $W_N$ is a reversible transition matrix, all its eigenvalues are real \citep[Section 12.1]{levin2017markov}. 
Assumption \ref{Graph1} is a  sufficient and necessary  condition for $W_N$ to be irreducible and aperiodic \citep[Chapter 7]{mitzenmacher2017probability}. This assumption assures 
$	\lim_{N\rightarrow \infty}P(\min_{1\leq i\leq N}\pi_i\geq (N(N-1))^{-1})=1$, and that the largest eigenvalue of $W_N$ is $1$ while 
all the other eigenvalues are in $(-1,1)$. We make the following assumption regarding these eigenvalues.
\begin{assumption}[Sparsity]\label{Graph2}
	Assume that the absolute spectral gap $\gamma_N^\star=1-\lambda_N$ satisfies 
	$\gamma_N^\star\geq \alpha(\log N)^{-\alpha_0}$ for some $\alpha_0, \alpha\geq 0$, 
	where $\lambda_N$ is the largest eigenvalue of $W_N$ in absolute value different from $1$. 
\end{assumption}
The absolute spectral gap $\gamma_N^\star$ is closely related to the bottleneck ratio of a transition matrix \cite[Chapter 7]{levin2017markov}. 
For any vertex set $S$ and its complement $S^c$, the bottleneck ratio of  $S$ is defined as
\begin{align}
	\Phi_B(S)=\frac{\sum_{i\in S, j\in S_c}\pi_iW_N^2(i,j)}{\sum_{i\in S}\pi_i}=
	\frac{\sum_{i\in S, j\in S_c}\sum_k a_{ik}a_{kj}/n_k}{\sum_{i\in S}n_i}=\frac{\sum_k (\sum_{i\in S, j\in S_c}a_{ik}a_{kj})/n_k}{\sum_{i\in S}n_i},
\end{align}
which measures 
{the total number of length-two paths between set $S$ and $S^c$ weighted by the degrees of the nodes that the paths pass through. The 
bottleneck ratio of $W_N^2$ itself is defined as 
$	\Phi_\star=\min_{S: \sum_{i\in S}\pi_i\leq 0.5} \Phi_B(S)$. 
Notice that $W^2_N$ is also a reversible transition matrix and that $\lambda_N^2$ is the second largest eigenvalue of $W_N^2$. By Theorem 13.10 of \cite{levin2017markov}  (see also \cite{lawler1988bounds} and \cite{sinclair1989approximate}), a sufficient condition for Assumption \ref{Graph2} to hold is 
$	\Phi^\star\geq \sqrt{2(1-[1-\alpha(\log N)^{-\alpha_0}]^2)}, $
which requires that the network is well connected having sufficiently many length-two paths. %
The introduction of Assumption \ref{Graph2} on the spectral gap strengthens Lemma 2 of \cite{zhu2017network} that is crucial to establish a theory for their NAR. In the proof of this lemma,  \cite{zhu2017network} 
requires that there exists a fixed integer $K>0$ such that 
$W_N^n\preceq C\mf 1_N \bar{\pi}_N^\top$ 
for some constant $C$ and all $n\geq K$, %
where $\mf 1_N$ is the $N$-dimensional column vector of $1$'s, and $\preceq$ denotes entrywise $\le$. This requirement  is appropriate when $N$, the size of a network, is bounded, but may be restrictive if one allows $N\rightarrow \infty$. To achieve more flexibility, \Cref{Prop-mix} of this paper establishes a probabilistic argument for this requirement to hold allowing $K$ to vary slowly with $N$  by using  Assumption \ref{Graph2}}.

\begin{assumption}[Influential points]\label{Graph3}
Assume	(i) $\sum_{i=1}^N\pi_i^2=\sum_{i=1}^N \frac{n^2_i}{4|E_N|^2}=o_p(1)$ as $N\rightarrow \infty$; (ii) There exists a polynomial $M(x)$ such that
	with probability approaching $1$, $|W_N^j|_F\leq N^{1/2}M(j)c_N$ as $N\rightarrow \infty$, where $c_N$ is a sequence satisfying $c_N=o(1)$. 
\end{assumption}

Condition (i) on  small probability mass $\pi_i$ excludes sparse graphs with few high degree nodes. {Condition (ii) is a mild technical assumption. When $j=1$ such that $|W_N|_F=\sqrt{\sum_{i=1}^N\frac{1}{n_i}}$,  (ii) requires that the network does not contain too many nodes with small degrees. When $j\rightarrow \infty$,  By the discussion below \Cref{Graph1}, the spectral radius of $W_N$ (the largest absolute eigenvalue), denoted by $\rho(W_n)$ here,  equals $1$.  By the Gelfand formula, for any $N\in \mathbb Z^+$, $\rho(W_N)=1=\lim_{j\rightarrow \infty} |W_N^j|_F^{1/j}$. Consider a gemoetric bound  such that $|W_N^j|_F\leq N^{1/2}a^j$. Clearly it is infeasible if $a<1$, too loose if $a>1$, and too restrictive if  $a=1$.  
Hence, the polynomial upper bound $M(j)$ in (ii) is a mild condition that allows $|W_N^j|_F$ to increase with $j$.}
\begin{Remark}\label{rmkzhucondition} %
For NAR in \cite{zhu2017network}, a uniformity assumption $\lambda_{\max}(W_N+W_N^\top)=O(\log N)$ is imposed. We replace this assumption with Assumption \ref{Graph2} on the sparsity of the graph and Assumption \ref{Graph3} on its  influential nodes, which are somewhat more intuitive. 
	\end{Remark}

\subsection{Regression identification}\label{sec:regression identification}
In this subsection, we provide assumptions needed for studying the estimation of the parameters.  
To allow a flexible parametrization of the network process, recall that we assume a graphon model $A_{ij}=\mathbbm 1( \eta_{ij}\leq g(C_i,C_j) )$ with $\eta_{ij}=\eta_{ji}$ in \eqref{graphon}. Apparently, the estimation of $f(C_i)$ and $\gamma$ is only feasible with appropriate restriction on the graphon $g$ which is a symmetric measurable function. We make the following assumption taken from  \cite{zhang2017estimating}.
\begin{assumption}[Graphon]\label{Assumption2.1}
{Assume that $g(u,v)$ is the piecewise Lipschitz graphon in the sense that  there exists an integer $K\geq 1$, positive constants $L$ and $l$, and a sequence $0=z_1<...<z_K=1$ with $\min_{0\leq s\leq K-1} (z_{s+1}-z_s)\geq l$, such that $|g(u_1,v)-g(u_2,v)|\leq L|u_1-u_2|$ for all $v\in[0,1]$ and all $u_1,u_2\in (z_t, z_{t+1})$, $0\leq t\leq K-1$. Moreover, $g(\cdot, \cdot)$ is symmetric satisfying $0\leq \inf_{u,v\in[0,1]}g(u,v)\leq \sup_{u,v\in[0,1]}g(u,v)\leq 1$.}
 \end{assumption}
 We also make the following modelling assumptions.
 \begin{assumption}[Model]\label{Assumption2.1_2}
In the graphon model for $A_{ij}$,  the random variables $\{\eta_{ij}\}_{i,j=1}^N$ and $\{C_i\}_{i=1}^N$are $i.i.d.$ within themselves respectively all having marginal distribution $U(0,1)$. In the RGAM, the random errors $\epsilon_{i,t}$'s are $i.i.d.$ with variance $\sigma^2$ and have finite eighth moments. The covariates $(U_i, C_i)$'s are $i.i.d.$ with $\sup_{u\in[0,1]}|E(U_i|C_i=u)|<\infty$. Finally, the random variables  $(U_i, C_i)$, $\eta_{ij}$ and $\epsilon_{i,t}$ are independent of each other.	\end{assumption}

We remark that the distribution and mutual independence assumption made on $\eta_{ij}$ and $C_i$ are the usual ones on networks. The assumptions on the covariate $(U_i, C_i)$ and its relationship with other random variables including the random errors $\epsilon_{i,t}$ are standard. 
By Assumption \ref{Assumption2.1_2}, $\|f(C_i)+\gamma^\top U_i\|_2<\infty$. Therefore if $|\alpha|+|\beta|<1$, then by Proposition \ref{prop1} and Assumptions \ref{Assumption2.1} and \ref{Assumption2.1_2}, it follows that 
$|E(\Y_t\bepsilon_{t+1}^\top|\GG_N)|_F=0$ almost surely 
for $1\leq t\leq T-1$. This is needed for the identification of $\alpha$ and $\beta$. %

\begin{assumption}[Variables]\label{Assumption2.2}
	We assume $\inf_{u\in [0,1]}\lambda_{min}(Cov(U_1|C_1=u))>0$ where 
	\begin{align}
		Cov(U_1|C_1=u)=E((U_i-E(U_i|C_i))^\top(U_i-E(U_i|C_i))|C_i=u).
	\end{align} 
\end{assumption}
	This assumption states that the conditional covariance of $U_i$ given $C_i=u$ is positive definite, which is a standard assumption made for partial linear models \citep{hardle2000partially}. It requires that the nodal covariates $U_i$ should reflect information different from  that carried by  $C_i$. If $U_i$ and $C_i$ are independent, the conditional covariance becomes unconditional. Since the network is determined by the latent variables, \Cref{Assumption2.2} also prevents strong dependence between the network and the nodal covariates in RGAM in  \eqref{singlebeta}. 

\begin{assumption}[Neighbourhood]\label{Assumption2.3}
	For every $\epsilon>0$ there exists $\delta>0$ such that for all $u, v$ satisfying $\|g(u,\cdot)-g(v,\cdot)\|_2=\{\int_0^1(g(u,x)-g(v,x))^2dx\}^{1/2}\leq \delta$, we have 
		$( f(u)-f(v))^2\leq \epsilon. $
	\end{assumption}\
This assumption is some kind of smoothness condition on $f$ in relationship to the graphon $g$ that enables us to estimate $f(C_i)$. To appreciate how it is useful for our estimator defined in Section \ref{sec:C and gamma}, we note that Assumption \ref{Assumption2.3} implies that as long as $g(C_i, \cdot)$ and $g(C_j, \cdot)$, the graphon slices of $C_i$ and $C_j$, are close, it holds $f(C_i)\approx f(C_j)$. Thus, to identify those $f(C_i)$'s that are close, we just need to examine $g(C_i, \cdot)$. \cite{zhang2017estimating} initiated this idea for the estimation of $\mathbb{E}(A)$, the probability matrix, by smoothing neigbours of nodes that are close in their graphon slices. \cite{auerbach2022identification} took this idea further by defining another closeness measure named codegree that can upper and lower bound the $\mathcal L^2$ distances between graphon slices.

Though $f(C_i)$ in our model is estimated by neighbourhood smoothing in the similar spirit as \cite{zhang2017estimating} and \cite{auerbach2022identification}, the objects for smoothing in our setting are the residuals obtained after plugging in the estimator of $\alpha$ and $\beta$ defined in Section \ref{sec:alpha and beta}. This rules out a direct application of the tools based on, for example, U-statistics in \cite{auerbach2022identification}. 
 To tackle the challenges that arise, we develop new theory based on delicate analysis of residual-based smoothing U-statistics.  We emphasize that results on U-statistics constructed by residuals are scarce, while in limited examples  \citep[cf.]{randles1982asymptotic, randles1984tests}, the assumptions needed are not satisfied under the scenario considered in this paper.

\section{Asymptotic Results}\label{sec:asymptotic}
We present the results for the estimation of $\theta=(\alpha, \beta)^\top$ first and that of $f(C_i)$ and $\gamma$ next. With some abuse of notation, we denote their true values as $\theta$, $f(C_i)$ and $\gamma$ respectively whenever no confusion arises.

\subsection{Asymptotic theory for $\hat \theta$ }
Intuitively for the consistent estimation of $\theta$, we just need $T \rightarrow \infty$ because there is enough information in the times series for its estimation. Our first theorem concerns this estimation  when $N$, the number of nodes in the network, is fixed. %
  \begin{theorem}\label{Theorem1}
Suppose that the  graph $G_N$  is connected and not a bipartite one, and that the matrices $\Sigma_{1,N}$ and $\Sigma_{2,N}$ defined in the Supplementary Materials are full rank. 
 If $|\alpha|+|\beta|<1$, $E \epsilon^4_{i,t}<\infty$,  and  \Cref{Assumption2.2} holds,  %
 then conditional on $\GG_N$, with probability going to $1$, it holds
\begin{align}
	\sqrt{T-1}\Sigma_N(\hat \theta-\theta)/\sigma\Rightarrow N(0,I_2)
	\end{align}
as $T\rightarrow \infty$, where  ``$\Rightarrow $" denotes the weak convergence and $\Sigma_N=\sqrt N(\Sigma_{2,N})^{-1/2} \Sigma_{1,N}$.

\end{theorem}
We note that with a fixed network size, $\theta$ is $\sqrt{T}$-consistent and asymptotically normal according to the theorem. In contrast, \cite{mcfowland2021estimating} only provided consistency results under the assumption that their network model and autoregessive model are both correctly specified.  \cite{zhu2017network} provided the asymptotic normality of their least-squares estimator in NAR requiring  $N$  to diverge. 

The next theorem concerns the estimation of $\theta=(\alpha,\beta)^\top$ when $\min(N,T)$ is diverging. For this purpose, we assume the following technical conditions.
\begin{assumption}\label{New.4.1}
Assume that
	\begin{description}
		\item (i)  there exists constants $m_0$ and $N_0$ such that for $m\geq m_0$ and $N\geq N_0$, $\Sigma^{(m)}_{1,N}$ and $\Sigma^{(m)}_{2,N}$ are full rank.  {Here $\Sigma^{(m)}_{1,N}$ and $\Sigma^{(m)}_{2,N}$ are some kind of truncated versions of $\Sigma_{1,N}$ and $\Sigma_{2,N}$ corresponding to the associated matrices calculated via $m$-dependent approximation series. The exact definition of those quantities can be found in   the Supplementary Materials.}
		\item (ii) there exists a sequence $\tilde m_n$ such that as $\tilde m_n\rightarrow \infty$, we have $\frac{\tilde m_n}{|\log \lambda_{\min}(\Sigma_{2,N})|+|\log \lambda_{\max}(\Sigma_{2,N})|}\rightarrow \infty$ and 
		$\frac{\tilde m_n^3}{T\lambda^2_{\min}(\Sigma_{2,N})}\rightarrow 0$.
		\item (iii) $T^{1/2}\lambda_{\min}^{1/2}(\Sigma_{2,N})|det(\Sigma_{1,N})|/\lambda_{\max}^{1/2}(\Sigma_{2,N})\rightarrow \infty$.
	\end{description} 
\end{assumption}
{When $N\rightarrow \infty$, as can be seen from the proof of Theorem \ref{diverging N}, $\Sigma_{1,N}$ and $\Sigma_{2,N}$ become degenerate.  \Cref{New.4.1} is introduced such that we can employ state-of-the-art CLT theorem developed for $m$-dependent random variables {\cite{janson2021central}} to establish the following asymptotic normality.}
  \begin{theorem}\label{diverging N}
  Assume that when $N\rightarrow \infty$, with probability approaching $1$, $\Sigma_{1,N}$ and $\Sigma_{2,N}$ are full rank.  
If $|\alpha|+|\beta|<1$, $E \epsilon^4_{1,1}<\infty$, Assumptions \ref{Graph1},  \ref{Graph2}, \ref{Graph3} and \ref{Assumption2.2}  hold, then conditional on $\GG_N$, as $\min(N,T)\rightarrow \infty$,
 with probability going to $1$,
\begin{align}
	\sqrt{N(T-1)}\tilde\Sigma_N(\hat \theta-\theta)/\sigma\Rightarrow N(0,I_2),
\end{align}
where $\tilde\Sigma_N=(\Sigma_{2,N})^{-1/2} \Sigma_{1,N}$.
\end{theorem}

\subsection{Consistency of $\hat\gamma$ and $\widehat{f(C_i)}$ }
We now study the properties of the estimators of $\gamma$ and $f(C_i)$ defined in Section \ref{sec:C and gamma}. Towards this,  we take the kernel $K(\cdot)$ as a nonnegative, twice continuously differentiable function supported on $[0,1)$, as is standard in the nonparametric smoothing literature. The kernel function is used to gather observations  close to each other with respect to the pseudo metric defined as
{\begin{align}\label{deltaxy}
	\delta(x,y)=\Big(\mathbb{E}_{U_1}\mathbb{E}^2_{U_2}\big[g(U_1,U_2)(g(x,U_1)-g(y,U_1))\big]\Big)^{1/2}, ~~x,y\in[0,1],
\end{align}
}where $U_1, U_2$ are independent  $U(0,1)$ random variables, and $\mathbb{E}_{U_1}$, $\mathbb{E}_{U_2}$ are expectations with respect to  $U_1$ and $U_2$ respectively. We assume that the bandwidth satisfies $h_N\rightarrow 0$, $N^{1-\gamma}h_N^2\rightarrow \infty$, $\inf_{u\in [0,1]}N^{\gamma/4}r_N(u)\rightarrow \infty $ as $N\rightarrow \infty$ for some $\gamma>0$ with %
 {$r_N(u)=\int K(\frac{\delta^{2}(u,v)}{h_N})dv$.}
 The assumptions on the bandwidth are made so that the estimator of $\gamma$ is consistent. We have the following consistency results.

\begin{theorem}\label{Thm3}
Under Assumptions \ref{Graph1}--\ref{Assumption2.3} and the conditions in \Cref{diverging N},  if  $NT\lambda_{min}(\Sigma^\top_{1,N}\Sigma^{-1}_{2,N}\Sigma_{1,N})\rightarrow \infty$, 
we have $\hat \gamma-\gamma=o_p(1)$ as $\min(N,T)\rightarrow \infty$.
\end{theorem}

\begin{theorem}\label{Thm4} Under Assumptions \ref{Graph1}--\ref{Assumption2.3} and the conditions in \Cref{diverging N},  if $N^{1/2}T\lambda_{\min}(\Sigma^\top_{1,N} \Sigma^{-1}_{2,N}\Sigma_{1,N})\rightarrow \infty$, we have  $\max_{1\leq i\leq N}|\widehat {f(C_i)}-f(C_i)|=o_p(1)$ as  $\min(N,T)\rightarrow \infty$,
\end{theorem}
{Comparing with the assumption $NT\lambda_{min}(\Sigma^\top_{1,N}\Sigma^{-1}_{2,N}\Sigma_{1,N})\rightarrow \infty$ in \Cref{Thm3}, we require larger $N$ and $T$ for \Cref{Thm4} to hold, due to the need to establish a {\it uniform} consistent  result on $\widehat {f(C_i)}$ for any $1\leq i\leq N$.}
The results in  \Cref{Thm3} and \Cref{Thm4} go substantially beyond the consistency results in \cite{mcfowland2021estimating}. The latter paper estimates $C_i$'s first thus requiring the correct specification of a network model. In contrast, we treat $f(C_i)$ as a single parameter and estimate it based on a general graphon model. 

\subsection{Asymptotic normality of $\hat\gamma$ and $\widehat{f(C_i)}$}\label{asynormal}
Having obtained the consistency of $\hat\gamma$ and $\widehat{f(C_i)}$, we present the results on their asymptotic normality. Establishing asymptotic normality requires assumptions in addition to those needed for consistency. Towards this, we make the following assumption on $g$. 
\begin{assumption}\label{furtherf}
 There exists $\delta>0$ such that for all $u,v$ satisfying $\|g(u,\cdot)-g(v,\cdot)\|_2=\{\int_0^1(g(u,x)-g(v,x))^2dx\}^{1/2}\leq \delta$, we have $f(u)=f(v)$.
\end{assumption}
 
\begin{theorem}\label{Thm5}
	Assume $\lambda_{min}(v_N)>0$ and $\lambda_{min}(\Sigma_{3,N})>0$ where $v_N$ and $\Sigma_{3,N}$ are defined in the Supplementary Materials. Assume that  $r_N=\int r_N(u)du>0$ for $N$ sufficiently large. If the conditions in \Cref{Thm3} and  
	\Cref{furtherf} hold
 	 with probability going to 1, we have that  as $\min(N,T)\rightarrow \infty$, given $\GG_N$,
	 \begin{align}
	 \sqrt{(T-1)N}(v_N+4\Sigma_{3,N}/r_N^2)^{-1/2}\hat \Gamma_1(\hat \gamma-\gamma)\Rightarrow N(0,I_p).
	 \end{align}
	\end{theorem}
We discuss the implication of this Theorem. When $g(C_i,\cdot)$ is piecewise constant for all $C_i$, then \Cref{furtherf} is implied by \Cref{Assumption2.3} since in this case $f(u)=f(v)$ if $g(u,\cdot)=g(v,\cdot)$. Thus Theorem \ref{Thm5} applies to stochastic block models, as their corresponding graphons are piecewise constant. There are other graphons beyond piece constancy ones for which we can apply this theorem. One example is what we call {\it piecewise {smooth}} graphons, as long as they have more than one piece of smooth parts and \Cref{furtherf} holds. {In this case, there exist $r>1$ and $c_1<...<c_r$, such that the graphon $g$ satisfies $\min_{j,k,j\not =k}\inf_{u\in (c_j,c_{j+1}],v\in (c_k,c_{k+1}] }\|g(u,\cdot)-g(v,\cdot)\|_2\geq \eta$ and $\max_j\sup_{u,v\in (c_j,c_{j+1}],u\not =v}\|g(u,\cdot)-g(v,\cdot)\|_2/|u-v|\leq M$ for some positive constant $M$ and $\eta$. Then \Cref{furtherf} holds if $f(\cdot)$ is a step function such that $f(u)=f(v)$ if $u, v\in (c_j,c_{j+1}], 1\leq j\leq r-1$. The following theorem provides the joint asymptotic normality of $\widehat {f(C_i)}'s$. }

\begin{theorem}\label{thm6}
	Assume the conditions of \Cref{Thm5} hold.%
	  Then given $\GG_N$, as  $\min(N,T)\rightarrow \infty$, for any fixed index set $\mathcal C=\{i_1,...,i_{|\mathcal C|}\}\subset \{1,...N\}$ (where $|\mathcal C|$ denotes the cardinality of $\mathcal C$), we have 
	\begin{align}\label{eq44-2023}
	\sqrt{N(T-1)}V_N^{-1/2}\Big((\widehat {f(C_i)}-f(C_i))\Big)_{i\in \mathcal C}\Rightarrow N(0,I_{|\mathcal C|}),
	\end{align}
where $V_{N}$ is defined in the Supplementary Materials.
\end{theorem}
{The proofs of \Cref{Thm5} and \Cref{thm6} are  based on state-of-the-art central limit theorems for $m$-dependent series, properties of U-statistics, and a delicate analysis using residuals since the influence of estimation errors of $\alpha$ and $\beta$ is non-negligible when deriving the asymptotic behavior of $\hat \gamma$ and $\widehat {f(C_i)}$. The results of \Cref{Thm5} and \Cref{thm6} can be used for constructing confidence intervals to quantify the uncertainty of the parameter estimates with their effectiveness justified by our simulation study. Notice that \cite{mcfowland2021estimating} does not provide approaches to construct confidence intervals for their estimators.

\section{Simulation}\label{sec:simulation}

We have conducted extensive simulation to verify our theory.  In this section, we consider the following true model
\[	Y_{i,t+1} =  \alpha Y_{i,t} +
	\beta\frac{\sum_{j}{\left( Y_{j,t}A^{(s)}_{ij}\right)}}{\sum_{j}{A^{(s)}_{ij}}} +
	f_{(s)}(C_{i})+\gamma^\top U_{i}+ \epsilon_{i,t+1}, \]
with $Y_{i,0}$ generated from a burn-in procedure by simulating from the above with $T=300$ starting from a  zero vector, and the innovations $\{\epsilon_{i,t}\}$ following $i.i.d.$ $N(0,\sigma^2)$ with $\sigma=1.5$.
The network $A$ and the function $f$ on the latent variable are generated under scenarios indexed by $s$ as $A^{(s)}$
and $f_{(s)}$, with, as a reminder,
	$A^{(s)}_{ij}=\mathbbm 1 \{\eta_{ij}\leq g_{(s)}(\Phi(C_i),\Phi(C_j)) \}\mathbbm  1(i\neq j), $
where $\{\eta_{ij}\}_{1\leq i<j\leq N}$  are $i.i.d.$ $U[0,1]$ random variables with $\eta_{ij}=\eta_{ji}$, and $\Phi(\cdot)$ is the CDF of the standard normal distribution,  $C_i$'s are $i.i.d.$ $N(0,1)$ and $g_{(s)}$ is the graphon function. Let $g(u,v)=\exp(u+v)/(1+\exp(u+v))$. 
The following three settings for the graphon and $f$ function  are considered.  {Note $C_i\overset{d}{=} \Phi^{-1}(U[0,1])$ so \Cref{Assumption2.1_2} holds. }
\begin{description}
	\item(Setting I):  $g_{(1)}(u,v)=\Big(0.6\mathbbm  1(0\leq u,v\leq 1/2)+0.1\mathbbm  1(u\leq 1/2)\mathbbm  1 (v>1/2)+0.1 \mathbbm  1(v\leq 1/2)\mathbbm  1 (u>1/2)+0.2 \mathbbm  1(1/2<u,v\leq 1)\Big)g(u,v)$,  and $f_{(1)}(C_i)=0.5\mathbbm 1(\Phi(C_i)\leq 1/2)$. %
	\item(Setting II): $g_{(2)}(u,v)=g(u,v)$, and $f_{(2)}(C_i)=C_i$.
	\item (Setting III): $g_{(3)}(u,v)=\Big(0.6\mathbbm  1(0\leq u,v\leq 1/2)+0.1\mathbbm  1(u\leq 1/2)\mathbbm  1 (v>1/2)+0.1 \mathbbm  1(v\leq 1/2)\mathbbm  1 (u>1/2)+0.2 \mathbbm  1(1/2<u,v\leq 1)\Big)$, and $f_{(3)}(C_i)=0.5\mathbbm 1(\Phi(C_i)\leq 1/2)$.
\end{description}
For the graphons in the settings above, $g_{(1)}$  is piecewise smooth, $g_{(2)}$ is  smooth, and $g_{(3)}$ is piecewise constant. In our simulation, we consider $\alpha=0.2$, and set $\beta=0.2$ or $0.5$ respectively for moderate and strong influence. The covariate parameter is set as $\gamma=1.5$.  For the bandwidth estimating $f(C_i)$ and $\gamma$ in Section \ref{sec:C and gamma}, we use a data-driven $h_N$ equal to the $a${th} sample quantile of those positive $\hat \delta_{ij}$'s defined as 
$h_N=Q_a(\{\hat \delta^2_{ij}:\hat \delta_{ij}>0\})$ 
with $a$ set as $a=h_0N^{-1/2}$, 
where  $h_0$ is  some  constant. We recommend $h_0=1.5$ in practice. We have also considered  other choices of $h_0$ in our simulation by setting $h_0=1.0, 1.5$, or  $2.0$, and found that the performance  of the estimators is rather insensitive to $h_0$.

We now discuss how to measure the performance. The estimation accuracy of $\hat \alpha$, $\hat \beta$, $\hat \gamma$ and $\widehat {f(C_i)}$ is evaluated by their mean absolute deviations. That is, in the case of estimating $\alpha$, the mean and standard error of $|\hat\alpha-\alpha|$. The normality of the estimators is evaluated by the coverage probability of the $95\%$ confidence intervals constructed by using our theory. 
 
 There are many unknown quantities in our theorems. Here we provide their consistent estimators. For $\Sigma_{1,N}$, $\Sigma_{2,N}$ and $\sigma^2$ in Theorem \ref{Theorem1} and \ref{diverging N}, as a consequence of \Cref{lemma1} and \Cref{lemma2} we use
\begin{align}
\hat \Sigma_{1,N}=\frac{1}{NT}\sum_{t=1}^T\Z_t^\top \X_t, ~~~~ \hat\Sigma_{2,N}=\frac{2}{NT}\sum_{t=1}^T\Z_t^\top\Z_t-\frac{1}{N(T-1)}\sum_{t=2}^T(\Z^\top_{t-1}\Z_t+\Z^\top_t\Z_{t-1}), 
\end{align}
and $\hat \sigma^2=\frac{1}{NT}\sum_{i=1}^N\sum_{t=0}^{T-1}\hat e^2_{it}$. For the unknown quantities in Theorem \ref{Thm5} and \ref{thm6}, we define their estimators in the Supplementary Materials as there are many.

We set the number of nodes in a network as $N=50, 250,$ or  $500$, and the number of time points as $T=10, 100,$ or $300$. All simulation results are based on $2000$ replications. The results are summarized in Table \ref{table:1:1} for estimation accuracy and Table \ref{table:2:1} for empirical coverage under the nominal level $95\%$.  Our results clearly show that the estimating accuracy improves while increasing the network size $N$ and/or time points $T$. The results also demonstrate the asymptotic normality of $\hat \alpha$ and $\hat \beta$ for all the settings, as well as that of $\hat \gamma$ and $\widehat {f(\cdot)} $ under setting I where the graphon is piecewise smooth and under setting III where the graphon is piecewise constant. Clearly the agreements between the nominal level and the empirical coverage become closer when $N$ and/or $T$ increases. Under setting II where the assumptions needed by Theorem \ref{Thm5} do not hold, we see undercoverage of the confidence intervals for $\gamma$ and ${f(\cdot)}$ under all parameter configurations, though the consistency of their estimates can still be seen from Table \ref{table:1:1}.

\begin{table}[!htb]
\centering \scriptsize
\begin{tabular}{crccccccc}\\
&	&\multicolumn{3}{c}{$\beta_1=0.2$ }&&\multicolumn{3}{c}{$\beta_1=0.5$} \\\cline{3-5}\cline{7-9}
 & N/T &  10 &100&300&&10 &100&300\\\hline\hline
&& \multicolumn{7}{c}{\underline{Setting I}}\\
\multirow{3}{*}{$\alpha$} 
& 50& 6.46 (1.09) & 1.74 (0.29) & 1.00 (0.17) && 8.46 (5.44) & 1.78 (0.31) & 1.00 (0.17)\\
&250& 2.90 (0.49)& 0.79 (0.14)& 0.46 (0.08) && 4.38 (3.65)& 0.80 (0.13)& 0.44 (0.07)\\ 
&500&2.06 (0.35) & 0.57 (0.10) & 0.33 (0.05) && 4.52 (10.75) & 0.57 (0.10) & 0.32 (0.05) \\ \hline
\multirow{3}{*}{$\beta$}

& 50& 18.45 (3.22) & 4.52 (0.78) & 2.68 (0.45)&& 30.30 (35.60) & 4.64 (0.81) & 2.68 (0.45)\\
&250&23.11 (4.26) & 5.23 (0.88) & 2.95 (0.50) &&77.02 (112.796) & 5.47 (0.94) & 2.99 (0.51) \\ 
&500& 28.35 (6.01) & 5.40 (0.92) & 2.88 (0.49) && 208.41 (775.44) & 5.97 (1.04) & 3.14 (0.52) \\ 
\hline
\multirow{3}{*}{$\gamma$} 
& 50& 15.25 (2.63) & 4.46 (0.76) & 2.81 (0.49) && 20.61 (8.41) & 4.75 (0.82) & 2.86 (0.50)\\	
&250& 7.16 (1.21) & 2.04 (0.34) & 1.17 (0.20) && 11.24 (11.51)& 2.05 (0.35) & 1.18 (0.192) \\ 
&500& 5.09 (0.85) & 1.50 (0.25) & 0.86 (0.14) && 11.83 (31.78) & 1.50 (0.26) & 0.87 (0.14) \\ 		 \hline
\multirow{3}{*}{$f(C_i)$}
& 50& 0.25 (2.34) & 0.086(0.49) & 0.061 (0.40) && 0.60 (78.60) & 0.10 (1.04) & 0.070 (0.60)\\ 
&250& 0.29 (1.82) & 0.085 (0.25) & 0.049 (0.14)&& 1.07 (183.4) & 0.10 (0.78) & 0.057 (0.38)\\ 		
&500& 0.31 (2.75) & 0.086 (0.23) & 0.049 (0.11)&& 2.21 (663.07) & 0.11 (0.83) & 0.059 (0.36) \\ 		
\hline\hline
&&\multicolumn{7}{c}{\underline{Setting II}}\\
\multirow{3}{*}{$\alpha$} 
& 50& 8.56 (10.67) & 1.81 (0.30) & 1.00 (0.17) && 9.54 (5.39) & 1.80 (0.30) & 1.00 (0.18) \\ 	
&250&  3.00 (0.54) & 0.79 (0.14) & 0.47  (0.08)&& 3.59 (1.78) & 0.79 (0.13) & 0.45 (0.08) \\ 
&500 & 2.15 (0.37) & 0.56 (0.09) & 0.32 (0.05) && 2.67 (1.45) & 0.56 (0.09) & 0.32 (0.06)  \\ \hline
\multirow{3}{*}{$\beta$} 
& 50& 98.01 (253.67) & 11.20 (1.90) & 6.44 (1.08) && 153.27 (153.29) & 13.29 (2.27) & 7.04 (1.16) \\ 	
&250& 73.63 (47.15) & 11.38(1.97) & 6.29 (1.04) && 212.89 (476.36) & 13.38 (2.40) & 7.17 (1.23) \\ 		
&500&89.88 (62.21) & 11.44 (1.97) & 6.56 (1.10) && 294.47 (548.11) & 14.56 (2.69) & 7.15 (1.24) \\ \hline
\multirow{3}{*}{$\gamma$} 
& 50& 20.49 (8.93) & 11.35 (1.92) & 11.12 (1.90) && 21.71 (8.79) & 11.56 (1.91) & 11.03 (1.91) \\ 	
&250&  7.66 (1.32) & 4.15 (0.71) & 3.87 (0.65) && 8.16 (2.56) & 4.06 (0.69) & 3.73 (0.64) \\ 	
&500& 5.30(0.92) & 2.68 (0.45) & 2.40 (0.41) && 5.96 (2.42) & 2.69 (0.46) & 2.52 (0.42) \\ 	\hline
\multirow{3}{*}{$f(C_i)$} 
& 50&  1.00 (220.24) & 0.48 (1.42) & 0.47 (1.41) && 1.78 (208.29) & 0.49 (2.22) & 0.47 (1.44) \\ 
&250&  0.44 (13.94) & 0.28 (0.41) & 0.27 (0.39) && 1.19 (166.13) & 0.29 (0.98) & 0.27 (0.51) \\ 
&500& 0.41 (16.33) & 0.21 (0.23) & 0.20 (0.22)&& 1.56 (361.35) & 0.22 (1.13) & 0.20 (0.30) \\ 	 \hline\hline
&&\multicolumn{7}{c}{\underline{Setting III}}\\
\multirow{3}{*}{$\alpha$} 
& 50& 6.67 (1.22) & 1.74 (0.30) & 1.02 (0.17) && 12.77 (24.9) & 1.85 (0.30) & 1.03 (0.17)\\ 
&250& 2.92 (1.01) & 0.78 (0.14) & 0.44 (0.08) && 5.08 (8.58) & 0.80 (0.14) & 0.455 (0.08) \\  
&500& 2.13 (0.47) & 0.56 (0.09) & 0.32 (0.05) & & 5.05 (11.60) & 0.57 (0.10) & 0.32 (0.05) \\ 	 \hline
\multirow{3}{*}{$\beta$} 
& 50& 23.85 (6.97) & 5.50 (0.93) & 3.15 (0.53)&& 71.44 (183.26) & 5.81 (0.98) & 3.14 (0.53)\\  
&250& 33.21 ( 48.64) & 6.01 (1.02) & 3.25 (0.55) && 133.44 (364.66) & 6.94 (1.21) & 3.49 (0.62) \\
&500& 48.15 (35.53) & 6.41 (1.06) & 3.36 (0.58) && 334.94 (1142.46) & 7.93 (1.42) & 3.84 (0.63) \\ 	\hline
\multirow{3}{*}{$\gamma$} 
& 50& 15.32 (2.72) & 4.22 (0.74) & 2.47 (0.42) && 29.28 (45.37) & 4.37 (0.76) & 2.45 (0.41)\\ 
&250& 7.03 (1.31) & 1.99 (0.33) & 1.17 (0.19) && 14.37 (9.03) & 2.04 (0.35) & 1.15 (0.20) \\ 
&500&  5.29 (1.22) & 1.45 (0.25) & 0.80 (0.14) && 13.96 (39.62) & 1.47 (0.24) & 0.82 (0.14) \\ 	\hline
\multirow{3}{*}{$f(C_i)$} 
& 50& 0.26 (6.47) & 0.07(0.58) & 0.04 (0.34) && 1.41 (439.00) & 0.100 (1.45) & 0.05 (0.81) \\  
&250& 0.34 ( 38.95) & 0.084 (0.35) & 0.048 (0.17) && 1.80  (513.75)& 0.11 (1.24) & 0.061 (0.54) \\ 
&500&0.40 (21.49) & 0.086 (0.30) & 0.049 (0.14) && 3.68 (1244.15) & 0.12 (1.37) & 0.063 (0.56) \\  \hline\hline
\end{tabular}
\caption{Average errors (multiplied by $100$) and in parentheses their standard errors (multiplied by $1000$).}
\label{table:1:1}
\end{table}

\begin{table}[!htb]
\centering
\begin{tabular}{crccccccc}\\
&	&\multicolumn{3}{c}{$\beta_1=0.2$ }&&\multicolumn{3}{c}{$\beta_1=0.5$} \\\cline{3-5}\cline{7-9}
 & N/T &  10 &100&300&&10 &100&300\\\hline\hline
&& \multicolumn{7}{c}{\underline{Setting I}}\\
\multirow{3}{*}{$\alpha$} 
& 50& 0.918 & 0.950 & 0.952 && 0.895 & 0.934 & 0.941 \\ 
&250& 	 0.920 & 0.943 & 0.941 && 0.882 & 0.942 & 0.957 \\ 
&500& 	  0.911 & 0.943 & 0.943 && 0.854 & 0.940 & 0.957\\  \hline
\multirow{3}{*}{$\beta$} 
& 50& 0.906 & 0.944 & 0.942 && 0.835 & 0.933 & 0.941 \\ 
&250&  0.838 & 0.927 & 0.940 && 0.672 & 0.924 & 0.935  \\ 
&500&    0.755 & 0.918 & 0.938 && 0.556 & 0.895 & 0.927\\ 
\hline
\multirow{3}{*}{$\gamma$} 
& 50& 	 0.928 & 0.933 & 0.899 && 0.906 & 0.926 & 0.913\\ 
&250&	 0.924 & 0.948 & 0.948 && 0.891 & 0.952 & 0.953 \\ 
&500&  	  0.938 & 0.949 & 0.954 && 0.885 & 0.944 & 0.952 \\ 
	 \hline
\multirow{3}{*}{$f(C_i)$} 
& 50&   0.915 & 0.860 & 0.823 && 0.877 & 0.868 & 0.844 \\  	
&250& 0.932 & 0.946 & 0.943 && 0.789 & 0.935 & 0.950 \\ 	
&500& 0.909 & 0.946 & 0.950 && 0.703 & 0.927 & 0.929 \\ 	
\hline\hline
&&\multicolumn{7}{c}{\underline{Setting II}}\\
\multirow{3}{*}{$\alpha$} 
& 50	 & 0.925 & 0.951 & 0.966 && 0.919 & 0.957 & 0.964  \\ 
&250&   0.919 & 0.947 & 0.944 && 0.908 & 0.951 & 0.961 \\  
&500&  0.910 & 0.951 & 0.960 && 0.900 & 0.953 & 0.953  \\  \hline
\multirow{3}{*}{$\beta$} 
& 50& 0.883 & 0.954 & 0.960 && 0.783 & 0.940 & 0.953  \\ 
&250& 0.849 & 0.948 & 0.961 && 0.738 & 0.927 & 0.949\\  
&500&  0.814 & 0.942 & 0.949 && 0.699 & 0.905 & 0.940 \\  \hline
\multirow{3}{*}{$\gamma$} 
& 50& 0.843 & 0.528 & 0.330 && 0.861 & 0.513 & 0.337\\  
&250& 0.882 & 0.626 & 0.408 && 0.889 & 0.636 & 0.428 \\  
&500&   0.891 & 0.672 & 0.463 && 0.889 & 0.663 & 0.449 \\   \hline
\multirow{3}{*}{$f(C_i)$} 
& 50& 0.594 & 0.276 & 0.181 && 0.620 & 0.371 & 0.244 \\  
&250&  0.635 & 0.357 & 0.276 && 0.645 & 0.411 & 0.315 \\ 
&500& 0.654 & 0.469 & 0.371 && 0.653 & 0.506 & 0.395 \\  \hline\hline
&&\multicolumn{7}{c}{\underline{Setting III}}\\
\multirow{3}{*}{$\alpha$} 
& 50  & 0.913 & 0.940 & 0.946 && 0.884 & 0.930 & 0.952 \\ 
&250  &0.892 & 0.944 & 0.942 && 0.867 & 0.940 & 0.949 \\ 
&500& 0.907 & 0.951 & 0.961 && 0.853 & 0.947 & 0.953 \\  \hline
\multirow{3}{*}{$\beta$} 
& 50  & 0.882 & 0.941 & 0.942 & &0.793 & 0.930 & 0.946\\ 
&250&   0.808 & 0.930 & 0.949& & 0.610 & 0.900 & 0.929  \\ 
&500& 0.714 & 0.927 & 0.945 && 0.498 & 0.851 & 0.916\\ \hline
\multirow{3}{*}{$\gamma$} 
& 50 & 0.932 & 0.933 & 0.941 && 0.896 & 0.935 & 0.945  \\ 
&250 & 0.923 & 0.952 & 0.950 && 0.882 & 0.939 & 0.946  \\ 
&500&0.924 & 0.946 & 0.955 && 0.864 & 0.948 & 0.949 \\ \hline
\multirow{3}{*}{$f(C_i)$} 
& 50 & 0.920 & 0.951 & 0.947 && 0.828 & 0.928 & 0.945 \\ 
&250& 0.912 & 0.944 & 0.939 && 0.723 & 0.913 & 0.940 \\ 
&500&0.869 & 0.946 & 0.942 && 0.647 & 0.900 & 0.930\\  \hline\hline
\end{tabular}
\caption{The empirical coverage of the sample confidence intervals at the nominal 95\% level.}
\label{table:2:1}
\end{table}	

In results not shown, the NAR approach in \cite{zhu2017network} gives biased estimates  since their method does not deal with latent variables. The approach of \cite{mcfowland2021estimating} also underperforms ours if one naively assumes a stochastic block model for the network, because either the graphon model is misspecified (Settings I and II) or the linear model for $C_i$ is misspecified (Setting I and III), or both.

\section{Data Analysis} \label{sec:data}
The emergence of socia media has provided many opportunities to test the existence of influence and homphily in a connected setting and to examine the extent to which they impact the entities in a network. Under this general setup,  we apply our method to evaluate how behaviours of Sina Weibo users are influenced by various factors. Sina Weibo (www.weibo.com) is the largest Twitter-style social media in China, which  had 255 million daily, close to 600 million monthly active  users  as of March 2023. Many of Sina Weibo's features resemble those of Twitter, for example, each user can post, re-post, comment and add hashtags. Inspired by the data analysis in \cite{zhu2017network} where a dataset was collected over $T=4$ consecutive weeks, we aimed to expand their idea to a larger dataset with longer time period so that the assumption $\min\{N, T \}\rightarrow \infty$ is more reasonable. 
 Towards this, we curated a  dataset consisting of weekly observations for the official accounts of Shen Zhou and his followers over $T=26$ consecutive weeks since April 4, 2022. Shen Zhou is one of the most popular young singers in China, having over 85 million direct followers as of 5 July, 2023. As in \cite{zhu2017network}, we define the network structure via the followee-follower relationship that is subsequently converted to an undirected network.  For analysis, we focus on  a sub-network consisting of those accounts with degrees larger or  equal than $6$ together with their links, giving rise to a network with  $N=342$ users. The resulting network is displayed in \Cref{network-overview}. 
\begin{figure}[!htb]
	\begin{center}
		\includegraphics[width=6in]{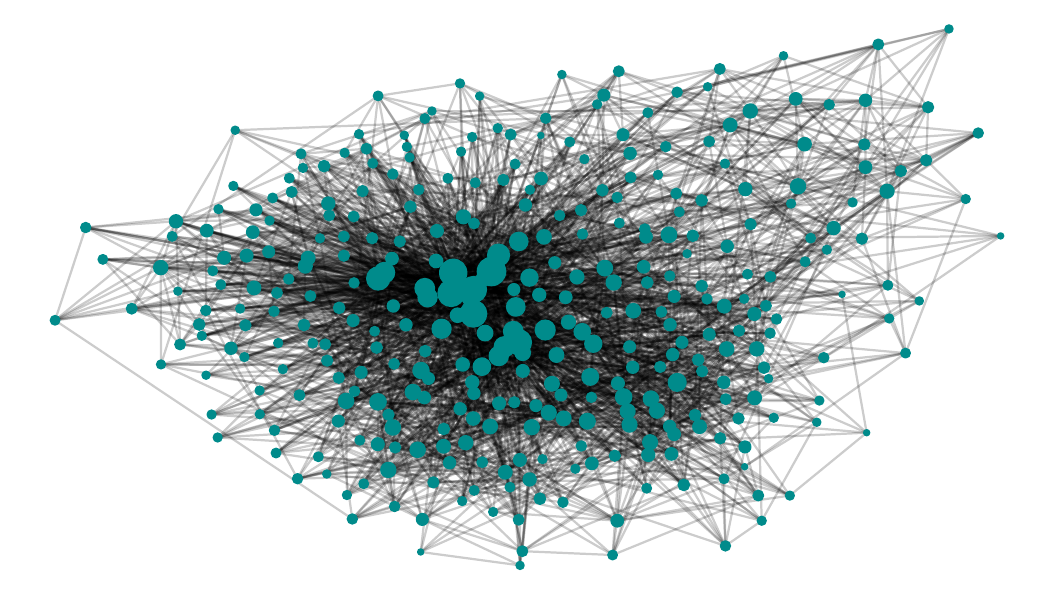}
		\caption{Sina Weibo data: the size of a node is proportional to its degree after discretization.}
		\label{network-overview}
	\end{center}
\end{figure}

With this network, we set out to explore the response variable defined as the post length after logarithmic transformation, the number of characters contained in the posts of  each week over these 26 weeks for each individual account. This response variable broadly reflects how active a user is on a weekly basis. On Sina Weibo,  each user's gender information and self-labelled interests are also available. We encode these variables as our observed covariates, for which we use $U_{i1}$ to denote the gender information ($U_{i1}=1$ if the $i${th} node is male or $0$ otherwise) and $U_{i2}$ for the the number of personal labels. By including these two covariates, we want to assess whether gender or the number of interesting labels have a role to play in determining the activeness of an account. Intuitively, there are other factors that may influence whether a user posts. One of them is how active their connected users are. The more active their linked users are, the more likely this user will participate in online social activities. On the other hand, as we have argued, it is also likely that users are connected because they share similar patterns in posting. Thus, our interest is to analyze how covariates, connected neighbours and homophily collectively decide the post pattern of users, in addition to the natural momentum effect that a user's activeness depends on their past posting pattern. We note that a similar scientific issue was investigated in \cite{zhu2017network} on a dataset with a smaller time window. 

We first employ preliminary data analysis to assess whether network effect exsits. Towards this, we plot  $Y_{i, t+1}$ versus $\frac{\sum_{j=1}^N A_{ij} Y_{j, t}}{\sum_{j=1}^N A_{ij}}$ for three users  in Figure \ref{NodevsNW} with the least-squares fitted lines added. Clearly we can see positive correlation patterns in the three plots in this figure, more or less with similar slopes. This suggests that the influence of neighbours on how much a user posts does exist and that it is appropriate to model this factor as a linear function of the average response of one's neighbours at a previous time.

\begin{figure}[!htb]
	\begin{center}
		\includegraphics[width=6.5in]{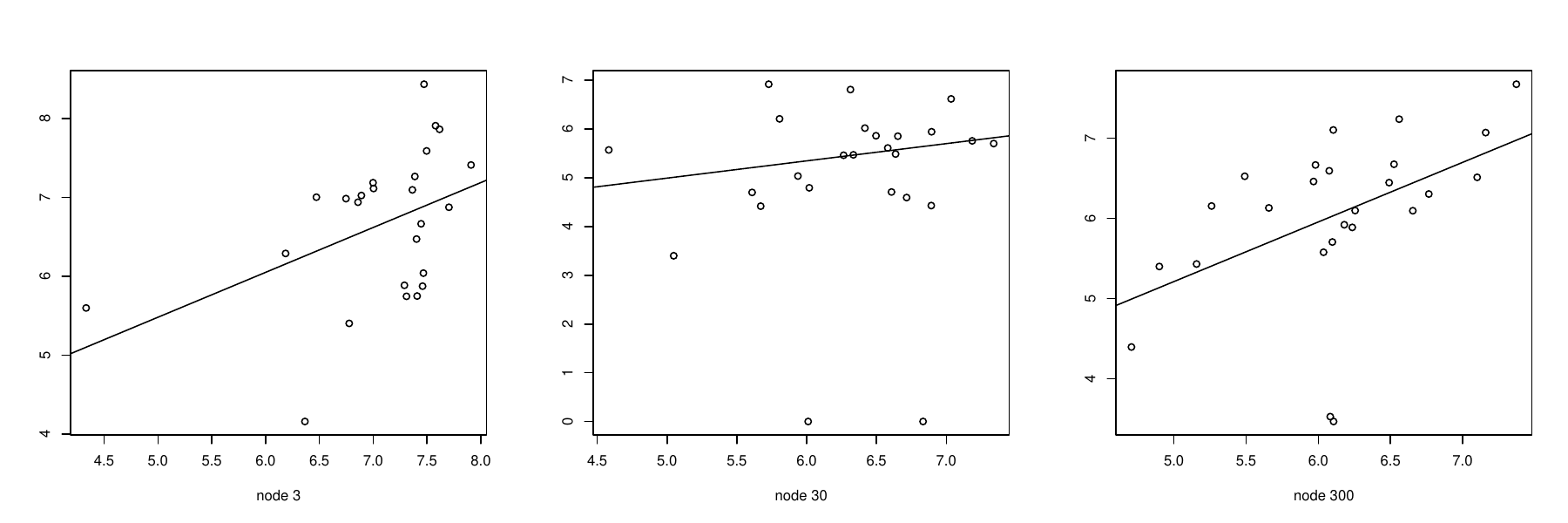}
		\caption{The (log) post length in each week versus the average of its neighbours at a previous time for nodes 3 (left), 30 (middle) and 300 (right).  For better visualization, least-squares lines are added.}
		\label{NodevsNW}
	\end{center}
\end{figure}

We first apply NAR in \cite{zhu2017network} to this dataset and summarize the results in the left part Table \ref{Table2}. It is found that all the parameters are estimated to have very small standard errors, resulting in very narrow $95\%$ confidence intervals. Surprisingly, the influence of neighbours is found to be significantly negative, while the gender and number of labels effects are positive as expected. 
\begin{table}[!htb]
	\centering
	\begin{tabular}{crrrrr}
		\hline
		& \multicolumn{2}{c}{NAR} && \multicolumn{2}{c}{RGAM}\\\cline{2-3}\cline{5-6}
		& Estimate & $z_{0.975}$*SE && Estimate &$z_{0.975}$*SE \\ \cline{2-3} \cline{5-6}
	Common baseline & $3.55$ & $0.03$ \\ 
		$\alpha$ (Momentum) & $0.5268$ & 0.0001 &&$0.27$ & $0.04$ \\ 
		$\beta$ (Influence) & $-0.0875$ & $0.0006$ &&$0.22$ & $0.21$  \\ 
		$\gamma_1$(Gender) & $0.152$ & $0.005$ && $0.18$ & $0.19$ \\ 
		 $\gamma_2$ (Number of labels)  & $0.0528$ & $0.0005$ &&$-0.078$ & $0.077$\\ 
		 $ f(C_{10})$ (Individual base-line) & &&&$3.04$ & $1.51$ \\ 
		\hline
	\end{tabular}
\caption{Estimates of the parameters in NAR (left) and RGAM (right) with $z_{0.975}$ times their standard errors (SEs). A $95\%$ confidence interval can be constructed as: Estimate $\pm$  $z_{0.975}$*SE.}
\label{Table2}
\end{table}

Next, we apply our RGAM to this dataset. For the estimation of $\alpha$ and $\beta$, we use the instrumental variable approach as discussed in Section \ref{sec:alpha and beta} that is also applicable to the estimation of $\alpha$ and $\beta$ in NAR in \eqref{eq:nar}. For the estimation of $\gamma$ and ${f(C_i)}$, we use $h_0=1.5$ recommended by our simulation. The estimates and their standard errors are presented in the right part of \Cref{Table2}, where we only show the estimated $f(C_{10})$ as an illustration.  

As seen from this table, both the momentum and influence parameters are significantly positive. While qualitatively the estimate of the momentum parameter is similar to that of the NAR model, quantitatively they are quite different. The momentum effect in RGAM is much smaller with a wider confidence interval. On the other hand, the estimates of the influence parameter of RGAM and NAR are both significant, but with opposite signs. Our estimate states that connected  neighbours have a positive influence on how a user posts, while counterintuitively the estimate of NAR indicates a negative one again with a narrow confidence interval. Here we remark that if the NAR model is correctly specified, the estimates of these parameters using the least-squares approach in \cite{zhu2017network}  and our instrumental variable approach will both be asymptotically normal. 
 The apparent discrepancy implies that the NAR model is likely misspecified. 
 
 In estimating $\gamma$, we found that the gender effect is marginally insignificant, indicating that gender may not have played a significant role in those users following Shen Zhou. On the other hand, the number of labels has a slightly negative and significant effect on how much a user posts. In 
the NAR model of \cite{zhu2017network}, there is a baseline term measured by the common intercept, while for our models each individual has their own baseline $f(C_i)$ which is determined by their latent attributes. In Table \ref{Table2} we display the estimated baseline for individual $10$ and find that it is strictly positive. We also display the histogram of estimated individual baseline $f(C_i)$'s in \Cref{hist_baseline} which shows clearly heterogeneous individual baseline effects. This plot may indicate the inadequacy of using a common baseline as in NAR. Interestingly, the average of the  estimated individual baselines in our model is $3.24$, which is close to the estimated common baseline effect $3.55$ in NAR. To confirm indeed there is a linear relation between $Y_{i,t}$ and $f(C_i)$, we plot in Figure \ref{VSlatent} three scatterplots for $t=24, 25$ and $26$. Again, these plots confirm a common linear pattern.
\begin{figure}[!htb]
	\begin{center}
		\includegraphics[width=4in]{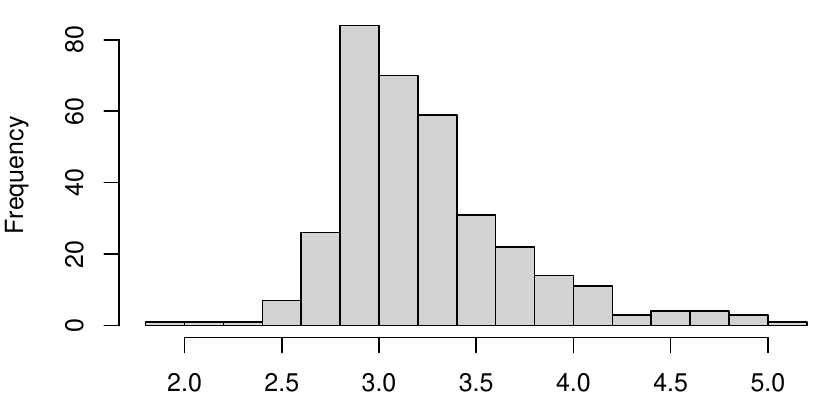}
		\caption{Histogram of the estimated individual baselines $\widehat{f(C_i)}$. }
		\label{hist_baseline}
	\end{center}
\end{figure}
\begin{figure}[!htb]
	\begin{center}
		\includegraphics[scale=0.55]{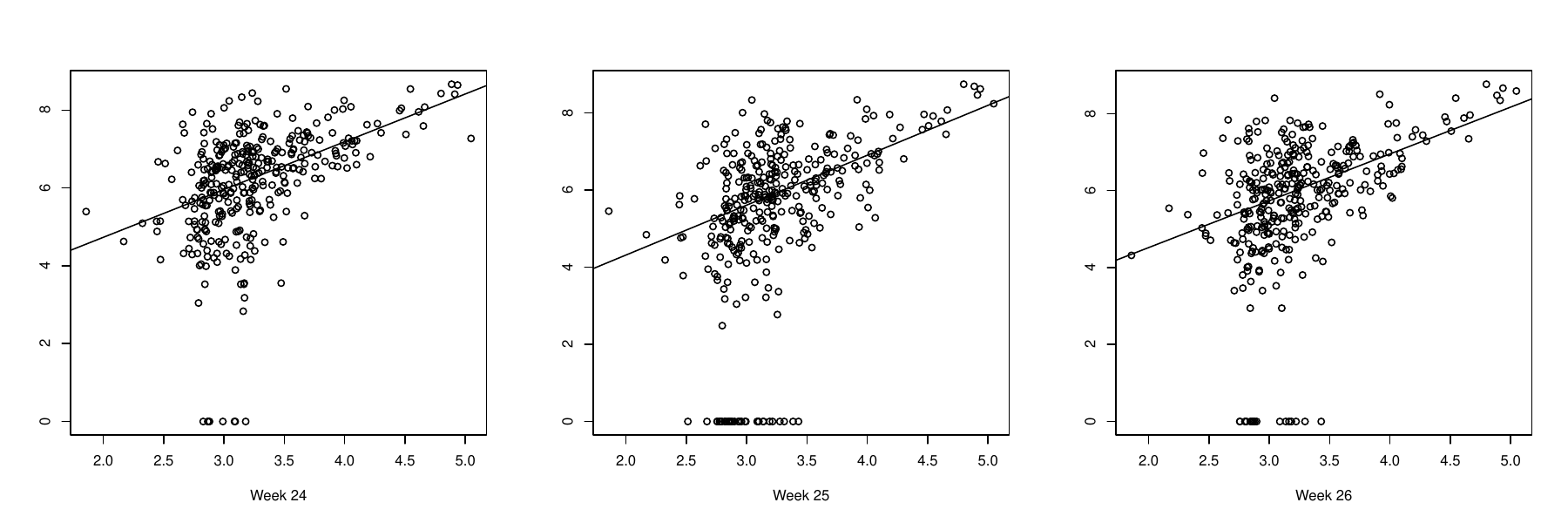}
		\caption{The responses in Week 24 (left), 25 (middle), and 26 (right) versus the estimated $f(C_i)$'s.}
		\label{VSlatent}
	\end{center}
\end{figure}

We also compare the performance of RGAM, NAR in \cite{zhu2017network} and the approach in \cite{mcfowland2021estimating} 
in terms of their prediction power by computing prediction errors. In particular, we calculate the {mean absolute prediction error between the observed post length at $t+1$ and the predicted one at this time point using the data from week $1$ to $t$. The errors for the weeks 20--26 are reported in  Table \ref{Tab:3}. Except week 22, our method outperforms its competitors in having smaller errors, suggesting that our method may be preferable for this dataset.

\begin{table}[!htb]
	\centering
	\begin{tabular}{crrrrrrr}
		\hline
	Method/Week&20&21 & 22 & 23 & 24 & 25 & 26 \\ 
		\hline
RGAM & ${\bf 1.075}$ & ${\bf 0.644}$ & $0.738$ & ${\bf 1.339}$ & ${\bf 0.840}$ & ${\bf 1.185}$ & ${\bf 0.846}$ \\ 
\cite{zhu2017network} & $1.186$ & $0.671$ & $0.670$ & $1.371$ & $0.862$ & $1.192$ & $0.877$ \\ 
\cite{mcfowland2021estimating} &	$1.225$&$ 0.673$&$ {\bf 0.657}$&$ 1.420$&$ 0.844$ &$1.222$&$ 0.880$\\	\hline
	\end{tabular}
\caption{Comparison of the mean absolute prediction errors of three methods for weeks 22--26. } 
\label{Tab:3}
\end{table}

\section{Conclusion}\label{sec:conclusion}
We have proposed RGAM, a random-graph based autoregressive model, that is more flexible than existing methods in the literature. Exploiting the first-order differencing operation, we have employed an instrumental variable approach for the estimation of the momentum and influence parameters that is less susceptible to model misspecification. Using latent variables to encode homophily, we bypass its estimation exploring the idea of neighbourhood smoothing when the effect of observed covariates is parametric. We show that our estimators are consistent and asymptotically normal thanks to new theories developed that {includes a new framework of analysis of network autoregression using central limit theorem and $m$-dependent approximation, new mathematical tools tackling the random network structures, and delicate analysis of residual based  U-statistics for homophilous effect and covariate effect.} Overall, our work contributes to the ever growing literature on using network for regression analysis.

We outline several directions for future research. In our work, we assume that the network is fully observed without error. If this is not the case, a need arises to develop robust estimation methods that can handle erroneous networks \citep{le2022linear}. Secondly, there is interest to explore a more general notion of neighbourhood, as ours is limited to immediate one-hop away nodes  when defining neighbours.  Thirdly, we have limited momentum and influence to time lag-one dependence but this dependence can be more persistent. It will be interesting to develop an autoregressive model incorporating longer time dependence. Lastly, we have chosen to model the nodal fixed effects $f(C_i)$ explicitly in this paper in the framework of partial linear models. It will be interesting to explore other approaches for their estimation. One promising approach is the network cohesion method in \cite{li2019prediction} that penalizes the differences between the $f(C_i)$'s of connected nodes. We leave these issues to future work.

\renewcommand{\baselinestretch}{1}
\normalsize
\bibliographystyle{apalike}
\bibliography{ref}

\newpage

\renewcommand{\thelemma}{\Alph{section}.\arabic{lemma}}
\renewcommand{\theproposition}{\Alph{section}.\arabic{proposition}}
\renewcommand{\theequation}{\Alph{section}.\arabic{equation}}
\renewcommand{\thetheorem}{\Alph{section}.\arabic{theorem}}
\renewcommand{\thetable}{\Alph{section}.\arabic{table}}
\renewcommand{\thecor}{\Alph{section}.\arabic{cor}}
\renewcommand{\thesection}{\Alph{section}}  

\renewcommand{\thefigure}{\Alph{section}.\arabic{figure}}  
\setcounter{section}{0}
\setcounter{equation}{0}
\section{Appendix}
\subsection{Notations}\label{Notation}
We collect all the additional notations used in the theory here. Define the $2\times 2$ matrix
\begin{align}
	\Sigma_{1,N}=\begin{pmatrix}
		\kappa_{11,N}&\kappa_{12,N}\\
		\kappa_{21,N}&\kappa_{22,N}
	\end{pmatrix}%
	,\ \ \Sigma_{1,N}^*=\frac{N}{\sigma^2}\Sigma_{1,N},
\end{align}
where
\begin{align}%
	\kappa_{11,N}&=\frac{\sigma^2}{N}\sum_{i=0}^\infty tr((G_1^i)^\top G_1^{i+1})-\frac{\sigma^2}{N}\sum_{i=0}^\infty tr((G_1^i)^\top G_1^{i}), 
	\\\kappa_{21,N}&= \frac{\sigma^2}{N}\sum_{i=0}^\infty  tr((G_1^i)^\top W_N^\top G_1^{i+1})-\frac{\sigma^2}{N}\sum_{i=0}^\infty  tr((G_1^i)^\top W_N^\top G_1^{i}),\\
	\kappa_{12,N}&=\frac{\sigma^2}{N}\sum_{i=0}^\infty  tr((G_1^{i+1})^\top W_N^\top G_1^{i})-\frac{\sigma^2}{N}\sum_{i=0}^\infty  tr((G_1^i)^\top W_N^\top G_1^{i}),\\
	\kappa_{22,N}&=\frac{\sigma^2}{N}\sum_{i=0}^\infty  tr((G_1^i)^\top W_N^\top W_N G_1^{i+1})-\frac{\sigma^2}{N}\sum_{i=0}^\infty  tr((G_1^i)^\top W_N^\top W_N G_1^{i}).
\end{align}%
Further define
\begin{align}%
	\kappa_{11,N}^-&=\frac{1}{N}((I-G_1)^{-1}G_0)^\top (I-G_1)^{-1}G_0+\frac{\sigma^2}{N}\sum_{i=0}^\infty tr((G_1^i)^\top G_1^i),\\
	\kappa_{11,N}^+&=\frac{1}{N}((I-G_1)^{-1}G_0)^\top (I-G_1)^{-1}G_0+\frac{\sigma^2}{N}\sum_{i=0}^\infty tr((G_1^i)^\top G_1^{i+1}),\\
	\kappa_{21,N}^-&= \frac{1}{N}G_0^\top [(I-G_1)^{-1}]^\top W_N^\top (I-G_1)^{-1}G_0+\frac{\sigma^2 }{N}\sum_{i=0}^\infty tr((G_1^i)^\top W_N^\top G_1^{i}),\\
	\kappa_{21,N}^+&= \frac{1}{N}G_0^\top [(I-G_1)^{-1}]^\top W_N^\top (I-G_1)^{-1}G_0+\frac{\sigma^2 }{N}\sum_{i=0}^\infty  tr((G_1^i)^\top W_N^\top G_1^{i+1}),
\end{align}%
and
\begin{align}%
	\kappa_{12,N}^-&= \frac{1}{N}G_0^\top [(I-G_1)^{-1}]^\top W_N^\top (I-G_1)^{-1}G_0+\frac{\sigma^2}{N}\sum_{i=0}^\infty  tr((G_1^i)^\top W_N^\top G_1^{i}),\\
	\kappa_{12,N}^+&= \frac{1}{N}G_0^\top [(I-G_1)^{-1}]^\top W_N^\top (I-G_1)^{-1}G_0+\frac{\sigma^2}{N}\sum_{i=0}^\infty  tr((G_1^{i+1})^\top W_N^\top G_1^{i}),\\
	\kappa_{22,N}^-&= \frac{1}{N}G_0^\top [(I-G_1)^{-1}]^\top W_N^\top W_N (I-G_1)^{-1}G_0+\frac{\sigma^2}{N}\sum_{i=0}^\infty  tr((G_1^i)^\top W_N^\top W_N G_1^{i}),\\
	\kappa_{22,N}^+&= \frac{1}{N}G_0^\top [(I-G_1)^{-1}]^\top W_N^\top W_N (I-G_1)^{-1}G_0+\frac{\sigma^2}{N}\sum_{i=0}^\infty  tr((G_1^i)^\top W_N^\top W_N G_1^{i+1}).
\end{align}%
By definition, we have $\kappa_{ij,N}=\kappa^+_{ij,N}-\kappa^-_{ij,N}$ for $i,j\in \{1,2\}$. 
Define \begin{gather}
	\Sigma_{2,N}^-=\begin{pmatrix}
		\kappa^-_{11,N}&\kappa^-_{12,N}\\
		\kappa^-_{21,N}&\kappa^-_{22,N}
	\end{pmatrix}, \quad \Sigma_{2,N}^+=\begin{pmatrix}
		\kappa^+_{11,N}&\kappa^+_{12,N}\\
		\kappa^+_{21,N}&\kappa^+_{22,N}
	\end{pmatrix}, \quad \tilde \Sigma_{2,N}^+=\begin{pmatrix}
		\kappa^+_{11,N}&\kappa^+_{21,N}\\
		\kappa^+_{12,N}&\kappa^+_{22,N}
	\end{pmatrix},\\ 
	\Sigma_{2,N}=2\Sigma_{2,N}^--\Sigma_{2,N}^+-\tilde \Sigma_{2,N}^+:=\frac{\sigma^2}{N}\Sigma^*_{2,N},
\end{gather}
where $\Sigma_{2N}^*=(\sigma^*_{2,i,j})_{1\leq i,j\leq 2}$ with \begin{align}
	\sigma^*_{2,1,1}&=2\sum_{i=0}^\infty tr((G_1^i)^\top G_1^{i})-2\sum_{i=0}^\infty tr((G_1^i)^\top G_1^{i+1}),\\
	\sigma^*_{2,2,2}&=2\sum_{i=0}^\infty  tr((G_1^i)^\top W_N^\top W_N G_1^{i})-2\sum_{i=0}^\infty  tr((G_1^i)^\top W_N^\top W_N G_1^{i+1}),\\
	\sigma^*_{2,1,2}&=\sigma^*_{2,2,1}=2\sum_{i=0}^\infty  tr((G_1^i)^\top W_N^\top G_1^{i})-\sum_{i=0}^\infty  tr((G_1^i)^\top W_N^\top G_1^{i+1})-\sum_{i=0}^\infty  tr((G_1^{i+1})^\top W_N^\top G_1^{i}).
\end{align}
Define $\Sigma_{1,N}^{(m)}$ and $\Sigma_{2,N}^{(m)}$ by 
truncating the $k_{th}$, $k\geq m+1$ summands in $\kappa^\pm_{ij,N}$. For example, 
\begin{align}
	\kappa_{11,N}^{(m)}=\frac{\sigma^2}{N}\sum_{i=0}^m tr((G_1^i)^\top G_1^{i+1})-\frac{\sigma^2}{N}\sum_{i=0}^m tr((G_1^i)^\top G_1^{i}).
\end{align}
Other $\kappa_{ij,N}$ and $\kappa_{ij,N}^\pm$ can be defined similarly. 

For the results in Thereom \ref{Thm5}, we define \begin{gather}
	v_N=\sigma^2(l_N^\top \Sigma_{1,N}^{-1}\Sigma_{2,N}\Sigma_{1,N}^{-\top}l_N),~~~l_N=(l_{1N},l_{2N})^\top,\\
	l_{1N}=\frac{1}{N(N-1)r_N}\sum_{i=1}^{N-1}\sum_{j=1,j\neq i}^N(U_i-U_j)(e_i-e_j)^\top (I-G_1)^{-1}G_0K(\frac{\hat \delta^2_{ij}}{h_N}),\\
	l_{2N}=\frac{1}{N(N-1)r_N}\sum_{i=1}^{N-1}\sum_{j=i,j\neq i}^N(U_i-U_j)(w_i-w_j)^\top K(\frac{\hat \delta^2_{ij}}{h_N}) (I-G_1)^{-1}G_0,\\
	\Sigma_{3N}=\frac{1}{N}\sum_{i=1}^N\big(\sum_{j=1}^N\frac{(U_i-U_j)}{N-1}K(\frac{\hat \delta_{ij}^2}{h_N})\big)\big(\sum_{j=1}^N\frac{(U_i-U_j)}{N-1}K(\frac{\hat \delta_{ij}^2}{h_N})\big)^\top,\\
	\hat \Gamma_1=\frac{2}{N(N-1)r_N}\sum_{i=1}^{N-1}\sum_{j=i+1}^N(U_i-U_j) (U_i-U_j)^\top K(\frac{\hat \delta^2_{ij}}{h_N}).\label{hatgamma1}
\end{gather}

For the results in Theorem \ref{thm6}, we further define
\begin{gather}
	l_{3N}(C_i)=(\sum_{t=1}^NK(\frac{\hat \delta^2_{it}}{h_N}))^{-1}\sum_{t=1}^NK(\frac{\hat \delta^2_{it}}{h_N})U_t,
	l_{4N}(C_i)=(\sum_{t=1}^NK(\frac{\hat \delta^2_{it}}{h_N}))^{-1}\sum_{t=1}^NK(\frac{\hat \delta^2_{it}}{h_N})e_t^\top (I-G_1)^{-1}G_0,\\
	l_{5N}(C_i)=(\sum_{t=1}^NK(\frac{\hat \delta^2_{it}}{h_N}))^{-1}\sum_{t=1}^NK(\frac{\hat \delta^2_{it}}{h_N})w_t^\top (I-G_1)^{-1}G_0,\\
	A_N(C_i)=-(\hat \Gamma_1^{-1}l_{1N})^\top l_{3N}(C_i)+l_{4N}(C_i),
	B_N(C_i)=-(\hat \Gamma_1^{-1}l_{2N})^\top l_{3N}(C_i)+l_{5N}(C_i).
\end{gather}
Let  $\Sigma'_{3N}$ be an $|\mathcal C|\times |\mathcal C|$ matrix with the $(i_1,i_2)${th} entry $\tilde \sigma_{i_1,i_2}$ given by
\begin{align}
	N\sum_{t=1}^N\Big(\big(\sum_{j=1}^N(-\hat \Gamma_1^{-1}\frac{2(U_t-U_j)}{N(N-1)r_N})^\top K(\frac{\hat \delta^2_{tj}}{h_N})III_1(C_{i_1})+[\sum_{t=1}^N K(\frac{\hat \delta_{i_1t}^2}{h_N})]^{-1}K(\frac{\hat \delta_{i_1t}^2}{h_N})\big)\times \\\big(\sum_{j=1}^N(-\hat \Gamma_1^{-1}\frac{2(U_t-U_j)}{N(N-1)r_N})^\top K(\frac{\hat \delta^2_{tj}}{h_N})III_1(C_{i_2})+[\sum_{t=1}^N K(\frac{\hat \delta_{i_2t}^2}{h_N})]^{-1}K(\frac{\hat \delta_{i_2t}^2}{h_N})\big)\Big).
\end{align}
Finally we define $V_{N}$ as the  $|\mathcal C|\times |\mathcal C|$ covariance matrix with the $(i,j)$th entry given by
\begin{align}
	\sigma^2((A_N(C_i), B_N(C_i))\Sigma_{1,N}^{-1}\Sigma_{2,N}\Sigma_{1,N}^{-1}	(A_N(C_j), B_N(C_j))^\top+\tilde \sigma_{i,j}),
\end{align}
provided that $\lambda_{\min}(V_N)>0$, $\Sigma'_{3N}$ and the matrix $(\sqrt{r_N(C_i)r_N(C_j)\tilde \sigma_{i,j}})_{1\leq i,j\leq N}$ are full rank. The following quantities will be used in the proofs. Write 
\begin{align}
	\bar \epsilon_i=\sum_{t=1}^T \epsilon_{i,t}/T, ~~~\bar Y_i=\sum_{t=1}^T Y_{i,t-1}/T, ~~~ \bar \Y=\sum_{t=1}^T \Y_{t-1}/T, 
\end{align}
and
\begin{align}
	I(C_i)=\big(\sum_{t=1}^NK(\frac{\hat \delta^2_{it}}{h_N})\big)^{-1}\big(\sum_{t=1}^N f(C_t)K(\frac{\hat \delta^2_{it}}{h_N})\big), ~II(C_i)=\big(\sum_{t=1}^NK(\frac{\hat \delta^2_{it}}{h_N})\big)^{-1}\big(\sum_{t=1}^N\bar \epsilon_t K(\frac{\hat \delta^2_{it}}{h_N})\big),\\
	III(C_i)=(\gamma-\hat \gamma)^\top III_1(C_i)+(\alpha-\hat \alpha)III_2(C_i)+(\beta-\hat \beta)III_3(C_i),
\end{align}
where
\begin{align} 
	III_1(C_i)=\big(\sum_{t=1}^NK(\frac{\hat \delta^2_{it}}{h_N})\big)^{-1}
	\sum_{t=1}^NU_tK(\frac{\hat \delta^2_{it}}{h_N}), ~III_2(C_i)=\big(\sum_{t=1}^NK(\frac{\hat \delta^2_{it}}{h_N})\big)^{-1}
	\sum_{t=1}^N\bar Y_tK(\frac{\hat \delta^2_{it}}{h_N}),\\ III_3(C_i)=\big(\sum_{t=1}^NK(\frac{\hat \delta^2_{it}}{h_N})\big)^{-1}
	\sum_{t=1}^Nw_t^\top\bar \Y K(\frac{\hat \delta^2_{it}}{h_N}).
\end{align}
\subsection{Estimators of the quantities needed in Theorem \ref{Thm5} and \ref{thm6}  for simulation}\label{Simulation-quantity}
Let $\hat \sigma_{\alpha}$ and $\hat \sigma_{\beta}$ be the first and second diagonal element of $(\hat \Sigma_{1N}^{-1}\hat \Sigma_{2N}\hat \Sigma_{1N}^{-1})\hat \sigma$ respectively.  
We construct the 95\% confidence interval for $\alpha$ and $\beta$ denoted as $CI_{\alpha}$ and $CI_\beta$ respectively, as the following
\begin{align}
	CI_{\alpha}=[\hat  {\alpha}_1-1.96(N(T-1))^{-1/2}\hat \sigma_{\alpha}, \hat {\alpha}_1+1.96(N(T-1))^{-1/2}\hat \sigma_{\alpha}], \\
	CI_{\beta}=[\hat  {\beta}-1.96(N(T-1))^{-1/2}\hat \sigma_{\beta}, \hat {\beta}+1.96(N(T-1))^{-1/2}\hat \sigma_{\beta}].
\end{align}
For the latent function, for brevity we only provide the results for $\widehat{ f(C_1)}$.  Define
\begin{align}
	\hat l_N=(\hat \Gamma_{2,2,\alpha},\hat \Gamma_{2,2,\beta})^\top,~~~\hat v_N=\hat \sigma^2(\hat l_N^\top \hat \Sigma_{1,N}^{-1}\hat \Sigma_{2,N}\hat \Sigma_{1,N}^{-\top}\hat l_N),
	~~~\hat \sigma_\gamma=(\hat v_N+4\Sigma_{3,N}/r_N^2)^{-1/2}\hat \Gamma_1,
\end{align}
where $\hat\Gamma_{2,2,\alpha}$, $ \hat\Gamma_{2,2,\beta}$ and $\Sigma_{3, N}$ as defined in \Cref{Thm5}, and $ \Gamma_1$ is  estimated by $$\hat \Gamma_1=\frac{2}{N(N-1)r_N}\sum_{i=1}^{N-1}\sum_{j=i+1}^N(U_i-U_j) (U_i-U_j)^\top K(\frac{\hat \delta^2_{ij}}{h_N}).$$
For the confidence interval of $\gamma$,   we calculate  
\[
CI_\gamma=[\hat \gamma -1.96(N(T-1))^{-1/2}\hat \sigma_{\gamma}, \hat \gamma+1.96(N(T-1))^{-1/2}\hat \sigma_{\gamma}].
\]
We shall mention that the final representation $\hat \sigma_\gamma$ does not depend on $r_N$. Therefore, we take $r_N=1$ in our simulation. For the latent function $f(C_i)$, we construct its confidence interval $CI_{c}$ as 
\begin{align}
	CI_c=[\widehat{f(C_1)}-1.96(N(T-1))^{-1/2}\hat \sigma_{c}, \widehat {f(C_1)}+1.96(N(T-1))^{-1/2}\hat \sigma_{c}],
\end{align}
where 
\begin{gather}%
	\hat \sigma_{c}=\hat \sigma((\hat A_N(C_1), \hat B_N(C_1))\hat \Sigma_{1,N}^{-1}\hat \Sigma_{2,N}\hat \Sigma_{1,N}^{-1}	(\hat A_N(C_1), \hat B_N(C_1))^\top+\hat {\tilde \sigma})^{1/2},\\
	A_N(C_1)=-(\hat \Gamma_1^{-1}\hat \Gamma_{2,2,\alpha})^\top \hat l_{3N}(C_1)+\hat l_{4N}(C_1),~~~
	B_N(C_1)=-(\hat \Gamma_1^{-1}\hat \Gamma_{2,2,\beta})^\top \hat l_{3N}(C_1)+\hat l_{5N}(C_1),\\
	\hat	l_{3N}(C_1)=III_1(C_1),~~~\hat l_{4N}(C_i)=III_2(C_1),~~~~\hat l_{5N}(C_i)=III_3(C_1),\\
	\hat {\tilde \sigma}=N\sum_{t=1}^N\Big(\big(\sum_{j=1}^N(-\hat \Gamma_1^{-1}\frac{2(U_t-U_j)}{N(N-1)r_N})^\top K(\frac{\hat \delta^2_{tj}}{h_N})III_1(C_{1})+[\sum_{t=1}^N K(\frac{\hat \delta_{i_1t}^2}{h_N})]^{-1}K(\frac{\hat \delta_{i_1t}^2}{h_N})\big)^2.
\end{gather}%
\subsection{Proofs}\label{Sec:Proofs}
We define the notations used in the following proofs. For a give matrix $A$ (not necessarily summetric), let $\lambda_i(A)$ be its $i${th} largest (in mode) eigenvalue.  For any matrix $A=(a_{ij})\in \R^{n\times p}$ let  $|A|_e=(|a_{ij}|)\in \R^{n\times p}$, and $|A|_\infty=\max_{1\leq i\leq n, 1\leq j\leq p} |a_{ij}|$. For $A=(a_{ij})\in \R^{n\times p}$ and $B=(b_{ij})\in \R^{n\times p}$, write $A\preceq B$ when $a_{ij}\leq b_{ij}$ for $1\leq i\leq n$ and $1\leq j\leq p$.   Let $\Phi(x)$ be the CDF of standard $N(0,1)$.  Let $\mf 1^N$ be the $N$ dimensional column vector with all entries $1$.  Let $\FF_N$ denote the $\sigma$ field generated by $\{(\eta_{ij})_{1\leq i<j\leq N}, (C_i)_{1\leq i\leq N}\}$. %
We omit the supscript $N$ for short when no confusion arises. For any random variable $X$, write $\|X\|_{2,\mathcal G_N}=E^{1/2}(X^2|\mathcal G_N)$, and $Var(X|\GG_N)=\|X-E(X|\GG_N)\|^2_{2,\GG_N}$. For any symmetric semi-positive definite matrix $A$, write its eigen-decomposition as 
$A=U\Lambda U^\top$  where $U$ is an unitary matrix and $\Lambda $ is a diagonal matrix of eigenvalues. Then $A^{1/2}=U \Lambda^{1/2}U^\top$.  Recall 	$\delta(x,y)=\Big(\mathbb{E}_{U_1}\mathbb{E}^2_{U_2}\big[g(U_1,U_2)(g(x,U_1)-g(y,U_1))\big]\Big)^{1/2}$, and let $\delta_{ij}=\delta(C_i,C_j)$. %

\noindent \textbf{ Proof of \Cref{Theorem1}} \\
The proof is similar to but easier than the proof of Theorem \ref{diverging N}. Details are omitted for the sake of brevity. \hfill $\Box$

\noindent \textbf{ Proof of \Cref{diverging N}}\\ 
It suffices to show that
for any $\eta\in \R^2$, $\eta\neq (0,0)^\top$, as $\min(N,T)\rightarrow \infty$,
\begin{align}
	\sup_{x\in \mathbb R}|P(\sqrt{N(T-1)}\eta^\top\Sigma_{2,N}^{-1/2} \Sigma_{1,N}^{}(\hat \theta-\theta)/\sigma\leq x|\GG_N)- \Phi(\frac{x}{\sqrt{\eta^\top \eta}})|\rightarrow_p 0,
\end{align}
then the theorem follows from Cramer-Wold device.
Notice that in \Cref{sec:estimation} we show  $$\hat \theta-\theta =(\sum_{t=1}^{T-1}\Z_t^\top\X_t)^{-1}\sum_{t=1}^{T-1}\Z^\top_t\Delta \bepsilon_{t+1}.$$
By straightforward calculation we have
\begin{align}\label{plugin}
	\sqrt{N(T-1)}\eta^\top\Sigma_{2,N}^{-1/2} \Sigma_{1,N}^{}(\hat \theta-\theta)/\sigma-\frac{\eta^\top}{\sigma\sqrt{N(T-1)}} \Sigma^{-1/2}_{2,N}\sum_{t=1}^{T-1}\Z^\top_t\Delta \bepsilon_{t+1}\\=
	\frac{\eta^\top}{\sigma\sqrt{N(T-1)}}\Big(\Sigma_{2N}^{-1/2}\big(\Sigma_{1N}(\frac{\sum_{t=1}^{T-1}\Z^{-1}_t\X_t}{N(T-1)})^{-1}-I_2\big)\sum_{t=1}^{T-1}\Z_t^\top \Delta \bepsilon_{t+1}\Big)
	\\=o_p(T^{-1/2}det^{-1/2}(\Sigma_{2N})|det(\Sigma_{1N})|^{-1}tr(\Sigma_{2N})) \text{~~~~~~~given $\GG_N$},
\end{align}
where we have used \Cref{lemma1} and Cauchy inequality. Since
$det(\mf A)=\Pi_i \lambda_i(\mf A)$ and $tr(\mf A)=\sum \lambda_i(\mf A)$, by \Cref{New.4.1} (iii) that $T^{1/2}\lambda_{\min}^{1/2}(\Sigma_{2N})|det(\Sigma_{1N})|/\lambda_{\max}^{1/2}(\Sigma_{2N})\rightarrow \infty$, we have \eqref{plugin} is $o_p(1)$ given $\GG_N$.
We therefore shall prove that  
\begin{align}\label{new.eq56}
	\sup_{x\in \mathbb R}|P(\frac{\eta^\top}{\sigma\sqrt{N(T-1)}} \Sigma^{-1/2}_{2,N}\sum_{t=1}^{T-1}\Z^\top_t\Delta \bepsilon_{t+1}\leq x |\GG_N)-\Phi(\frac{x}{\sqrt{\eta^\top\eta}})|\rightarrow_p 0,
\end{align} as $T\rightarrow \infty$. Then combining Lemma \ref{lemma1} the theorem will follow.
Recall $\Z_{t}^{(m)}$ defined in Proposition \ref{prop2}. Conditional on $\GG_N$,
$\Z_t^{(m)}\Delta \bepsilon_{t+1}$ is an $m$-dependent vector with finite forth moment. Let $V^{(m)}_{N,T}=\frac{1}{\sqrt{N(T-1)}}\sum_{t=1}^{T-1}(\Z^{(m)}_t)^\top \Delta \bepsilon_{t+1}.$  In the following, if we can show the asymptotic normality of $V_{N,T}^{(m)}$, then by \Cref{prop2} the theorem will follow.

\textbf{Step I}. We first show that  for all $N$, $T$, it holds

\begin{align}\label{eq19}
	\lim_{m\rightarrow \infty} |E(V^{(m)}_{N,T}(V^{(m)}_{N,T})^\top|\GG_N)-\sigma^2\Sigma_{2,N}|_F=o_{a.s}(\chi^{m})%
\end{align}
for some constant $\chi=(|\alpha|+|\beta/r|)^2\in (0,1)$ and $r$ defined in \Cref{lemma4}.
To see this, notice that since $|E(\Y_t\bepsilon_{t+1}^\top|\GG_N)|_F=0$ almost surely,
\begin{align}
	E(V^{(m)}_{N,T}(V^{(m)}_{N,T})^\top|\GG_N)&=I+II+III,
\end{align}
where \begin{align}
	I=\frac{1}{N(T-1)} \sum_{t=1}^{T-1} E( (\Z_t^{(m)})^\top \Delta \bepsilon_{t+1}\Delta \bepsilon_{t+1}^\top\Z_t^{(m)}|\GG_N) =\frac{2\sigma^2}{N(T-1)} \sum_{t=1}^{T-1}  E((\Z_t^{(m)})^\top  \Z_t^{(m)} |\GG_N),
	\\II=\frac{1}{N(T-1)} \sum_{t=1}^{T-2}  E( (\Z_t^{(m)})^\top\Delta \bepsilon_{t+1}\Delta \bepsilon_{t+2}^\top \Z_{t+1}^{(m)} |\GG_N)=\frac{-\sigma^2}{N(T-1)} \sum_{t=1}^{T-2}  E((\Z_t^{(m)})^\top \Z_{t+1}^{(m)} |\GG_N),
	\\III=\frac{1}{N(T-1)} \sum_{t=1}^{T-2} E( (\Z_{t+1}^{(m)})^\top \Delta \bepsilon_{t+2}\Delta \bepsilon_{t+1}^\top\Z_t^{(m)}|\GG_N )=\frac{-\sigma^2}{N(T-1)} \sum_{t=1}^{T-2} E( (\Z_{t+1}^{(m)})^\top \Z_t^{(m)} |\GG_N).
\end{align}
Therefore, by Lemma \ref{lemma2}, it suffices to show that as $m\rightarrow \infty$, uniformly for $1\leq t\leq T-1,$ it holds
\begin{align}\label{eq27}
	\frac{1}{N}|E((\Z^{(m)}_t)^\top \Z^{(m)}_t |\GG_N)-E(\Z_t^\top \Z_t |\GG_N)|_F=o_{a.s.} (\chi^{m}),
\end{align}
and uniformly for $1\leq t\leq T-2$,
\begin{align}\label{eq28}
	\frac{1}{N}|E((\Z^{(m)}_{t+1})^\top \Z^{(m)}_t|\GG_N)-E(\Z_{t+1}^\top \Z_t|\GG_N)|_F=o_{a.s.}(\chi^m),\\\label{eq22} \frac{1}{N}|E((\Z^{(m)}_t)^\top \Z^{(m)}_{t+1}|\GG_N)-E(\Z_t^\top \Z_{t+1}|\GG_N)|_F=o_{a.s.}(\chi^m).
\end{align}
Equations \eqref{eq27}, \eqref{eq28} and \eqref{eq22} can be verified by checking each element of the matrices $E((\Z^{(m)}_t)^\top \Z^{(m)}_t|\GG_N)$, $E((\Z^{(m)}_{t+1})^\top \Z^{(m)}_t|\GG_N)$ and $E((\Z^{(m)}_{t})^\top \Z^{(m)}_{t+1}|\GG_N)$. For example, to see that $N^{-1}|\sum_{i=1}^N E(w_i^\top \Y^{(m)}_{t}w_i^\top\Y^{(m)}_{t-1}|\GG_N)-\sum_{i=1}^N E(w_i^\top \Y^{}_{t}w_i^\top\Y^{}_{t-1}|\GG_N)|=o_{a.s.} (\chi^{m})$ uniformly over $1\leq t\leq T$, just notice that by direct calculation,
\begin{align}\label{argue1}
	N^{-1}|\sum_{i=1}^N E(w_i^\top \Y^{(m)}_{t}w_i^\top\Y^{(m)}_{t-1}&|\GG_N)-\sum_{i=1}^N E(w_i^\top \Y^{}_{t}w_i^\top\Y^{}_{t-1}|\GG_N)|\\
	&=|\frac{\sigma^2}{N}(\sum_{i=0}^{m-1} tr((G_1^i)^\top W_N^\top W_N G_1^{i+1}))-\sum_{i=0}^\infty tr((G_1^i)^\top W_N^\top W_N G_1^{i+1}))|%
	\\	&\preceq \frac{\sigma^2}{N}\sum_{j=m}^{\infty}(|\alpha|+|r^{-1}\beta|)^{2j+1}tr(M_N^\top W_N^\top W_N M_N),
\end{align}
where we have used \Cref{lemma4}. On the other hand by   \Cref{lemma4} $$ tr(M_N\top W_N^\top W_N M_N))/N\rightarrow 0$$ almost surely. Notice that   \Cref{argue1} holds for all $t$. Hence together with \Cref{argue1} we have for any $N$ and $T$
\begin{align}
	\sup_{1\leq t\leq T}N^{-1}|\sum_{i=1}^N E(w_i^\top \Y^{(m)}_{t}w_i^\top\Y^{(m)}_{t-1}&|\GG_N)-\sum_{i=1}^N E(w_i^\top \Y^{}_{t}w_i^\top\Y^{}_{t-1}|\GG_N)|=o_{a.s.}(\chi^m).
\end{align}

\noindent The corresponding results of other components, which include $E((\Z^{(m)}_t)^\top \Z^{(m)}_t/N|\GG_N)-E(\Z_t^\top \Z_t/N|\GG_N)$, $E((\Z^{(m)}_{t+1})^\top \Z^{(m)}_t/N|\GG_N)-E(\Z_{t+1}^\top \Z_t/N|\GG_N)$ and $E((\Z^{(m)}_{t})^\top \Z^{(m)}_{t+1}|\GG_N)-E(\Z_{t}^\top \Z_{t+1}/N|\GG_N)$ can be verified similarly. Hence we show \eqref{eq19}.

\textbf{Step II}. Write $Z_{t,N,T}^{(m)}=Z_t^{(m)}/\sqrt{N(T-1)}$ and $\sigma^2_N(\eta)=Var(\sum_{t=1}^T\eta^\top(Z_{t,N,T}^{(m)})^\top \Delta{\bepsilon_{t+1}}|\GG_N)$. We shall prove that for every $N$, as $T\rightarrow \infty$, there exists $m=m_n$ such that for any $\eta\neq (0,0)^\top$,
\begin{align}
	\label{conidtionjason}\frac{m^3}{\sigma^4_N(\eta)}\sum_{t=1}^{T-1}E((\eta^\top (Z_{t,N,T}^{(m)})^\top \Delta \bepsilon_{t+1})^4|\GG_N)\rightarrow_p 0.
\end{align} Then by Theorem 1.4 of \cite{janson2021central} and the fact that 
$\sigma_N^2(\eta)=	E(\eta^\top V^{(m)}_{N,T}(V^{(m)}_{N,T})^\top\eta)$, we then have given $\GG_N$
\begin{align}\label{mdependent}
	(E(\eta^\top V^{(m)}_{N,T}(V^{(m)}_{N,T})^\top\eta |\GG_N))^{-1/2}\frac{1}{\sqrt{N(T-1)}}\sum_{t=1}^{T-1}\eta^\top(\Z^{(m)}_t)^\top \Delta \bepsilon_{t+1}\Rightarrow_{T\rightarrow \infty} N(0,1).
\end{align}

To show \cref{conidtionjason},  we write  $Z_{t,N}^{(m)}=Z_t^{(m)}/\sqrt{N}=\sqrt{T-1}Z_{t,N,T}^{(m)}$ and shall show that
\begin{description}
	\item(a)$E((\eta^\top (Z_{t,N}^{(m)})^\top \Delta \bepsilon_{t+1})^4|\GG_N)$ is bounded with probability going to $1$, uniformly for $1\leq t\leq T-1$ and for all $N$. 
\end{description}	
Once (a)  is proved, then the LHS of \cref{conidtionjason} is bounded with probability going to $1$ by
\begin{align}
	\frac{Mm^3(T-1)}{\sigma_N^4(\eta)(T-1)^2}=	\frac{Mm^3}{\sigma_N^4(\eta)(T-1)}
\end{align}
for some large constant $M$. By Step I, we have that
\begin{align}\label{SigmaNeta}
	|\sigma_N^2(\eta)-\eta^\top \Sigma_{2N}\eta|\leq \chi^m|\eta|^2.
\end{align}
Take  $m=\tilde m_n$. Since  $\tilde m_n\rightarrow \infty$,  $\frac{\tilde m_n}{|\log \lambda_{\min}(\Sigma_{2N})|}\rightarrow \infty$ and 
$\frac{\tilde m_n^3}{T\lambda^2_{\min}(\Sigma_{2N})}\rightarrow 0$,  \cref{conidtionjason} holds.

We now prove (a). To save notation, we use the term `bounded' in the remaining of the proof for `bounded uniformly for $1\leq t\leq T-1$ and for all $N$' if no confusion arises. To see this, notice that by Jansen's inequality, it suffices to show that $E((\eta^\top (\Z_{t,N}^{(m)})^\top \bepsilon_{t+1})^4|\GG_N)$ and $E((\eta^\top (\Z_{t,N}^{(m)})^\top \bepsilon_{t})^4|\GG_N)$ are bounded with probability going to $1$. We only show the former for the sake of brevity. The latter follows similarly. Notice that by Lemma \ref{fourth},   using the fact that $\Z_t^{(m)}$ and $\bepsilon_{t+1}$ are independent, then almost surely
\begin{align}
	E((\eta^\top (\Z_t^{(m)})^\top \bepsilon_{t+1})^4|\GG_N)=E(E((\bepsilon_{t+1}^\top \Z_t^{(m)}\eta \eta^\top(\Z_t^{(m)})^\top\bepsilon_{t+1})^2|\Z_t^{(m)},\GG_N)|\GG_N)\\\leq
	(2\sigma^4+|E\epsilon_{1,1}^4-3\sigma^4|)E((\eta^\top(Z_t^{(m)})^\top Z_t^{(m)}\eta)^2|\GG_N)+E(E^2(\bepsilon_{t+1}^\top \Z_t^{(m)}\eta \eta^\top(\Z_t^{(m)})^\top\bepsilon_{t+1}
	|\Z_t^{(m)},\GG_N)|\GG_N)
	\\=(3\sigma^4+|E\epsilon_{1,1}^4-3\sigma^4|)E((\eta^\top(\Z_t^{(m)})^\top \Z_t^{(m)}\eta)^2|\GG_N).\label{new.27}\end{align}
Hence by similar arguments to Lemma \ref{lemma2} and Cauchy inequality,
$E((\eta^\top (\Z_{t,N}^{(m)})^\top \bepsilon_{t+1})^4|\GG_N)$ is bounded with probability tending to 1, if with probability tending to 1
\begin{align}\label{new.28}
	E((\sum_{i=1}^N(Y^{(m)}_{i,t-1})^2)^2|\GG_N)/N^2,~~~~~~
	E((\sum_{i=1}^N(w_i^\top \Y^{(m)}_{t-1})^2)^2|\GG_N)/N^2
\end{align} 
are bounded.
For the sake of brevity, we shall show 	$E((\sum_{i=1}^N(w_i^\top \Y^{(m)}_{t-1})^2)^2|\GG_N)/N^2$ is bounded with probability going to $1$. The boundedness of $E((\sum_{i=1}^N(Y^{(m)}_{i,t-1})^2)^2|\GG_N)/N^2$ will follow similarly. Write $\tilde  Y_{t}^{(m)}=(I-G_1)^{-1}G_0$, $\bar  Y_{t}^{(m)}=\sum_{j=0}^mG_1^j\bepsilon_{t-j}$. Let $C$ be a generic large constant which varies from line to line. Thus $\Y_{t}^{(m)}=\tilde \Y_{t}^{(m)}+\bar \Y_{t}^{(m)}$. Using the inequality that for any two vectors $a$ and $b$, it holds $2|a^\top b|\leq a^\top a+b^\top b$, we have%
\begin{align}\label{new.eq67}
	E((\sum_{i=1}^N(w_i^\top \Y^{(m)}_{t-1})^2)^2|\GG_N)=
	E((\Y^{(m)}_{t-1})^\top W_N^\top W_N\Y^{(m)}_{t-1}\Y^{(m)}_{t-1})^\top W_N^\top W_N\Y^{(m)}_{t-1}|\GG_N)	\\\leq C
	E((\bar \Y^{(m)}_{t-1})^\top W_N^\top W_N\bar \Y^{(m)}_{t-1}(\bar \Y^{(m)}_{t-1})^\top W_N^\top W_N\bar \Y^{(m)}_{t-1}|\GG_N)\\+C
	E((\tilde \Y^{(m)}_{t-1})^\top W_N^\top W_N\tilde \Y^{(m)}_{t-1}(\tilde \Y^{(m)}_{t-1})^\top W_N^\top W_N\tilde \Y^{(m)}_{t-1}|\GG_N):=C(A+B),
\end{align}
where $A$ and $B$ are defined in an obvious manner. Notice that $A= A_1+A_2$ where
\begin{align}
	A_1=var(\bar \Y^{(m)}_{t-1})^\top W_N^\top W_N\bar \Y^{(m)}_{t-1}|\GG_N),~~A_2=(E(\bar \Y^{(m)}_{t-1})^\top W_N^\top W_N\bar \Y^{(m)}_{t-1}|\GG_N))^2.
\end{align}
By similar but easier argument to the argument of evaluating quantity $IV$ in $\mathcal M_{11}$ in the proof of \Cref{lemma1}, with probability approaching $1$ we have
\begin{align}
	(A_1)^{1/2}/N\leq \frac{C}{N}\sum_{i,j\geq 0} tr^{1/2}((G_1^i)^
	\top G_1^jW_N^\top W(G_1^j)^\top W_N^\top W_N G_1^i)\leq C_0,
\end{align}
where the limit is due to \Cref{lemma4}. Straightforward calculations using \Cref{lemma4} show that with probability tending to $1$, \begin{align} \label{eq71}
	A_2/N^2\leq \frac{C^2tr^2(M_N^\top W_N^\top W_N M_N )}{N^2}\leq C_0,
\end{align}
where the matrix $M_N$ is determined by the network structure with the exact formula defined in \Cref{lemma4}, and the zero limit is due to the similar  argument to the proof of \Cref{new.eq43} in the proof of \Cref{lemma4}.

For $B$, notice  that $G_0={\mf f}_1+{\mf f}_2$, where \begin{align}%
	\mf f_1=( f (C_1),...., f (C_N))^\top,~~
	\mf f_2=(\gamma^\top U_1,....,\gamma^\top U_N^\top)^\top.
\end{align} Similarly to \eqref{new.eq67}, we have
\begin{align}
	B\leq C B_1+CB_2+CB_3,
\end{align}
where 
\begin{align}
	B_1=	E((((I-G_1)^{-1}\mf f_1)^\top((I-G_1)^{-1}\mf f_1))^2|\GG_N),\\
	B_2=	E((((I-G_1)^{-1}( {\mf f_2}-E{\mf f}_2))^\top((I-G_1)^{-1}( {\mf f_2}-E{\mf f}_2))^2|\GG_N),\\
	B_3=	E((((I-G_1)^{-1}E {\mf f}_2)^\top((I-G_1)^{-1}E {\mf f}_2))^2|\GG_N).
\end{align}

Using \Cref{lemma4} and the fact that $(I-G_1)^{-1}=\sum_{j=0}^\infty G_1^j$, we have that
$B_1\leq (\mf f_1^\top M_N^\top M_N\mf f_1)^2$. By the definition of $M_N$, we shall see that every element of  $M_N$ is positive, and also $M_N\mf 1\preceq C \mf 1$ for some large constant $C$.
Using the boundedness of $\mf f_1$, we shall see that
$B_1\leq CN^2$ with probability tending to $1$. Similarly $B_3\leq CN^2$ with probability tending to $1$. To shorten the notation write $G_2=(I-G_1)^{-\top}( I-G_1)^{-1}$. For $B_2$, notice that $E(B_2|W)=B_2'$  since $\sigma(W)\subset \GG_N$,
where
\begin{align} B_2'=E((((I-G_1)^{-1}( {\mf f_2}-E{\mf f}_2))^\top((I-G_1)^{-1}( {\mf f_2}-E{\mf f}_2))^2|W)
	\\\leq E^2((\mf f_2-E\mf f_2)^\top G_2 (\mf f_2-E\mf f_2)|G_2)+C tr(G_2G_2^\top)\leq C((tr(G_2))^2+tr(G_2G_2^\top)),
\end{align}
where  we have used \Cref{Assumption2.1_2}, Lemma \ref{fourth} and Fubini theorem (which is used to close the gap between conditioning on $W$ instead of conditioning on $G_2$). Notice  that $G_2=\sum_{j_1=0}^\infty \sum_{j_2=0}^\infty (G_1^{j_1})^\top G_1^{j_2}$, and  $tr(G_2G_2^\top)\leq (tr(G_2))^2$ due to the symmetric and semi-positive definiteness of $G_2$. Then using similar argument to \Cref{eq71} using \Cref{lemma4}, it follows that
$B_2'/N^2\rightarrow_p 0$. Since $B_2$ is always non-negative, we show that with probability tending to $1$, $B_2\leq CN^2$. Hence step II is proved.

\textbf{Step III}. 
We shall show with  probability tending to $1$, for any $\eta\neq (0,0)^\top$,
condition on $\GG_N$,
\begin{align}\label{Normal1}
	(\eta^\top \Sigma_{2N}\eta)^{-1/2}\frac{1}{\sqrt{N(T-1)}}\sum_{t=1}^{T-1}\eta^\top(\Z_t)^\top \Delta \bepsilon_{t+1}\Rightarrow_{T\rightarrow \infty} N(0,1).
\end{align}

To see this, notice that 
\begin{align}
	&	(\eta^\top \Sigma_{2N}\eta)^{-1/2}\frac{1}{\sqrt{N(T-1)}}\sum_{t=1}^{T-1}\eta^\top(\Z_t)^\top \Delta \bepsilon_{t+1}-	\notag\\&(E(\eta^\top V^{(m)}_{N,T}(V^{(m)}_{N,T})^\top\eta |\GG_N))^{-1/2}\frac{1}{\sqrt{N(T-1)}}\sum_{t=1}^{T-1}\eta^\top(\Z^{(m)}_t)^\top \Delta \bepsilon_{t+1}:	=A+B,
\end{align}
where 
\begin{align}
	A=(N(T-1))^{-1/2}\Big((\eta^\top \Sigma_{2N}\eta)^{-1/2}-(E(\eta^\top V^{(m)}_{N,T}(V^{(m)}_{N,T})^\top\eta |\GG_N))^{-1/2}\Big)
	\Big(\sum_{t=1}^{T-1}\eta^\top(\Z_t)^\top \Delta \bepsilon_{t+1}\Big),
	\\
	B=(N(T-1))^{-1/2}E(\eta^\top V^{(m)}_{N,T}(V^{(m)}_{N,T})^\top\eta |\GG_N))^{-1/2}
	\Big(\sum_{t=1}^{T-1}\eta^\top(\Z_t)^\top \Delta \bepsilon_{t+1}-\sum_{t=1}^{T-1}\eta^\top(\Z^{(m)}_t)^\top \Delta \bepsilon_{t+1}\Big).
\end{align}
By \Cref{SigmaNeta},
\begin{align}
	\Big((\eta^\top \Sigma_{2N}\eta)^{-1/2}-(E(\eta^\top V^{(m)}_{N,T}(V^{(m)}_{N,T})^\top\eta |\GG_N))^{-1/2}\Big)=O(\chi^m/\sqrt{\lambda_{\min}(\Sigma_{2N})}).
\end{align} 
By evaluating the variance of $A$ we shall see that conditional on $\GG_N$, $A=O_p(\chi^m\frac{\lambda^{1/2}_{\max}(\Sigma_{2N})}{\lambda^{1/2}_{\min}(\Sigma_{2N})})$. Using \Cref{prop2}, it follows that
$B=O_p(\chi^{m/2}/\lambda^{1/2}_{\min}(\Sigma_{2N}))$. By taking $m=\tilde m_n$, \Cref{mdependent} and Slutsky theorem, \cref{Normal1} follows. Furthermore, by considering $\eta=\Sigma_{2,N}^{-1/2}\eta'$. 
and using Cramer Wold device, assertion \eqref{new.eq56} holds. 

\hfill $\Box$\\

\ \\
\noindent\textbf{Proof of \Cref{Thm3}}\\
Recall that 	\begin{align}\label{tildee}
	\tilde e_{i,t}=Y_{i,t}-\hat \alpha Y_{i,t-1}-\hat \beta \frac{\sum_j(Y_{j,t-1}A_{ij})}{\sum_j A_{ij}}.
\end{align}
Then by \Cref{tildee} and \Cref{singlebeta}, it follows  that
\begin{align}
	\tilde e_{i,t}=\epsilon_{i,t}+\gamma^\top U_i+ f(C_i)+(\alpha-\hat \alpha)Y_{i,t-1}+(\beta-\hat \beta)\frac{\sum_j Y_{j,t-1}A_{ij}}{\sum_j A_{ij}}.
\end{align}
then
\begin{align}\label{eq143}
	\hat e_i=\bar \epsilon_i+\gamma^\top U_i+ f(C_i)+(\alpha-\hat \alpha)\bar Y_{i}+(\beta-\hat \beta) w_i^\top \bar \Y.
\end{align}
Therefore by the definition of $\hat \gamma$, we have that
\begin{align}
	\hat \gamma=\hat \Gamma_1^{-1}\hat \Gamma_2,
\end{align}
where 	$\hat \Gamma_1$ is defined in \cref{hatgamma1}
and
\begin{align}
	\hat \Gamma_{2}=\hat \Gamma_{2,1}+\hat \Gamma_{2,2},
\end{align}
where 
\begin{align}
	&	\hat \Gamma_{2,1}=\frac{2}{N(N-1)r_N}\sum_{i=1}^{N-1}\sum_{j=i+1}^N(U_i-U_j) [(\bar \epsilon_i+\gamma^\top U_i+ f(C_i))-(\bar \epsilon_j+\gamma^\top U_j+ f(C_j))]K(\frac{\hat \delta^2_{ij}}{h_N})\\&=\hat \Gamma_1\gamma +\frac{2}{N(N-1)r_N}\sum_{i=1}^{N-1}\sum_{j=i+1}^N(U_i-U_j) [(\bar \epsilon_i+ f(C_i))-(\bar \epsilon_j+ f(C_j))]K(\frac{\hat \delta^2_{ij}}{h_N}):=\hat \Gamma_1\gamma+\tilde \Gamma_{2,1},\\
	&	\hat \Gamma_{2,2}=\frac{2(\alpha-\hat \alpha)}{N(N-1)r_N}\sum_{i=1}^{N-1}\sum_{j=i+1}^N(U_i-U_j)(\bar Y_i-\bar Y_j)K(\frac{\hat \delta^2_{ij}}{h_N})\\&\quad \quad \quad \quad \quad\quad\quad\quad+\frac{2(\beta-\hat \beta )}{N(N-1)r_N}\sum_{i=1}^{N-1}\sum_{j=i+1}^N(U_i-U_j)(w_i-w_j)^\top K(\frac{\hat \delta^2_{ij}}{h_N})  \bar  \Y.
\end{align}
Therefore %
\begin{align}\label{hatgammaforumla}
	\hat \gamma-\gamma=\hat \Gamma_1^{-1}(\tilde \Gamma_{2,1}+\hat \Gamma_{2,2}).
\end{align}
Following the proof of Proposition 2 of \cite{auerbach2022identification} (see the quantity of $\Gamma_n$ there),
$\hat \Gamma_1-\Gamma_n\rightarrow_p  0$ where $\Gamma_n=r_N^{-1}E[(U_i-U_j) (U_i-U_j)^\top K(\frac{\delta^2_{ij}}{h_N})]$, and the smallest eigenvalues of $\Gamma_n$ are bounded away from $0$. Also, by the proof of Proposition 2 of \cite{auerbach2022identification} (See the quantity of $U_n$ there) it follows that $\tilde \Gamma_{2,1}=o_p(1)$ since $\bar \epsilon_i$ are $i.i.d.$ random variables with variance $O(1/T)$ with finite eighth moments. Hence $\hat \Gamma_1^{-1} \tilde \Gamma_{2,1}=o_p(1)$. Therefore it suffices to show that
$\hat \Gamma_{2,2}=o_p(1)$.  By mean value theorem,
\begin{align}
	\frac{1}{N(N-1)r_N}\sum_{i=1}^{N-1}\sum_{j=i+1}^N&(U_i-U_j)(\bar Y_i-\bar Y_j)K(\hat \delta^2_{ij}/h_N)\\&= \frac{1}{N(N-1)r_N}\sum_{i=1}^{N-1}\sum_{j=i+1}^N(U_i-U_j)(\bar Y_i-\bar Y_j)K( \delta^2_{ij}/h_N)\\&+
	\frac{1}{N(N-1)r_N}\sum_{i=1}^{N-1}\sum_{j=i+1}^N(U_i-U_j)(\bar Y_i-\bar Y_j)K'(\frac{\iota_{ij}}{h_N})\frac{\hat \delta^2_{ij}-\delta^2_{ij}}{h_N},
\end{align}
where $\iota_{ij}'s$ are numbers between $\delta^2_{ij}$ and $\hat \delta^2_{ij}$.
Notice that for piecewise Lipschitz graphon, we have for every number $\epsilon>0$\begin{align}
	\inf_{u\in [0,1]}\int \mathbbm 1(v\in [0,1]: \sup_{\tau\in[0,1]}|g(u,\tau)-g(v,\tau)<\epsilon|)dv>0.
\end{align}	
Hence by Lemma B1 of \cite{auerbach2022identification},  $\max_{i\neq j}\frac{|\hat \delta^2_{ij}-\delta^2_{ij}|}{h_N}=o_p(N^{-\gamma/4})$. Together with \Cref{eq84}, using Cauchy inequality and kernel and bandwidth conditions we have
\begin{align}
	\frac{1}{N(N-1)r_N}\sum_{i=1}^{N-1}\sum_{j=i+1}^N(U_i-U_j)(\bar Y_i-\bar Y_j)K'(\frac{\iota_{ij}}{h_N})\frac{\hat \delta^2_{ij}-\delta^2_{ij}}{h_N}=o_p(1),
\end{align}
where $o_p(1)$ represents a $p$ dimensional vector with each entry $o_p(1)$.
On the other hand, notice that by \Cref{eq84} and Cauchy inequality,
\begin{align}
	\frac{1}{N(N-1)r_N}\sum_{i=1}^{N-1}\sum_{j=i+1}^NE(|U_i-U_j||\bar Y_i-\bar Y_j||\FF_N)K( \delta^2_{ij}/h_N)\leq  \frac{M}{N(N-1)r_N}\sum_{i=1}^{N-1}\sum_{j=i+1}^NK( \delta^2_{ij}/h_N)
\end{align}
with probability tending to $1$  for some large constant $M$. Since $U_i$ and $\delta_{ij}$ are $\FF_N$ measurable,
by taking expectation on both side of the above inequality and use kernel and bandwidth conditions it follows that 
$$\frac{1}{N(N-1)r_N}\sum_{i=1}^{N-1}\sum_{j=i+1}^N(U_i-U_j)(\bar Y_i-\bar Y_j)K( \delta^2_{ij}/h_N)=O_p(1).$$ Since \Cref{diverging N} shows that $\hat \alpha-\alpha=O_p(\frac{1}{\sqrt {N\lambda_{\min}(\Sigma^\top_{1N}\Sigma^{-1}_{2N}\Sigma_{1N})T}})$, we shall see that the first summand in $\hat \Gamma_{2,2}$ is of the order $O_p(\frac{1}{\sqrt{N\lambda_{\min}(\Sigma^\top_{1N}\Sigma^{-1}_{2N}\Sigma_{1N})T}})$. Similarly using  kernel and bandwidth conditions, mean value theorem, \Cref{eq84}, we shall see that the second summand of  $\hat \Gamma_{2,2}$ is $2(\beta-\hat \beta)$ times
\begin{align}%
	\frac{1}{N(N-1)r_N}\sum_{i=1}^{N-1}\sum_{j=i+1}^N(U_i-U_j)(w_i-w_j)^\top K(\frac{\hat \delta^2_{ij}}{h_N})  \bar  \Y+o_p(1),
\end{align}%
where by kernel and bandwidth conditions the leading term is $O_p(1)$. Since by \Cref{diverging N} $\beta-\hat \beta=O_p(\frac{1}{\sqrt{N\lambda_{\min}(\Sigma^\top_{1N}\Sigma^{-1}_{2N}\Sigma_{1N})T}})$,   we shall see that the second summand in $\hat \Gamma_{2,2}$ is of the order $O_p(\frac{1}{\sqrt{N\lambda_{\min}(\Sigma^\top_{1N}\Sigma^{-1}_{2N}\Sigma_{1N})T}})$. Therefore
\begin{align}
	\hat \Gamma_{2,2}=O_p(\frac{1}{\sqrt{N\lambda_{\min}(\Sigma^\top_{1N}\Sigma^{-1}_{2N}\Sigma_{1N})T}}),
\end{align}
which proves (i).

\noindent\textbf{ Proof of \Cref{Thm4}.} By the definition of $\widehat{ f(C_i)}$ and \Cref{eq143}, we have
\begin{align}
	\widetilde {\widehat {f(C_i)} }=I(C_i)+II(C_i)+III(C_i).
\end{align}

By the proof of Proposition 2 of \cite{auerbach2022identification}, %
\begin{align}
	\max_i|I(C_i)+II(C_i)+(\gamma-\hat \gamma)^\top III_1(C_i)- f(C_i)|=o_p(1).
\end{align}
By \Cref{diverging N}, it suffices to show that $$\max_{i}|III_2(C_i)|=o_p((N\lambda_{\min}(\Sigma^\top_{1N}\Sigma^{-1}_{2N}\Sigma_{1N})T)^{1/2})$$ and 
$$\max_{i}|III_3(C_i)|=o_p((N\lambda_{\min}(\Sigma^\top_{1N}\Sigma^{-1}_{2N}\Sigma_{1N})T)^{1/2}).$$ 
By  \Cref{eq84} and Jansen's inequality, we shall see that for $1\leq i\leq N$, $E|\bar Y_i|^4\leq C$ and $E|w_i^\top \bar \Y|\leq C$ for some large constant $C$. By using the inequality 
$\max_{1\le i\leq N}|X_i|^4\leq \sum_{i=1}^N|X_i^4|$ for any random variables $X_i$ and any integer $N$, 
we have \begin{align}\label{eq161}
	\max_i| III_2(C_i)|\leq \max_{1\leq i\leq N}| \bar Y_i|=O_p(N^{1/4}), ~~\max_i| III_3(C_i)|\leq \max_{1\leq i\leq N}|  w_i^\top \bar \Y|=O_p(N^{1/4}).
\end{align}
By \Cref{eq161} and the fact that $N^{1/4}=o(N\lambda_{\min}(\Sigma^\top_{1N}\Sigma^{-1}_{2N}\Sigma_{1N})T)^{1/2})$ the theorem follows. \hfill $\Box$

\ \\
\noindent\textbf{Proof of \Cref{Thm5}}\\\ \\
Recall from \Cref{Thm3} that
\begin{align}
	\hat \gamma-\gamma=\hat \Gamma_1^{-1}(\tilde \Gamma_{2,1}+\hat \Gamma_{2,2}).
\end{align}
Notice that 
\begin{align}
	E(\tilde \Gamma_{2,1}|\GG_N)=\frac{2}{N(N-1)r_N}\sum_{i=1}^{N-1}\sum_{j=i+1}^N(U_i-U_j)( f(C_i)- f(C_j))K(\frac{\hat \delta^2_{ij}}{h_N}),\\
	\tilde \Gamma_{2,1}-E (\tilde \Gamma_{2,1}|\GG_N)=\frac{2}{N(N-1)r_N}\sum_{i=1}^N\sum_{j=i+1}^N(U_i-U_j)(\bar \epsilon_i-\bar \epsilon_j)K(\frac{\hat \delta^2_{ij}}{h_N}).\label{new.172}
\end{align}

Consider $\delta$ in \Cref{furtherf}. Then by Lemma 1 of \cite{auerbach2022identification}, there exists $\delta'>0$ such that if $\delta_{ij}\leq \delta'$, then $\{\int_0^1(g(C_i,x)-g(C_j,x))^2dx\}^{1/2}\leq \delta$ which implies $f(C_i)=f(C_j)$ due to \Cref{furtherf}. Now consider the event $\{\max_{i\neq j}|\hat \delta^2_{ij}-\delta_{ij}^2|\leq MN^{-\gamma/4}h_N\}$. On this event, for sufficiently large $N$ we have that if $K(\frac{\hat \delta_{ij}^2}{h_N})\neq 0$ then $\delta_{ij}^2\leq 2h_N\leq \delta'^2$ due to the fact that $K(\cdot)$ is supported on $[0,1)$ and $h_N\rightarrow 0$. Therefore, by Lemma B1 of \cite{auerbach2022identification}
\begin{align}\label{new.eq176}
	P\Big(E(\tilde \Gamma_{2,1}|\GG_N)=0\Big)\geq P\Big(\max_{i\neq j}|\hat \delta^2_{ij}-\delta^2_{ij}|\leq MN^{-\gamma/4}h_N\Big)\rightarrow 1.
\end{align}
On the other hand, 
notice that \begin{align}\sum_{i=1}^{N-1}\sum_{j=i+1}^N(U_i-U_j) [\bar \epsilon_i -\bar \epsilon_j]K(\frac{\hat \delta^2_{ij}}{h_N})
	=\sum_{i=1}^N\sum_{j=1}^N(U_i-U_j)K(\frac{\hat \delta^2_{ij}}{h})\bar \epsilon_i,
\end{align} 
thus
\begin{align}
	\tilde \Gamma_{2,1}-E (\tilde \Gamma_{2,1}|\GG_N)=\frac{2}{N(N-1)r_N}\sum_{i=1}^N\sum_{j=1}^N(U_i-U_j)K(\frac{\hat \delta^2_{ij}}{h_N})\bar \epsilon_i\\
	=\frac{2}{r_N}\sum_{t=1}^T \sum_{i=1}^N\frac{1}{TN(N-1)}\sum_{j=1}^N(U_i-U_j)K(\frac{\hat \delta^2_{ij}}{h_N})\epsilon_{i,t}.
\end{align}
Let $\nu=%
(\tilde \Gamma_{2,1}-E (\tilde \Gamma_{2,1}|\GG_N))r_N/2$.  Then at the end of this proof of \Cref{Thm5} we shall  verify that
conditional on $\GG_N$, with probability going to $1$,
\begin{align}\label{new.180}
	\sqrt{N(T-1)}\tilde \Sigma_{3N}(\hat \alpha-\alpha, \hat \beta-\beta,\nu)^\top/\sigma\Rightarrow N(0,I_{2+p}),
\end{align}
where 
\begin{align}
	\tilde	\Sigma_{3N}=\begin{pmatrix}
		\Sigma_{2,N}^{-1/2}\Sigma_{1,N}& 0\\
		0&\Sigma^{-1/2}_{3N}
	\end{pmatrix}.
\end{align}
Note that the zero covariance are due to the fact that $\Delta \epsilon_{i_1,t}$ and $\bar \epsilon_{i_2}$ are uncorrelated for $1\leq i_1,i_2\leq N$ and $1\leq t\leq T-1$.

Now we analyze $\hat \Gamma_{2,2}$. Recall that
\begin{align}\label{new.178}
	\hat \Gamma_{2,2}=(\alpha-\hat \alpha)\hat \Gamma_{2,2,\alpha}+(\beta-\hat \beta )\hat \Gamma_{2,2,\beta},
\end{align}
where
\begin{align}
	\hat \Gamma_{2,2,\alpha}=\frac{1}{N(N-1)r_N}\sum_{i=1}^{N-1}\sum_{j=1,j\neq i}^N(U_i-U_j)(\bar Y_i-\bar Y_j)K(\frac{\hat \delta^2_{ij}}{h_N}),\\
	\hat \Gamma_{2,2,\beta}=\frac{1}{N(N-1)r_N}\sum_{i=1}^{N-1}\sum_{j=1,j\neq i}^N(U_i-U_j)(w_i-w_j)^\top K(\frac{\hat \delta^2_{ij}}{h_N})  \bar  \Y.
\end{align}

Notice that by Proposition \ref{prop1}, given $\GG_N$, 
$\bar Y_i=e_i^\top (I-G_1)^{-1}G_0+e_i^\top \sum_{j=0}^\infty G_1^j\sum_{t=1}^T\bepsilon_{t-j-1}/T$, $a.s.$. %
Then 
\begin{align}
	E(\hat \Gamma_{2,2,\alpha}|\GG_N)=\frac{1}{N(N-1)r_N}\sum_{i=1}^{N-1}\sum_{j=1,j\neq i}^N(U_i-U_j)(e_i-e_j)^\top (I-G_1)^{-1}G_0K(\frac{\hat \delta^2_{ij}}{h_N}),\\
	E(\hat  \Gamma_{2,2,\beta}|\GG_N)=\frac{1}{N(N-1)r_N}\sum_{i=1}^{N-1}\sum_{j=1,j\neq i}^N(U_i-U_j)(w_i-w_j)^\top K(\frac{\hat \delta^2_{ij}}{h_N}) (I-G_1)^{-1}G_0.
\end{align}
By \Cref{Conditional-Constant}, \cref{new.180}, equation \eqref{new.178},\Cref{diverging N} and Slutsky theorem, we shall see that given $\GG_N$,
\begin{align}
	\sqrt{N(T-1)}v_N^{-1/2}\hat \Gamma_{2,2}\Rightarrow N(0, I_p).
\end{align}
As a consequence, the theorem follows from the fact that $\hat \Gamma_1$ is $\GG_N$ measurable and that
$$\hat \Gamma_{1}-r_N^{-1}E[(U_i-U_j) (U_i-U_j)^\top K(\frac{\delta^2_{ij}}{h_N})]=o_p(1),$$ where the smallest  eigenvalue of  $r_N^{-1}E[(U_i-U_j) (U_i-U_j)^\top K(\frac{\delta^2_{ij}}{h_N})]$ is bounded away from $0$, which is discussed in below \eqref{hatgammaforumla} of the proof of \Cref{Thm3}.

We now verify \eqref{new.180}. We first verify the asymptotic normality of the $q_{th}$ $(1\leq q\leq p)$ coordinate
of	$r_N/2(\tilde \Gamma_{2,1}-E (\tilde \Gamma_{2,1}|\GG_N))$, i.e., the asymptotic normality of 
\begin{align}\label{122}\sum_{t=1}^T \frac{1}{\sqrt{NT}}\sum_{i=1}^N\frac{1}{(N-1)}\sum_{j=1}^N(U_{i,q}-U_{j,q})K(\frac{\hat \delta^2_{ij}}{h_N})\epsilon_{i,t}
\end{align}
given $\GG_N$. Then  we can justify \eqref{new.180} using similar argument and also the similar argument in the proof of \Cref{diverging N} as well as the results of \Cref{diverging N}, Cauchy's, Jansen's inequality and Cramer-Wold device.
Let $\zeta_i=\frac{1}{(N-1)}\sum_{j=1}^N(U_{i,q}-U_{j,q})K(\frac{\hat \delta^2_{ij}}{h_N})$ then \eqref{122} has the representation of $\sum_{t=1}^T\sum_{i=1}^N(NT)^{-1/2}\zeta_i \epsilon_{i,t}$. By Lyapunov CLT, it suffices to show that 
the following quantity is bounded with probability tending to $1$
\begin{align}\label{eq123}
	\frac{1}{N}\sum_{i=1}^N\big(\sum_{j=1}^N\frac{(U_{i,q}-U_{j,q})}{N-1}K(\frac{\hat \delta_{ij}^2}{h_N})\big)^2
\end{align}
and that
\begin{align}\label{eq124}
	\sum_{t=1}^T \frac{1}{N^2T^2} E((\sum_{i=1}^N\zeta_i\epsilon_{i,t})^4|\GG_N)\rightarrow_p 0.
\end{align}  
To prove the boundedness of \eqref{eq123}, notice that by the boundedness of $K(\cdot)$, \eqref{eq123} is bounded by 
\begin{align}
	\frac{C}{N}\sum_{i=1}^N(\sum_{j=1}^N
	|U_{i,q}-U_{j,q}|/(N-1))^2\leq 
	\frac{CN}{(N-1)^2}\sum_{i=1}^N(\sum_{j=1}^N
	(|U_{i,q}|+|U_{j,q}|)/N)^2
	\\	=\frac{CN}{(N-1)^2}\sum_{i=1}^N(|U_{i,q}|+\frac{1}{N}\sum_{j=1}^N|U_{j,q}|)^2\leq \frac{2NC}{(N-1)^2}\sum_{i=1}^N|U_{i,q}|^2+\frac{2CN^2}{(N-1)^2}(\frac{1}{N}\sum_{j=1}^N|U_{j,q}|)^2\\\rightarrow_{a.s.} 2C(E|U_{1,q}|^2+(E|U_{1,q}|)^2)
\end{align}
for some large constant $C$ where for the convergence we have used the strong law of large number.

To verify \eqref{eq124}, by using conditional  version of Rosenthal inequalities (\cite{rosenthal1970subspaces}) and the fact that $\epsilon_{i,t}$ are independent  random variables independent of $\GG_N$ with finite fourth moments and $\zeta_i's$ are $\GG_N$ measurable, it suffices to show that
\begin{align}
	\frac{1}{N^2T^2}\sum_{t=1}^T (\sum_{i=1}^N \zeta_i^4)\rightarrow_p 0, ~~~	\frac{1}{N^2T^2}\sum_{t=1}^T (\sum_{i=1}^N \zeta_i^2)^2\rightarrow_p 0. 
\end{align}
For the sake of brevity we only prove the second argument. The first can be shown similarly.  Notice that by the boundedness of $K(\cdot)$, we have $\zeta_i\leq \frac{C}{N}\sum_{j=1}^N(|U_{i,q}|+|U_{j,q}|)$ when $N$ is sufficiently large and for some large constant $C$. Hence we only need to show that
\begin{align}\label{127}
	\frac{1}{N^2T}\Big(\sum_{i=1}^N  (\frac{1}{N}\sum_{s=1}^N(|U_{i,q}|+|U_{s,q}|))^2 \Big )^2\rightarrow_p 0
\end{align}
as $\min(N,T)\rightarrow \infty$. Notice that the LHS of \eqref{127} is further bounded by 
\begin{align}
	\frac{C}{N^2T}\Big(\sum_{i=1}^N \frac{1}{N}(\sum_{s=1}^N|U_{s,q}|^2+\sum_{s=1}^N|U_{i,q}^2|) \Big)^2
	= \frac{C}{N^4T}(2N\sum_{s=1}^N|U_{s,q}|^2)^2\\	=\frac{4C}{N^2T}(\sum_{s=1}^N|U_{s,q}|^2)^2=O_p(\frac{1}{T}),
\end{align}
which finishes the proof, where the $O_p$ can be shown by taking expectation.
\hfill $\Box$
\ \\\ \\
\noindent \textbf{Proof \Cref{thm6}.}\\
Recall the definition of $III_1(C_i)$, $III_2(C_i)$,
$III_3(C_i)$ in the proof of \Cref{Thm4}. Then
\begin{align}\label{new.187}
	\hat{ f}(C_i)-f(C_i)=[\sum_{t=1}^N K(\frac{\hat \delta_{it}^2}{h_N})]^{-1}[\sum_{t=1}^N( f(C_t)- f(C_i)+\bar \epsilon_t)K(\frac{\hat \delta_{it}^2}{h_N})]
	+(\gamma-\hat \gamma)^\top III_1(C_i)\\
	+(\alpha-\hat \alpha)III_2(C_i)+(\beta-\hat \beta)III_3(C_i).
\end{align}
By the proof of \Cref{Thm5}, 
\begin{align}\label{new.188}
	\gamma-	\hat \gamma=\hat \Gamma_1^{-1}
	((\hat \alpha-\alpha)\hat \Gamma_{2,2,\alpha}+(\hat \beta-\beta)\hat \Gamma_{2,2,\beta})-\hat \Gamma_1^{-1}\tilde \Gamma_{2,1}.
\end{align}

Via similar argument to the proof of \Cref{new.eq176}, since $r_N>0$,
\begin{align}
	P([\sum_{t=1}^N K(\frac{\hat \delta_{it}^2}{h_N})]^{-1}[\sum_{t=1}^N( f(C_t)- f(C_i))K(\frac{\hat \delta_{it}^2}{h_N})]\mathbbm 1 \{\max_{i\neq j}\frac{|\hat \delta^2_{ij}-\delta^2_{ij}|}{h_N}\leq N^{-\gamma/4}\} =0|\GG_N)\rightarrow 1.
\end{align}
By Lemma B1 of \cite{auerbach2022identification},  $\max_{i\neq j}\frac{|\hat \delta^2_{ij}-\delta^2_{ij}|}{h_N}=o_p(N^{-\gamma/4})$. Hence 
given $\GG_N$, 
\begin{align}\label{new.190}\sqrt{N(T-1)}[\sum_{t=1}^N K(\frac{\hat \delta_{it}^2}{h_N})]^{-1}[\sum_{t=1}^N(f(C_t)- f(C_i))K(\frac{\hat \delta_{it}^2}{h_N})]=o_p(1).
\end{align}
It is easy to see that a.s.,
$E ([\sum_{t=1}^N K(\frac{\hat \delta_{it}^2}{h_N})]^{-1}\sum_{t=1}^N\bar \epsilon_t K(\frac{\hat \delta_{it}^2}{h_N})|\GG_N]=0$. 
Define 
\begin{align}
	\nu'(C_i)=[\sum_{t=1}^N K(\frac{\hat \delta_{it}^2}{h_N})]^{-1}\sum_{t=1}^T\bar \epsilon_t K(\frac{\hat \delta_{it}^2}{h_N})-[\hat \Gamma^{-1}(\tilde \Gamma_{2,1}-E (\tilde \Gamma_{2,1}|\GG_N))]^\top III_1(C_i)		\\=\sum_{t=1}^N\big(\sum_{j=1}^N(-\hat \Gamma_1^{-1}\frac{2(U_t-U_j)}{N(N-1)r_N})^\top K(\frac{\hat \delta^2_{tj}}{h_N})III_1(C_i)+[\sum_{t=1}^N K(\frac{\hat \delta_{it}^2}{h_N})]^{-1}K(\frac{\hat \delta_{it}^2}{h_N})\big)\bar \epsilon_t.
\end{align} Since $\mathcal C$ is a fixed finite set, using similar argument to the proof of \Cref{Thm5},
we can verify that
conditional on $\GG_N$, with probability going to $1$, as $\min(N,T)\rightarrow \infty$,
\begin{align}\label{new.196}
	\sqrt{N(T-1)}\tilde \Sigma_{3N}(\hat \alpha-\alpha, \hat \beta-\beta,(\sqrt{r_N(C_i)}\nu'(C_i))_{i\in \mathcal C})^\top/\sigma\Rightarrow N(0,I_{2+|\mathcal C|}),
\end{align}
where 
\begin{align}
	\tilde	\Sigma_{3N}=\begin{pmatrix}
		\Sigma_{2,N}^{-1/2}\Sigma_{1,N}& 0\\
		0&(\Sigma^*_{3N})^{-1/2}
	\end{pmatrix}
\end{align}
and $\Sigma^*_{3N}$ is a $p\times p$ matrix with the $(i,j)$th entry $\sqrt{r_N(C_i)r_N(C_j)\tilde \sigma_{i,j}}$.

Combing with \eqref{new.eq176}, \eqref{new.187}, \eqref{new.188} and \eqref{new.190} we have that
given $\GG_N$, %
\begin{align}\label{New.132}
	\sqrt{N(T-1)}(\hat { f}(C_i)-f(C_i))_{i\in \mathcal C}=
	\sqrt{N(T-1)}\Big( (\alpha-\hat \alpha)\big[-(\hat \Gamma_1^{-1}\hat \Gamma_{2,2,\alpha})^\top III_1(C_i)+III_2(C_i)\big]\\+(\beta-\hat \beta)\big[-(\hat \Gamma_1^{-1}\hat \Gamma_{2,2,\beta})^\top III_1(C_i)+III_3(C_i)\big]
	+\nu'(C_i)%
	\Big)_{i\in \mathcal C}+o_p(1).
\end{align}
By the proof of \Cref{Conditional-Constant}, given $\GG_N$, it can be verified similarly and more easily that for $i\in \mathcal C$ and $a=1,2,3$, as $\min(N,T)\rightarrow \infty$,
\begin{align}
	E( (III_a(C_i)-E(III_a(C_i)|\GG_N))^2|\GG_N)\rightarrow_p 0,
\end{align}
which together with \Cref{Conditional-Constant}, \eqref{new.196}, \eqref{New.132}, Slutsky theorem,  and the fact that a.s.,
\begin{align}
	E(\hat \Gamma_{2,2,\alpha}|\GG_N)=l_{1N},~~	E(\hat \Gamma_{2,2,\beta}|\GG_N)=l_{2N},\\ E(III_1(C_i)|\GG_N)=l_{3N}(C_i), ~~E(III_2(C_i)|\GG_N)=l_{4N}(C_i), 
	E(III_3(C_i)|\GG_N)=l_{5N}(C_i)
\end{align}
proves the theorem. \hfill $\Box$ 
\subsection{Auxiliary Propositions and Lemmas}

\begin{proposition}\label{Prop-mix}
	Assume Assumptions \ref{Graph1} and \ref{Graph2} hold. 
	Then there exists a constant $C_0$ depending on $\alpha$ such that for  $n\geq C_0\log^{1+\alpha_0} N$, $N\geq 3$,
	\begin{align}
		\lim_{N\rightarrow \infty}P(	W_N^n \preceq 2\mf 1^N \bar \pi_N)=1	\end{align}
	where $\bar \pi=(\pi_1,...,\pi_N)$ is defined in \Cref{sec:assumption}.
\end{proposition}
{\it Proof.}
By (12.13) of \cite{levin2017markov}, we shall see that for $1\leq i,j\leq N$, $n\in \mathbb Z^+$,
\begin{align}
	\left|\frac{(W_N)^n_{ij}}{\pi_j}-1\right|\leq \frac{\lambda_N}{\min_{1\leq i\leq N}\pi_i}=\frac{(1-\gamma^\star_N)^n}{\min_{1\leq i\leq N}\pi_i}
	\leq \frac{(1-\alpha\log^{-\alpha_0} N)^n}{\min_{1\leq i\leq N}\pi_i},
\end{align}
where for the last inequality we use \Cref{Graph2}. Therefore, the proposition holds if we show there exists a constant $C_0$, such that for $n\geq C_0\log^{1+\alpha_0} N$, $N\geq 3$, 
\begin{align}\label{new.eq34}
	\frac{(1-\log^{-\alpha_0} N)^n}{\min_{1\leq i\leq N}\pi_i}\leq 1.
\end{align}
On the other hand, using Taylor expansion of $\log(1-x)$ we can verify that 
\begin{align}\label{new.eq35}
	\frac{(1-\log^{-\alpha_0} N)^n}{N^{-2}}\leq 1
\end{align}
for $n\geq C_0\log^{1+\alpha_0}N$ for some sufficiently large constant $C_0$ and $N\geq 3$.
By  \Cref{Graph1} and the corresponding discussions, \eqref{new.eq35} implies  \eqref{new.eq34}, hence the proposition follows. \hfill $\Box$
\begin{lemma}\label{lemma4}
	Let Assumptions \ref{Graph1}, \ref{Graph2} hold. 
	(a)  Let $r$ be a positive constant, with $r<1$ and $|\alpha|+|\beta|/r<1$. Then with probability approaching one as $N$ diverges, for all $j\in \mathbb Z, j\geq 0$,
	\begin{align}\label{neweq36}
		|G_1^j|_e\preceq (|\alpha|+|r^{-1}\beta|)^{j}M_N,
	\end{align}
	and
	\begin{align}\label{eq14}
		|G_1^j (G_1^\top)^j|_e  \preceq (|\alpha|+|\beta/r|)^{2j}M_NM_N^\top , ~~ |(G_1^\top )^j G_1^j|_e  \preceq (|\alpha|+|\beta/r|)^{2j}M_N^\top M_N,\end{align}
	where $M_N=2\mf 1 \bar \pi_N+\sum_{j=0}^{\lceil C_0\log^{1+\alpha_0} N\rceil}r^jW_N^j$ and  $C_0$ is defined in Proposition \ref{Prop-mix}. Here we use $A^0=I$ for any  square matrix  $A$  where $I$ is the identity matrix with the same dimension as $A$.\\\ \\
	(b) Further assume Assumption \ref{Graph3}. Let $g_{j,k_1,k_2}(G_1,W_N)=| W_N^{k_1}\{G_1^j(G_1^\top)^j\}^{k_2}(W_N^\top)^{k_1}|_e$, $f(M_N,V)=M_NM_N^\top V M_NM_N^\top V^\top$.%
	Then for integers $0\leq  k_1,k_2, m_0, m_1,m _2\leq 1$, we have that as $N\rightarrow \infty$, 
	\begin{align}\label{new.eq37}
		N^{-1}\sum_{i,j=0}^\infty\{tr\{ g_{i,k_1,k_2}( G_1,W_N)g_{j,m_1,m_2}(G_1,W_N)\}\}^{1/2}\rightarrow_p 0,\\\label{new.eq38}
		\frac{1}{N^2}\mf 1^\top(W_N)^{k_1} M_NM_N^\top (W_N^\top)^{k_1} \mf 1\rightarrow_p 0,\\
		\frac{1}{N^2}tr(f(M_N,I_N)+f(M_N,W_N)+f(M_N, W_N^\top W_N))\rightarrow_p 0,\label{new.eq39}\\
		\frac{1}{N}(tr(M_NM_N^\top)+tr(M_N^\top W_N^\top W_N M_N))\rightarrow_p 0. \label{new.eq40}
	\end{align}
\end{lemma}
{\it Proof.} We first prove (a). To save the notation write $K_N=\lceil C_0\log^{1+\alpha_0} N\rceil$. Recall that $G_1=\alpha I+\beta W_N$. For any integer $n$ such that $n\geq K_N$,  by binomial expansion we have that, with probability approaching 1
\begin{align}
	|G_1^n|_e\preceq \sum_{j=K_N+1}^n{n \choose j}|\alpha|^{n-j}|\beta|^j W_N^j+\sum_{j=0}^{K_N} {n\choose j}|\alpha|^{n-j}|\beta|^jW_N^j\\
	\preceq 2\sum_{j=K_N+1}^n{n \choose j}|\alpha|^{n-j}|\beta|^j \mf  1 \bar \pi_N+\sum_{j=0}^{K_N} {n\choose j}|\alpha|^{n-j}|\beta/r|^jr^jW_N^j\\
	\preceq  2\mf 1 \bar \pi_n(|\alpha|+|\beta|/r)^n+(|\alpha|+|\beta|/r)^n\sum_{j=0}^{K_n}r^jW_N^j=(|\alpha|+|r^{-1}\beta|)^{n}M_N \label{eq149},
\end{align}
where for the second $\preceq$  we have used Proposition \ref{Prop-mix} and for the third $\preceq$ we have used the binomial expansion of $(|\alpha|+|\beta|/r)^n$.%
\ Thus we show \eqref{neweq36}, and \eqref{eq14} directly follows from \eqref{neweq36}. For $0\leq n\leq K_N-1$,  \eqref{eq149} holds trivially.

To show (b), we consider $k_0=k_1=k_2=m_1=m_2=1$ for expression \eqref{new.eq37} and \eqref{new.eq38}. Other situations could be proved similarly. For \eqref{new.eq39}, we shall show $\frac{1}{N^2}tr(f(M_N,W_N^\top W_N))\rightarrow_p 0$, and the results that $\frac{1}{N^2}tr(f(M_N,I_N))\rightarrow_p 0$ and  $\frac{1}{N^2}tr(f(M_N,W_N^\top ))\rightarrow_p 0$ will follow mutatis mutandis. \Cref{new.eq40} will follow similar arguments for proving \eqref{new.eq39} and details will be omitted.  In summary, we shall show
\begin{align}
	N^{-1}\sum_{i,j=0}^\infty tr^{1/2}\{|W_NG_1^i(G_1^\top)^iW_N^\top|_e| W_NG_1^j(G_1^\top)^jW_N^\top|_e\}\rightarrow_p 0,\label{new.eq42}\\
	N^{-2}\mf 1^\top W_NM_NM_N^\top W_N^\top \mf 1 \rightarrow _p 0, \label{new.eq43-1}\\
	N^{-2} tr(M_NM_N^\top W_N^\top W_NM_NM_N^\top W_N^\top W_N)\rightarrow_p 0. \label{new.eq44}
\end{align}
Let $\mathcal{M}_N=W_NM_NM_N^\top W_N^\top$. By using \eqref{eq14}, for \eqref{new.eq42} it suffice to show
\begin{align}\label{new.eq43}
	N^{-2}  tr(\mathcal M_N^2)\rightarrow_p 0,
\end{align}
which also implies \eqref{new.eq44}. It is easy to verify that
$W_NM_N=2\mf 1 \bar \pi_N+\sum_{j=0}^{K_N}r^jW_N^{j+1}$. Using Cauchy-Schwartz inequality, it suffices to show that as $N\rightarrow \infty$,
\begin{align}
	\frac{1}{N^2}tr(\mf 1 \bar \pi_N(\mf 1 \bar \pi_N)^\top\mf 1 \bar \pi_N(\mf 1 \bar \pi_N)^\top)\rightarrow_p 0, \\
	\frac{1}{N^2}tr(\sum_{j=0}^{K_N}r^jW_N^{j+1}(\sum_{j=0}^{K_N}r^jW_N^{j+1})^\top\sum_{j=0}^{K_N}r^jW_N^{j+1}(\sum_{j=0}^{K_N}r^jW_N^{j+1})^\top)\rightarrow_p 0.\label{eq43}
\end{align}
The first convergence is due to (i) of Assumption \ref{Graph3}. To see the second convergence, notice that
\begin{align}\label{eq44}
	tr(\sum_{j=1}^{K_N}r^jW_N^{j+1}(\sum_{j=1}^{K_N}r^jW_N^{j+1})^\top\sum_{j=1}^{K_N}r^jW_N^{j+1}(\sum_{j=1}^{K_N}r^jW_N^{j+1})^\top)\\=|\sum_{j=1}^{K_N}r^jW_N^{j+1}(\sum_{j=1}^{K_N}r^jW_N^{j+1})^\top|^2_F\leq |\sum_{j=1}^{K_N+1}r^jW_N^{j+1}|^4_F.
\end{align}
By the properties of Frobenius  norm and (ii) of Assumption \ref{Graph3},  we have with probability approaching $1$,
\begin{align}\label{new.45}
	|\sum_{j=1}^{K_N}r^jW_N^{j+1}|_F\leq  \sum_{j=1}^{K_N}r^j|W_N^{j+1}|_F\leq  (\sum_{j=1}^{K_N}r^jM(j+1))N^{1/2}c_N.
\end{align}
Combining the last equality and  \eqref{eq44} and the fact that $r\in(0,1)$ (such that $\sum_{j=1}^\infty r^jM(j+1)<\infty$) and $c_N=o(1)$,
we prove \eqref{eq43}. Therefore \eqref{new.eq42} and \eqref{new.eq44} hold. For \eqref{new.eq43-1}, notice  that 
\begin{align}\label{new.51}
	N^{-2}\mf 1^\top W_NM_NM_N^\top W_N^\top\mf 1=N^{-2}\mf 1^\top (2\mf 1 \bar\pi_N+\sum_{j=0}^{K_N}r^jW_{N}^{j+1})(2\mf 1 \bar\pi_N+\sum_{j=0}^{K_N}r^jW_{N}^{j+1})^\top \mf 1.
\end{align}
By Cauchy inequality, the above will converge to $0$ in probability if $\bar \pi_N\bar \pi_N^\top \rightarrow 0$, and \begin{align} \label{new.eq49}
	N^{-2}\mf 1^\top  (\sum_{j=0}^{K_N} r^jW_N^{j+1})(\sum_{j=0}^{K_N} r^jW_N^{j+1})^\top \mf 1 \rightarrow_p 0.
\end{align}
By Assumption \ref{Graph3}  $\bar \pi_N\bar \pi_N^\top \rightarrow 0$. Inequality \eqref{new.eq49} holds because the left hand side is bounded by
\begin{align}\label{new.53}
	\frac{|\mf 1\mf 1^\top|_F}{N^2}|\sum_{j=0}^{K_N} r^jW_N^{j+1}|_F^2=\frac{|\sum_{j=0}^{K_N} r^jW_N^{j+1}|_F^2}{N}%
	\leq_p (\sum_{j=1}^{K_N}r^jM(j+1))^2g^2_N\rightarrow 0.
\end{align}
where $\leq_p$ represents that ``with probability tending to $1$ less than and equal to".
\hfill $\Box$

Lemma \ref{lemma4} implies the following corollary for $N$ fixed but $T$ diverging, with weaker assumptions.
\begin{cor}\label{fixN} Assume the  network $G=(V,E)$ has $N$ nodes, and is almost surely fully connected and not bipartite. Then (a)   equations \eqref{neweq36} and \eqref{eq14} of Lemma \ref{lemma4} hold with $M_N=2\mf 1 \bar \pi_N+\sum_{j=0}^{M'}r^jW_N^j$ for some constant $M'$, and the term `with probability approaching $1$' being replaced by the term 'almost surely', and that (b) the left hand side of \eqref{new.eq37}-\eqref{new.eq40} is finite almost surely.
\end{cor}
{\it Proof.} Since the  network is  $a.s.$ fully connected, then $a.s.$ $\min_{1\leq i\leq N}\pi_i=\min_{1\leq i\leq N}\frac{n_2}{2|E_N|}>0$. Since the network is $a.s.$  not bipartite, $W_N$ is $a.s.$ aperiodic therefore the absolute spectral gap $\gamma_N^\star>0$ $a.s.$. (see Lemma 12.1 of \cite{levin2017markov}). By the proof of Proposition \ref{Prop-mix}, it follows that there exists a constant $M'$ such that when $n\geq M'$, $W_N^n\preceq 2 \mf 1\bar \pi_N$ $a.s.$. Therefore (a) follows. (b) follows trivially from (a) and  the proof of  \Cref{lemma4} (b). %

\begin{proposition}\label{prop2}
	Define for positive integer $m$,
	\begin{align}
		\Y^{(m)}_t=(I-G_1)^{-1}G_0+\sum_{j=0}^{m} G_1^j\bepsilon_{t-j}:=(Y_{1,t}^{(m)},...,Y_{N,t}^{(m)})^\top.
	\end{align}
	Let $Z^{(m)}_{i,t}=(Y^{(m)}_{i,t-1}, w_i^\top\Y^{(m)}_{t-1})^\top$, and $\Z^{(m)}_t=(Z^{(m)}_{1,t},...,Z^{(m)}_{N,t})^\top\in \R^{N\times 2}$. 
	Assume $|\alpha|+|\beta|<1$, $E \epsilon^4_{i,t}<\infty$.  Let $\chi=(|\alpha|+|\beta/r|)^{2}\in (0,1)$ where $r$ is defined in \Cref{lemma4}.
	Then (i) Assume Assumption \ref{Assumption2.1_2}, and that the network is fully connected and not bipartite with fixed size $N$. Then we have as $m\rightarrow \infty$
	\begin{align}
		\limsup_{T\rightarrow \infty }\frac{1}{\sqrt{N(T-1)}}\|\sum_{t=1}^{T-1}\Z_t^\top\Delta \bepsilon_{t+1}
		-\sum_{t=1}^{T-1}(\Z_t^{(m)})^\top\Delta \bepsilon_{t+1}\|_{2,\GG_N}=o_p(\chi^{m/2}),
	\end{align}
	and (ii)  Assume  Assumptions \ref{Graph1}, \ref{Graph2}, \ref{Graph3} and  Assumption \ref{Assumption2.1_2} hold. Further assume condition \ref{New.4.1}. 
	Then  as $m\rightarrow \infty$,	\begin{align}
		\limsup_{T,N\rightarrow \infty }\frac{1}{\sqrt{N(T-1)}}\|\Sigma_{2,N}^{-1/2}\big(\sum_{t=1}^{T-1}\Z_t^\top\Delta \bepsilon_{t+1}
		-(\Sigma_{2,N}^{(m)})^{-1/2}\sum_{t=1}^{T-1}(\Z_t^{(m)})^\top\Delta \bepsilon_{t+1}\big)\|_{2,\GG_N}=o_p(\chi^{m/2}).
	\end{align}
\end{proposition}
{\it Proof.} %
We shall show (ii) and (i) will follow from similar but easier arguments. Write $W_N$ as $W$ for short. Straightforward calculations using Proposition \ref{prop1} show that 
\begin{align}
	&\frac{1}{\sqrt{N(T-1)}}\big(\sum_{t=1}^{T-1}\Z_t^\top\Delta \bepsilon_{t+1}
	-\sum_{t=1}^{T-1}(\Z_t^{(m)})^\top\Delta \bepsilon_{t+1}\big)
	\\&=\begin{pmatrix}
		\frac{1}{\sqrt{N(T-1)}}\sum_{t=1}^{T-1}\sum_{i=1}^N(Y_{i,t-1}-Y_{i,t-1}^{(m)})\Delta \epsilon_{i,t+1}
		\\\frac{1}{\sqrt{N(T-1)}}\sum_{t=1}^{T-1}\sum_{i=1}^Nw_i^\top(\Y_{t-1}-\Y_{t-1}^{(m)}) \Delta \epsilon_{i,t+1}
	\end{pmatrix}=\begin{pmatrix}
		\frac{1}{\sqrt{N(T-1)}}\sum_{t=1}^{T-1}(\Y_{t-1}-\Y_{t-1}^{(m)})^\top\Delta \bepsilon_{t+1}
		\\
		\frac{1}{\sqrt{N(T-1)}}\sum_{t=1}^{T-1}(\Y_{t-1}-\Y_{t-1}^{(m)})^\top W_N^\top\Delta \bepsilon_{t+1}
	\end{pmatrix}\\
	&=\begin{pmatrix}
		\frac{1}{\sqrt{N(T-1)}}\sum_{t=1}^{T-1}(\sum_{j=m+1}^\infty G_1^j\bepsilon_{t-j})^\top \Delta \bepsilon_{t+1}\\
		\frac{1}{\sqrt{N(T-1)}}\sum_{t=1}^{T-1}(\sum_{j=m+1}^\infty G_1^j\bepsilon_{t-j})^\top W_N^\top\Delta \bepsilon_{t+1}
	\end{pmatrix}.
\end{align}
Therefore it sufficies to show	\begin{align}\label{eq49}
	\limsup_{T,N\rightarrow \infty} \frac{1}{\sqrt{N(T-1)}}\|\sum_{t=1}^{T-1}\sum_{j=m+1}^\infty(G_1^j\bepsilon_{t-j})^\top \Delta \bepsilon_{t+1}\|_{2,\GG_N}=o_p(\chi^{m/2}),
	\\\label{eq50}
	\limsup_{T,N\rightarrow \infty}  \frac{1}{\sqrt{N(T-1)}}\|\sum_{t=1}^{T-1}\sum_{j=m+1}^\infty(G_1^j\bepsilon_{t-j})^\top W_N^\top \Delta \bepsilon_{t+1}\|_{2,\GG_N}=o_p(\chi^{m/2}).
\end{align}
To see \eqref{eq49}, notice that since for $j\geq 1$, by Assumption \ref{Assumption2.1_2},
\begin{align}
	\|\sum_{t=1}^{T-1}(G_1^j\bepsilon_{t-j})^\top\bepsilon_{t+1}\|_{2,\GG_N}^2=(T-1)\sigma^4tr((G_1^j)^\top G_1^j),
\end{align}
we have that by the triangle inequality and Corollary \ref{fixN}, with probability approaching $1$,
\begin{align}
	\| \frac{1}{\sqrt{N(T-1)}}\sum_{t=1}^{T-1}\sum_{j=m+1}^\infty(G_1^j\bepsilon_{t-j})^\top \bepsilon_{t+1}\|_{2,\GG_N}\leq \frac{\sigma^2\sum_{j=m+1}^\infty tr^{1/2}((G_1^j)^\top G_1^j)}{\sqrt N}\\
	\leq	\sigma^2N^{-1/2}\sum_{j=m+1}^\infty(|\alpha|+|\beta/r|)^{j}tr^{1/2}(M_N^\top M_N)=o_p(\chi^{m/2}),
\end{align}
where we have used \eqref{new.eq40}.
Similarly argument yields that $ \|\frac{1}{\sqrt{N(T-1)}}\sum_{t=1}^{T-1}\sum_{j=m+1}^\infty(G_1^j\bepsilon_{t-j})^\top \bepsilon_{t}\|_{2,\GG_N}=o_p(\chi^{m/2})$. Therefore 
\eqref{eq49} follows.
On the other hand, since by the triangle inequality and Corollary  \ref{fixN},  with probability approaching $1$, \begin{align}
	\| \frac{1}{\sqrt{N(T-1)}}\sum_{t=1}^{T-1}\sum_{j=m+1}^\infty(G_1^j\bepsilon_{t-j})^\top W_N^\top \bepsilon_{t+1}\|_{2,\GG_N}\leq \frac{\sigma^2\sum_{j=m+1}^\infty tr^{1/2}((G_1^j)^\top W_N^\top W_N G_1^j)}{\sqrt N}\\
	\leq	\sigma^2N^{-1/2}\sum_{j=m+1}^\infty(|\alpha|+|\beta/r|)^{j}tr^{1/2}(M_N^\top W_N^\top W_N M_N)
	=o_p(\chi^m),
\end{align}
where we have used \cref{new.eq40} again
and similarly $\frac{1}{\sqrt{N(T-1)}}\|\sum_{t=1}^{T-1}\sum_{j=m+1}^\infty(G_1^j\bepsilon_{t-j})^\top W_N^\top \bepsilon_{t}\|_{2,\GG_N}=o_p(\chi^{m/2})$ hence \eqref{eq50} follows which completes the proof of (i).\hfill $\Box$%

\begin{lemma}\label{lemma1}
	Assume $|\alpha|+|\beta|<1$, $E \epsilon^4_{i,t}<\infty$. Assume  Assumptions \ref{Graph1}, \ref{Graph2}, \ref{Graph3} and  Assumption \ref{Assumption2.1_2} hold. Then (i) or as $N\rightarrow \infty$ and $T\rightarrow \infty$, we have
	\begin{align}E(|\frac{1}{NT}\sum_{t=1}^T\Z_t^\top \X_t- \Sigma_{1,N}|_F|\mathcal G_N)=o_p(T^{-1/2}) .\end{align} 
	The above also holds for $N$ fixed but $T\rightarrow \infty$, if we assume the network is fully connected and not bipartite instead of  Assumptions \ref{Graph1}, \ref{Graph2} and \ref{Graph3}. %
\end{lemma}
{\it Proof.} 
We only show the case when $N\rightarrow \infty$ and  $T\rightarrow \infty$.
When $T\rightarrow \infty$ but $N$ fixed, the results follow similarly by using Corollary \ref{fixN} instead of \Cref{lemma4}.  We also only prove (i). %

We now prove (i) under the assumption of finite eighth moment of $U_i$.
Notice that
\begin{align} 
	\frac{1}{NT}\sum_{t=1}^T\Z_t^\top \X_t=\frac{1}{NT}\sum_{t=1}^T\sum_{i=1}^N\begin{pmatrix}
		Y_{i,t-1}\Delta Y_{i,t}& Y_{i,(t-1)}w_i^\top \Delta \Y_t\\
		w_i^\top\Y_{t-1}\Delta Y_{i,t} & w_i^\top \Y_{t-1}w_i^\top \Delta \Y_t
	\end{pmatrix}:=(\mathcal M_{ij})_{1\leq i\leq 2, 1\leq j\leq  2}.
\end{align}
For $\mathcal M_{11}$, notice that by Proposition \ref{prop1}, given $\mathcal G_N$, 
\begin{align}
	\frac{1}{NT}\sum_{t=1}^T\sum_{i=1}^NY_{i,t-1} Y_{i,t-1}=\frac{1}{NT}\sum_{t=1}^T\Y_{t-1}^\top \Y_{t-1}=I+II+III+IV, a.s.,
\end{align}
where 
\begin{align}
	I=\frac{1}{N}((I-G_1)^{-1}G_0)^\top (I-G_1)^{-1}G_0,~
	II=\frac{1}{NT}\sum_{t=1}^T((I-G_1)^{-1}G_0)^\top \sum_{j=0}^\infty G_1^j\bepsilon_{t-1-j},\\
	III=\frac{1}{NT}\sum_{t=1}^T \sum_{j=0}^\infty (G_1^j\bepsilon_{t-1-j})^\top(I-G_1)^{-1}G_0,~
	IV=\frac{1}{NT}\sum_{t=1}^T\sum_{j=0}^\infty (G_1^j\bepsilon_{t-1-j})^\top \sum_{j=0}^\infty G_1^j\bepsilon_{t-1-j}.\label{New.92}
\end{align}
By Assumption \ref{Assumption2.1_2},
\begin{align}\label{checkstar}
	\|II\|_{2,\GG_N}&\leq \frac{1}{NT}\sum_{j=0}^\infty\|\sum_{t=1}^T((I-G_1)^{-1} {\mf f})^\top  G_1^j\bepsilon_{t-1-j}\|_{2,\GG_N}\\
	&\leq \frac{1}{N\sqrt T}\sum_{j=0}^\infty\|((I-G_1)^{-1} {\mf f})^\top  G_1^j\bepsilon_{t-1-j}\|_{2,\GG_N}\\&=\frac{\sigma}{N\sqrt T}\sum_{j=0}^\infty
	(((I-G_1)^{-1} {\mf f})^\top G_1^j(G_1^\top)^j((I-G_1)^{-1} {\mf f}))^{1/2}.
\end{align}

Meanwhile, since $(I-G_1)^{-1}=\sum_{j=0}^{\infty} G_1^j$, by using (a) of Lemma \ref{lemma4} and Cauchy inequality, the above is bounded by 
\begin{align}\label{control1}
	\frac{K_0 ( {\mf f}^\top \sum_{u=0}^\infty  (G_1^\top)^u M_NM_N^\top \sum_{v=0}^\infty G_1^v {\mf f})^{1/2}}{N\sqrt T}&\leq \frac{K_0' ( {\mf f}^\top M_N^\top M_NM_N^\top  M_N  {\mf f})^{1/2}}{N\sqrt T}
	\\&\leq
	\frac{\tilde K_0' ( {\mf f_1}^\top M_N^\top M_NM_N^\top  M_N  {\mf f_1})^{1/2}}{N\sqrt T}\\&+
	\frac{\tilde K_0' ( {\mf f_2}^\top M_N^\top M_NM_N^\top  M_N  {\mf f_2})^{1/2}}{N\sqrt T}
\end{align} for some large constants $K_0, K_0', \tilde K_0'$,
where 
\begin{align}
	\mf f_1=(f (C_1),...., f (C_N))^\top,~~
	\mf f_2=(\gamma^\top U_1,....,\gamma^\top U_N^\top)^\top.
\end{align}
Observe that $\bar \pi \mf 1=1$ and $W^j\mf 1=\mf 1$ since $W^j$ is also a transition probability matrix. Since $0<r<1$, there exists a large constant $K_0''$ such that 
$M_N\mf 1\preceq  K_0''\mf 1$. %
Further by the boundedness of $ f(\cdot)$,
we have by \Cref{lemma4} (b)
\begin{align}\label{eq161.new}
	\frac{\tilde K_0' ( {\mf f_1}^\top M_N^\top M_NM_N^\top  M_N  {\mf f_1})^{1/2}}{N\sqrt T}\leq \frac{\tilde K_0'|\mf f_1|_\infty ( {\mf 1}^\top M_N^\top M_NM_N^\top  M_N  {\mf 1})^{1/2}}{N\sqrt T}=o_p(T^{-1/2}).%
\end{align}
On the other hand,  write $\bar {\mf f}_2=\mf f_2-E\mf f_2$ and another application of Cauchy inequality yield that
\begin{align}\label{eq162}
	\frac{ ( {\mf f_2}^\top M_N^\top M_NM_N^\top  M_N  {\mf f_2})^{1/2}}{N\sqrt T}\leq 
	\frac{\tilde C_0 ( {(E \mf f_2)}^\top M_N^\top M_NM_N^\top  M_N  {(E\mf  f_2)})^{1/2}}{N\sqrt T}\\+	\frac{\tilde C_0 ( (\bar {\mf f}_2^\top M_N^\top M_NM_N^\top  M_N  \bar {\mf f}_2)^{1/2}}{N\sqrt T}
\end{align}
for some large constant $\tilde C_0$. Since $U_i$ are $i.i.d.$ random variables, $E\mf f_2=\gamma^\top E U_1 \mf 1$ and the first term of RHS of \cref{eq162} is $o_p(T^{-1/2})$ due to the same  arguments to \eqref{eq161.new}. For the second term, notice that 
\begin{align}\label{eq163.new}
	\frac{ E( \bar {\mf f}_2^\top M_N^\top M_NM_N^\top  M_N  \bar {\mf f}_2|W_N)}{N^2 T}
	=\frac{ E( tr({M_N^\top M_NM_N^\top  M_N  \bar {\mf f}_2 \bar{\mf f}}_2^\top |W_N)}{N^2 T}=o_p(T^{-1}),
\end{align}
due to the fact that $E(\bar {\mf f}_2 \bar{\mf f}_2^\top |W_N)=Var(\gamma^\top U_1)I$ where $I$ is the $N\times  N$ identity matrix, and (b) of \Cref{lemma4}. Hence the second term  of RHS of \Cref{eq162} is $o_p(T^{-1/2})$. 
Therefore it follows that%
\begin{align}\label{IIGN}
	\|II\|_{2,\GG_N}=o_p(T^{-1/2}).
\end{align} Since $III=II$ we have $\|III\|_{2,\GG_N}=o_p(T^{-1/2})$. %
For $IV$, straightforward calculation shows that
\begin{align}%
	E(IV|\GG_N)=\frac{1}{NT}\sum_{t=1}^T\sum_{i,j}E((G^i_1\bepsilon_{t-1-i})^\top (G_1^j\bepsilon_{t-1-j})|\GG_N)=\frac{\sigma^2}{N}\sum_{i=0}^\infty tr((G_1^i)^\top G_1^i),\\
	Var(IV|\GG_N)=\frac{1}{N^2T^2}\|\sum_{t=1}^T[\sum_{i=0}^\infty[(G_1^i\bepsilon_{t-1-i})^\top G_1^i\bepsilon_{t-1-i}-\sigma^2tr((G_1^i)^\top G_1^i)]\\+\sum_{i\neq j,i,j\geq 0}(G_1^i\bepsilon_{t-1-i})^\top G_1^j\bepsilon_{t-1-j}]\|^2_{2,\GG_N}.
\end{align}%
Notice that $E(IV|\GG_N)<\infty$ with probability approaching $1$ due to \eqref{eq14}. By Lemma \ref{fourth} we have
\begin{align}%
	\|\sum_{t=1}^T\sum_{i=0}^\infty[(G_1^i\bepsilon_{t-1-i})^\top G_1^i\bepsilon_{t-1-i}&-\sigma^2tr((G_1^i)^\top G_1^i)]\|_{2,\GG_N}\\&\leq \sum_{i=0}^\infty\|\sum_{t=1}^T[(G_1^i\bepsilon_{t-1-i})^\top G_1^i\bepsilon_{t-1-i}-\sigma^2tr((G_1^i)^\top G_1^i)]\|_{2,\GG_N}
	\\&=  \sum_{i=0}^\infty T^{1/2}Var^{1/2}((G_1^i\bepsilon_{t-1-i})^\top G_1^i\bepsilon_{t-1-i}|\GG_N)\\&\leq \tilde C^{1/2}T^{1/2} \sum_{i=0}^\infty tr^{1/2}((G_1^i)^
	\top G_1^i(G_1^i)^\top G_1^i),
\end{align}%
where $\tilde C=2\sigma^4+|E(\epsilon_{i,t}^4)-3\sigma^4|$.
On the other hand,
\begin{align}%
	&\|\sum_{t=1}^T\sum_{i\neq j,i,j\geq 0}(G_1^i\bepsilon_{t-1-i})^\top G_1^j\bepsilon_{t-1-j}\|_{2,\GG_N}\\&\leq  \sum_{i\neq j,i,j\geq 0} T^{1/2}E^{1/2}(((G_1^i\bepsilon_{t-1-i})^\top G_1^j\bepsilon_{t-1-j})^\top (G_1^i\bepsilon_{t-1-i})^\top G_1^j\bepsilon_{t-1-j}|\GG_N)\\&= \sum_{i\neq j,i,j\geq 0} T^{1/2}E^{1/2}(tr((G_1^j)^\top G_1^i\bepsilon_{t-1-i}\bepsilon_{t-1-i}^\top (G_1^i)^\top G_1^j \bepsilon_{t-1-j}\bepsilon_{t-1-j}^\top )|\GG_N)\\&\leq \sigma^2 T^{1/2} \sum_{i\neq j,i,j\geq 0}^\infty tr^{1/2}((G_1^i)^
	\top G_1^j(G_1^j)^\top G_1^i),
\end{align}%
where for the last inequality we have used Assumption \ref{Assumption2.1_2} and the linearity of trace.
Therefore, by Lemma \ref{lemma4},
\begin{align}
	(Var(IV|\GG_N))^{1/2}\leq \frac{C_0}{N\sqrt T}\sum_{i,j\geq 0} tr^{1/2}((G_1^i)^
	\top G_1^j(G_1^j)^\top G_1^i)=o_p(T^{-1/2}),
\end{align}
for some sufficiently large constant $C_0$
which implies 
\begin{align}
	\|IV- \frac{\sigma^2}{N}\sum_{i=0}^\infty tr((G_1^i)^\top G_1^i)\|_{2,\GG_N}=o_p(T^{-1/2}).
\end{align}
Combining the results of $I$, $II$, $III$, $IV$, we obtain 
\begin{align}
	\|\frac{1}{NT}\sum_{t=1}^T\sum_{i=1}^NY_{i,t-1} Y_{i,t-1}- \frac{1}{N}((I-G_1)^{-1}G_0)^\top (I-G_1)^{-1}G_0-\frac{\sigma^2}{N}\sum_{i=0}^\infty tr((G_1^i)^\top G_1^i)\|_{2,\GG_N}=o_p(T^{-1/2}),\end{align}
and similarly
\begin{align}
	\|\frac{1}{NT}\sum_{t=1}^T\sum_{i=1}^NY_{i,t-1} Y_{i,t}- \frac{1}{N}((I-G_1)^{-1}G_0)^\top (I-G_1)^{-1}G_0-\frac{\sigma^2}{N}\sum_{i=0}^\infty tr((G_1^i)^\top G_1^{i+1})\|_{2,\GG_N}=o_p(T^{-1/2}).
\end{align}
Combining the above two equation and apply the triangle inequality, we have
\begin{align}
	\|\mathcal M_{11}-\kappa_{11,N}\|_{2,\GG_N}=o_p(T^{-1/2}).
\end{align}

For $\mathcal M_{21}$, notice that
\begin{align}
	\frac{1}{NT}\sum_{t=1}^T\sum_{i=1}^Nw_i^\top\Y_{t-1}Y_{i,t}=\frac{1}{NT}\sum_{t=1}^T \Y_{t-1}^\top W_N^\top \Y_t,~~ \frac{1}{NT}\sum_{t=1}^T\sum_{i=1}^Nw_i^\top\Y_{t-1}Y_{i,t-1}=\frac{1}{NT}\sum_{t=1}^T \Y_{t-1}^\top W_N^\top \Y_{t-1}.
\end{align}
Observe that
\begin{align}
	\frac{1}{NT}\sum_{t=1}^T\Y_{t-1}^\top W_N^\top \Y_t:=A+B+C+D,
\end{align}
where 
\begin{align}\label{new.eq115}
	A=\frac{1}{N}G_0^\top [(I-G_1)^{-1}]^\top W_N^\top (I-G_1)^{-1}G_0,~~
	B=\frac{1}{NT}\sum_{t=1}^T\sum_{j=0}^\infty  ((I-G_1)^{-1}G_0)^\top W_N^\top G_1^j\bepsilon_{t-j},\\
	C=\frac{1}{NT}\sum_{t=1}^T\sum_{j=0}^\infty (G_1^j\bepsilon_{t-1-j})^\top W_N^\top (I-G_1)^{-1}G_0,~~
	D=\frac{1}{NT}\sum_{t=1}^T[\sum_{j=0}^\infty G_1^j\bepsilon_{t-1-j}]^\top W_N^\top \sum_{j=0}^\infty G_1^j\bepsilon_{t-j}.
\end{align}
Notice that by the similar argument to \eqref{checkstar},
\begin{align}
	\|B\|_{2,\GG_N}\leq \frac{\sigma}{N\sqrt T}\sum_{j=0}^\infty
	(((I-G_1)^{-1}\mf { f})^\top W_N^\top G_1^j(G_1^\top)^jW((I-G_1)^{-1}\mf { f}))^{1/2}.
\end{align}
Using the  the fact that there exists a large constant $K_0''$ such that 
$M_N\mf 1\preceq  K_0''\mf 1$, $W_N\mf 1=\mf 1$, and  (b) of Lemma \ref{lemma4},  we have 
$\frac{ \mf 1^\top M_N^\top W_N^\top M_N M_N^\top W M_N\mf 1}{N^2}\rightarrow_p 0$. By similar  argument to \eqref{control1}-\eqref{eq163.new}, we obtain $\|B\|_{2,\GG_N}=o_p(T^{-1/2})$. Similarly,  $\|C\|_{2,\GG_N}=o_p(T^{-1/2}) $.
For $D$,  it is easy to see that
\begin{align}
	E(D|\GG_N)=\frac{1}{N}\sum_{i=0}^\infty \sigma^2 tr((G_1^i)^\top W_N^\top G_1^{i+1}).
\end{align}
By similar argument to the evaluation of $IV$ in \eqref{New.92} and using Lemma \ref{lemma4}, we have that with probability approaching $1$,
\begingroup\allowdisplaybreaks
\begin{align}
	Var(D|\GG_N)&=\frac{1}{N^2T^2}\|\sum_{t=1}^T\sum_{i=0}^\infty[(G_1^i\bepsilon_{t-1-i})^\top W_N^\top G_1^{i+1}\bepsilon_{t-1-i}-\sigma^2tr((G_1^i)^\top W_N^\top G_1^{i+1})]\\&+\sum_{j\neq i+1,i,j\geq 0}(G_1^i\bepsilon_{t-1-i})^\top W_N^\top G_1^j\bepsilon_{t-j}\|^2_2\\
	&\leq \frac{1}{N^2T}(\sum_{i,j\geq 0}C_0 tr^{1/2}(W_N^\top G_1^j(G_1^j)^\top WG_1^i(G_1^i)^\top))^2,
\end{align}
\endgroup
where $C_0$ is a sufficiently large constant and the upperbound is $o_p(T^{-1})$ by (b) of \Cref{lemma4}.
Hence we have that
\begin{align}
	\|\frac{1}{NT}\sum_{t=1}^T\Y_{t-1}^\top W_N^\top \Y_t-\frac{1}{N}G_0^\top [(I-G_1)^{-1}]^\top W_N^\top (I-G_1)^{-1}G_0-\frac{1}{N}\sum_{i=0}^\infty \sigma^2 tr((G_1^i)^\top W_N^\top G_1^{i+1})\|_{2,\GG_N}=o_p (T^{-1/2}).
\end{align}
Similarly, 
\begin{align}
	\|\frac{1}{NT}\sum_{t=1}^T\Y_{t-1}^\top W_N^\top \Y_{t-1}- \frac{1}{N}G_0^\top [(I-G_1)^{-1}]^\top W_N^\top (I-G_1)^{-1}G_0- \frac{1}{N}\sum_{i=0}^\infty \sigma^2 tr((G_1^i)^\top W_N^\top G_1^{i})\|_{2,\GG_N}=o_p(T^{-1/2}).
\end{align}
Therefore
\begin{align}
	\|\mathcal M_{21}-\kappa_{21,N}\|_{2,\GG_N}=o_p(T^{-1/2}).
\end{align} 
For $\mathcal M_{12}$, we shall see that
\begin{align}
	\mathcal M_{12}=\frac{1}{NT}\sum_{t=1}^T\Y_t^\top W_N^\top \Y_{t-1}-\frac{1}{NT}\sum_{t=1}^T\Y_{t-1}^\top W_N^\top \Y_{t-1}.
\end{align}
Applying similar argument to  $\mathcal M_{21}$, we shall see that
\begin{align}
	\|\mathcal M_{12}-\kappa_{12,N}\|_{2,\GG_N}=o_p (T^{-1/2}).
\end{align}
Finally, 
\begin{align}
	\mathcal M_{22}=\frac{1}{NT}\sum_{t=1}^T\Y_{t-1}^\top W_N^\top W_N \Y_{t}-\frac{1}{NT}\sum_{t=1}^T\Y_{t-1}^\top W_N^\top W_N \Y_{t-1}.
\end{align}
Notice that by Lemma \ref{lemma4},
\begin{align}\label{new.197}
	\frac{1}{N^2}\mf 1^\top W_NM_NM_N^\top W_N^\top \mf 1\rightarrow_p0, ~~~\text{as}~~ N\rightarrow \infty,
\end{align}
which leads to 
\begin{align}\label{new.198}
	\frac{1}{N^2 }(\mf 1^\top M_N^\top W_N^\top W_N M_NM_N^\top W_N^\top W_N M_N\mf 1)\rightarrow _p 0 ~~~\text{as}~~ N\rightarrow \infty.
\end{align}
Also by \Cref{lemma4}%
\begin{align}\label{new.199}
	\frac{1}{N^2}(\sum_{i,j} tr^{1/2}(WG_1^j(G_1^j)^\top W_N^\top WG_1^i(G_1^i))^\top W_N^\top 
	)^2\rightarrow_p 0,\\
	\frac{1}{N^2 }tr( M_N^\top W_N^\top W_N M_NM_N^\top W_N^\top W_N M_N)\rightarrow _p 0,
\end{align}
as $N\rightarrow \infty$.
Now following the argument of $\mathcal M_{21}$ using Lemma \ref{lemma4}, similar argument to the evaluation of $B$ in \eqref{new.eq115}, as well as formula \eqref{new.197}, \eqref{new.198} and \eqref{new.199}, we have
\begin{align}
	\|\mathcal M_{22}-\kappa_{22,N}\|_{2,\GG_N}=o_p(T^{-1/2}),
\end{align}
which finishes the proof.
\hfill $\Box$
\begin{lemma}\label{lemma2}
	Assume $|\alpha|+|\beta|<1$, $E \epsilon^4_{i,t}<\infty$.  Assume  Assumptions \ref{Graph1}, \ref{Graph2}, \ref{Graph3} and  Assumption \ref{Assumption2.1_2} hold. Then (i) as $N\rightarrow \infty$ and $T\rightarrow \infty$,  we have that
	\begin{align}
		E(|\frac{1}{NT}\sum_{t=1}^T\Z^\top_t\Z_t -\frac{1}{N(T-1)}\sum_{t=2}^T\Z^\top_{t-1}\Z_t-(\Sigma_{2,N}^{-}-\Sigma_{2,N}^+ )|_F^2|\GG_N)\rightarrow_p 0,\\ %
		E(|\frac{1}{NT}\sum_{t=1}^T\Z^\top_t\Z_t-\sum_{t=2}^T\frac{1}{N(T-1)}\Z^\top_{t}\Z_{t-1}-(\Sigma_{2,N}^--\tilde \Sigma_{2,N}^{+} )|_F^2 |\GG_N)\rightarrow_p 0. %
	\end{align}
	(ii)The results of (i) also hold as $T\rightarrow \infty$ with $N$ fixed, if we assume the network is fully connected and not bipartite instead of  Assumptions \ref{Graph1}, \ref{Graph2} and \ref{Graph3}.
\end{lemma}

{\it Proof.}
Elementary calculations show that 
\begin{align}
	\Z_t^\top\Z_t=\begin{pmatrix}
		\sum_{i=1}^N Y_{i,t-1}^2&\sum_{i=1}^NY_{i,t-1}w_i^\top \Y_{t-1}\\
		\sum_{i=1}^N w_i^\top \Y_{t-1}Y_{i,t-1}& \sum_{i=1}^N (w_i^\top\Y_{t-1})^2
	\end{pmatrix},\\
	\Z_{t-1}^\top\Z_{t}=\begin{pmatrix}
		\sum_{i=1}^N Y_{i,t-2}Y_{i,t-1}&\sum_{i=1}^NY_{i,t-2}w_i^\top \Y_{t-1}\\
		\sum_{i=1}^N w_i^\top \Y_{t-2}Y_{i,t-1}& \sum_{i=1}^N w_i^\top \Y_{t-2}w_i^\top\Y_{t-1}
	\end{pmatrix},\\
	\Z_{t+1}^\top\Z_{t}=\begin{pmatrix}
		\sum_{i=1}^N Y_{i,t-1}Y_{i,t}&\sum_{i=1}^NY_{i,t}w_i^\top \Y_{t-1}\\
		\sum_{i=1}^N w_i^\top \Y_{t}Y_{i,t-1}& \sum_{i=1}^N w_i^\top \Y_{t}w_i^\top\Y_{t-1}.
	\end{pmatrix}
\end{align}
Obverse that the lemma  follows from the arguments of \Cref{lemma1}. Details are omitted  for the sake of brevity. \hfill $\Box$
\begin{lemma}\label{fourth}
	Let $X_1,...,X_n$ be $i.i.d$ random variables with mean zero, variance $\sigma_X^2$ and finite fourth moments. Let $X=(X_1,...,X_n)^\top\in \mathbb R^n$. The for any matrix $A=(a_{ij})\in \mathbb R^{n\times n}$, we have
	\begin{align}
		E((X^\top AX)^2)\leq 	E^2(X^\top AX)+(2\sigma_X^4+|EX_1^4-3\sigma_X^4|)tr(AA^\top).
	\end{align}
	If further $A$ is random but independent of $\{X_i\}_{i=1,...n}$, then almost surely
	\begin{align}
		E((X^\top AX)^2|A)\leq 	E^2(X^\top AX|A)+(2\sigma_X^4+|EX_1^4-3\sigma_X^4|)tr(AA^\top).
	\end{align}
\end{lemma}
{\it Proof.} The lemma follows from the proof of Lemma 1 of \cite{zhu2017network}. \hfill $\Box$ 
\begin{lemma}\label{Auxillary}
	For random variables $X_i$, $i\in\mathbb Z$ with finite fourth moments, and a sequence of non-negative numbers $r_i$ such that $\sum_i r_i<\infty$, there exists a constant $M$ such that $E((\sum_i r_iX_i)^4)\leq M \max_i E(X^4_i)$.
\end{lemma}
{\it Proof.} Notice that 
\begin{align}
	(\sum_i r_iX_i)^4  \leq \sum_{i_1}\sum_{i_2}\sum_{i_3}\sum_{i_4}r_{i_1}r_{i_2}r_{i_3}r_{i_4}|X_1X_2X_3X_4|.
\end{align}
Taking expectation on both sides, using Fubini Theorem and Cauchy-Schwartz equality, the Lemma follows. \hfill $\Box$

\begin{proposition}\label{eq84}
	Recall that $\FF_N=\{(\eta_{ij})_{1\leq i<j\leq N}, (C_i)_{1\leq i\leq N}\}$.
	Under conditions of \Cref{Thm3} (i), there exists some large constant $M_0$ such that given $\FF_N$, with probability tending to $1$ that  uniformly for  $1\leq i\leq N, 0\leq t\leq T-1$, 
	\begin{align}
		E( Y_{it}^4|\FF_N)\leq M_0, ~~~E((w_i^\top \Y_t)^4|\FF_N)\leq M_0.
	\end{align}
\end{proposition}
{\it Proof.} By lemma \ref{Auxillary}, it suffices to show $E(Y_{i,t}^4|\FF_N)\leq M_0$ with probability approaching $1$. %
Notice that $Y_{i,t}=e_i^\top Y_t$. Let $C$ be a generic large constant which may vary from line to line. Using Proposition \ref{prop1}, it suffices to show that with probability tending to $1$, for some large constant $C$ (in the following proofs, $C$ is a generic constant which may vary from line to line) uniformly for all $i,t$,
\begin{align}\label{FjN}
	E((e_i^\top \sum_{j=0}^\infty G_1^j\bepsilon_{t-j})^{4}|\FF_N)\leq C, ~~E((e_i^\top(I-G_1)^{-1}G_0)^4|\FF_N)\leq C.
\end{align} 
Via Lemma \ref{lemma4}, we have with probability approaching $1$, for all $i$, $t$,
\begin{align}
	|e_i^\top \sum_{j=0}^\infty G_1^j\bepsilon_{t-j}|\leq
	e_i^\top \sum_{j=0}^\infty| G_1^j|_e|\bepsilon_{t-j}|_e
	\leq \sum_{j=0}^\infty  (|\alpha|+|r^{-1}\beta|)^{j}e_i^\top M_N|\bepsilon_{t-j}|_e,
\end{align}
which implies that with probability tending to $1$, we have for all $i$, $t$,
\begin{align}
	E((e_i^\top \sum_{j=0}^\infty G_1^j\bepsilon_{t-j})^{4}|\FF_N)\leq C E((e_i^\top M_N|\bepsilon_{t-j}|_e)^4|\FF_N).
\end{align}
Notice that for all $j\geq 1$, $W_N^j$ is a transition matrix. Notice that $G_1$ and $M_N$ are fully determined by $W_N$, and also by the definition of $M_N$ in \Cref{lemma4}, the sum of each row of $M_N$ is uniformly bounded over $N$ by a constant only depending  on $r$. Then by Lemma \ref{Auxillary}, almost surely $E((e_i^\top M_N|\bepsilon_{t-j}|_e)^4|\FF_N)\leq C E\epsilon^4_{11}$, which shows the first term of \cref{FjN}. For the second term, recall in the proof of \Cref{diverging N}, we have the following decomposition $G_0={\mf f}_1+{\mf f}_2$, with $
\mf f_1=( f (C_1),...., f (C_N))^\top,~~
\mf f_2=(\gamma^\top U_1,....,\gamma^\top U_N^\top)^\top$. Hence by Jansen's inequality, to show the second term  of \cref{FjN}, it suffices to show for all $i,t$, a.s.,
\begin{align}\label{eq211}
	E((e_i^\top(I-G_1)^{-1}\mf f_1)^4|\FF_N)\leq C, ~~
	E((e_i^\top(I-G_1)^{-1}\mf f_2)^4|\FF_N)\leq C.
\end{align} 
Notice that $(I-G_1)^{-1}=\sum_{j=0}^\infty G_1^j$, By \Cref{Assumption2.1} and \Cref{Assumption2.1_2}, each element of $\mf f_1 $ is bounded, so $a.s.$,
\begin{align}
	E((e_i^\top(I-G_1)^{-1}\mf f_1)^4|\FF_N)
	=(e_i^\top(I-G_1)^{-1}\mf f_1)^4
	= (\mf f_1^\top M_N^\top e_ie_i^\top M_N\mf f_1)^4\\
	\leq C(\mf 1^\top M_N^\top e_ie_i^\top M_N\mf 1 )^2\leq C^5,
\end{align}
where for the first inequality we have used the fact that all the elements in $M_N$ is positive, while for the second inequality we used  the fact that $M_N\mf 1\preceq C\mf 1$. Moreover, to see $E((e_i^\top(I-G_1)^{-1}\mf f_2)^4|\FF_N)\leq C$ holds just notice that for all $i$,
$$
e_i^\top (I-G_1)^{-1}\mf 1=(1-\alpha-\beta),
$$
since it can be verified that $\mf 1$ is the eigenvector of $(I-G_1)^{-1}$ with eigenvalue $1-\alpha-\beta$. Hence given $\FF_N$, 
$e_i^\top(I-G_1)^{-1}\mf f_i$ can be written as
$\sum_{i=1}^Nr_{i,N}\mf f_{2,i}$ with $\sum_{i=1}^N r_{i,N}=1-\alpha-\beta$, where $r_{i,N}$ and $\mf f_{2,i}$ are the $i_{th}$ entry of $e_i^\top(I-G_1)^{-1}$ and $\mf f_2$, respectively. Then by \Cref{Auxillary} and \Cref{Assumption2.1_2}, the second inequality of \Cref{eq211} holds, which finishes the proof. \hfill $\Box$\\

\begin{proposition}\label{Conditional-Constant}
	Under conditions of (ii) of  \Cref{Thm3}, then as $\min(N,T)\rightarrow \infty$,  we have that
	\begin{align}
		E((\hat \Gamma_{2,2,\alpha}-E(\hat \Gamma_{2,2,\alpha}|\GG_N))^2|\GG_N)\rightarrow_p 0,\label{Claim1}\\
		E((\hat \Gamma_{2,2,\beta}-E(\hat \Gamma_{2,2,\beta}|\GG_N))^2|\GG_N)\rightarrow_p 0,\label{Claim2}
	\end{align}
	where $\hat \Gamma_{2,2,\alpha}$ and $\hat \Gamma_{2,2,\alpha}$ are defined in the proof of \Cref{Thm5}.
\end{proposition}
{\it Proof.} 
We first show \eqref{Claim1}.  Notice that
\begin{align}
	\hat \Gamma_{2,2,\alpha}-E(	\hat \Gamma_{2,2,\alpha}|\GG_N)=
	\frac{1}{N(N-1)r_N}\sum_{i=1}^N\sum_{j=1}^N(U_i-U_j)(e_i-e_j)^\top K(\frac{\hat \delta_{ij}^2}{h_N})\sum_{s=0}^\infty G_1^s\sum_{t=1}^T\bepsilon_{t-1-s}/T.
\end{align}
Let
\begin{align}
	\hat \Gamma_{2,2,\alpha,1}=\frac{1}{N(N-1)r_N}
	\sum_{i=1}^N\sum_{j=1}^N K(\frac{\hat \delta_{ij}^2}{h_N})U_ie_i^\top \sum_{s=0}^\infty G_1^s\sum_{t=1}^T\bepsilon_{t-1-s}/T,\\
	\hat \Gamma_{2,2,\alpha,2}=\frac{1}{N(N-1)r_N}
	\sum_{i=1}^N\sum_{j=1}^N K(\frac{\hat \delta_{ij}^2}{h_N})U_ie_j^\top \sum_{s=0}^\infty G_1^s\sum_{t=1}^T\bepsilon_{t-1-s}/T.
\end{align}
Then it suffices to show that
\begin{align}
	|E(\hat\Gamma_{2,2,\alpha,1}\hat\Gamma_{2,2,\alpha,1}^\top|\GG_N)|_F\rightarrow_p 0,~~~
	|E(\hat\Gamma_{2,2,\alpha,2}\hat\Gamma_{2,2,\alpha,2}^\top|\GG_N)|_F\rightarrow_p 0.
\end{align}
Recall from the proof of \Cref{Thm3} (ii) that $r_N>0$.	Let $\kappa_i=\sum_{j=1}^NK(\frac{\hat\delta_{ij}^2}{h_N})/((N-1)r_N)$. Then 
\begin{align}
	\hat \Gamma_{2,2,\alpha,1}=\frac{1}{N}\sum_{i=1}^N \kappa_iU_ie_i^\top \sum_{s=0}^\infty G_1^s\sum_{t=1}^T\bepsilon_{t-1-s}/T.
\end{align}
Elementary calculations show that
\begin{align}
	\tilde \Sigma:=E ((\sum_{s=0}^\infty G_1^s\sum_{t=1}^T\bepsilon_{t-1-s}/T)(\sum_{s=0}^\infty G_1^s\sum_{t=1}^T\bepsilon_{t-1-s}/T)^\top|\GG_N)\\=\sigma^2 \sum_{l=0}\sum_{s=0}G_1^s(G_1^l)^\top (T-|l-s|)\mf 1 (|s-l|\leq T)/T^2.\end{align}
Then by \Cref{lemma4}, there exists  a constant $M_0$ independent of $N$ such that
$|\tilde \Sigma|_e	\preceq  M_0M_NM_N^\top/T$. Using this fact, and the fact that $\kappa_i$ is bounded, we have that
\begin{align}\label{circlestar}
	|	E(\hat\Gamma_{2,2,\alpha,1}\hat\Gamma_{2,2,\alpha,1}^\top|\GG_N)|_e	\preceq  \frac{M_0'}{N^2T}\sum_i\sum_j|U_i|_ee_i^\top M_NM_N^\top e_j|U_j|_e^\top
\end{align}
for some sufficiently large constant $M_0'$. Consider the $(q,l)_{th}$, $1\leq q,l\leq p$, element of the matrix in the  RHS of the above inequality, which can be written as
\begin{align}\label{circ0}
	\frac{M_0'}{N^2T}\sum_{i=1}^N\sum_{j=1}^N|U_{i,q}||U_{j,l}|e_i^\top M_NM_N^\top e_j:=\frac{M_0'}{N^2T}\tilde U_q^\top M_NM_N^\top \tilde U_l,
\end{align}  
where $\tilde U_q=(U_{1,q},...,U_{N,q})$. Observe that
for $1\leq q,l\leq p$,  by \Cref{Assumption2.1} and \Cref{Assumption2.1_2},
\begin{align}
	|\tilde U_q\tilde U_l^\top|_F/N^2=O_p(1).
\end{align}
By the inequality $tr(A^\top B)\leq |A|_F|B|_F$ and \Cref{Graph3} we have as $\min(N,T)\rightarrow \infty$,
\begin{align}
	\frac{\tilde U_q^\top \mf 1 \bar \pi \bar\pi^\top \mf 1^\top \tilde U_l}{N^2T}\rightarrow_p 0,~~~~
	\frac{|\tilde U_q\tilde U_l|_F}{N^2T}|\sum_{j=0}^{K_N}r^j W_N^j|_F^2=o_p(1),
\end{align}
where  $K_N=\lceil C_0\log^{1+\alpha_0} N\rceil$.  Recall in  \Cref{lemma4} the quantity $M_N=2\mf 1 \bar \pi_N+\sum_{j=0}^{K_N}r^jW_N^j$.
Using  the fact that $2|a^\top b|\leq a^\top a+b^\top b$ for any vectors $a,b$ with same  length, we shall see that
\begin{align}\label{circ1}
	\frac{1}{N^2T}\tilde U_q^\top M_NM_N^\top \tilde U_l\rightarrow_p 0.
\end{align}
Together with \eqref{circlestar} and \eqref{circ0} we show that 	$|E(\hat\Gamma_{2,2,\alpha,1}\hat\Gamma_{2,2,\alpha,1}^\top|\GG_N)|_F\rightarrow_p 0$. Define 
\begin{align}
	\tilde \kappa_j=\frac{\sum_{i=1}^NU_iK(\frac{\hat \delta_{ij}^2}{h_N})}{(N-1)r_N}.
\end{align}
Then 	\begin{align}
	\hat \Gamma_{2,2,\alpha,2}=\frac{1}{N}\sum_{j=1}^N\tilde  \kappa_je_j^\top \sum_{s=0}^\infty G_1^s\sum_{t=1}^T\bepsilon_{t-1-s}/T.
\end{align}
Following the argument leading to \eqref{circlestar} we shall see that
\begin{align}
	|E(\hat\Gamma_{2,2,\alpha,2}\hat\Gamma_{2,2,\alpha,2}^\top|\GG_N)|_e\preceq \frac{M_0'}{N^2T}\sum_{i=1}^N\sum_{j=1}^N |\tilde \kappa_i|e_i^\top M_NM_N^\top e_j|\tilde \kappa_j|^\top
\end{align}
for some large constant $M_0'$ independent of $N$. Let $\tilde \kappa_{iq}$ be the $q_{th}$ ($1\leq q\leq p$) element of 
$\tilde \kappa_i$. Then  for $1\leq q\leq p$,
$\max_{1\leq j\leq N}|\tilde \kappa_{jq}|\leq C_0 \frac{\sum_{i=1}^N|U_{i,q}|}{N}$ for some large constant $C_0$, which leads to that for $1\leq q,l\leq p$, the $(q,l)_{th}$ entry of $|E(\hat\Gamma_{2,2,\alpha,2}\hat\Gamma_{2,2,\alpha,2}^\top|\GG_N)|_e$ is bounded by
\begin{align}
	&	\frac{M_0'}{N^2T}\sum_{i=1}^N\sum_{j=1}^N |\tilde \kappa_{iq}|e_i^\top M_NM_N^\top e_j|\tilde \kappa_{jl}|^\top\\&\leq C_1(\frac{1}{N^2T}\sum_{i=1}^N\sum_{j=1}^N e_i^\top M_NM_N^\top e_j)(\frac{\sum_{i=1}^N|U_{iq}|}{N})(\frac{\sum_{i=1}^N|U_{il}|}{N})=o_p(1)
\end{align}
for some large constant $C_1$, where for $\leq$ we have used \Cref{lemma4}. Hence we show that
$$|E(\hat\Gamma_{2,2,\alpha,2}\hat\Gamma_{2,2,\alpha,2}^\top|\GG_N)|_F\rightarrow_p 0$$ and \eqref{Claim1} follows. Observing that $w_i^\top M_NM_N^\top w_j$ is the $(i,j)_{th}$ entry of $W_NM_NM_N^\top W_N^\top$. Using similar argument to the proof of \eqref{circ1} and the fact that 
$W_NM_N=2\mf 1 \bar \pi_N+\sum_{j=0}^{K_N}r^jW_N^{j+1}$, we shall see that  for $1\leq q,l\leq p$,
$$\frac{1}{N^2T}\tilde U_q^\top W_N M_NM_N^\top W_N^\top \tilde U_l\rightarrow_p 0.$$
Therefore the assertion \eqref{Claim2} follows similarly.  Hence the proposition holds. \hfill $\Box$

\end{document}